# ABSTRACT

DICKSON, DAVID HARPER. Quantifying Roche Lobe Overflow in the Formation of Merging Black Hole Binaries. (Under the direction of John Blondin).

We demonstrate a new methodology to model Roche lobe overflow (RLO) systems to unprecedented resolution simultaneously across the envelope, donor wind, tidal stream, and accretion disk regimes without reliance upon previously-universal symmetry, mass flux, and angular momentum flux assumptions. We have applied this method to the semi-detached high-mass X-ray binary (HMXB) M33 X-7 in order to provide a direct comparison to recent observations of an RLO candidate system at two overflow states of overfilling factors $f = 1.01$ and $f = 1.1$. We found extreme overflow ($f = 1.1$) to be entirely conservative in both mass and angular momentum transport, forming a conical L1 tidal stream of density and deflected angle comparable to existing predictions. This case lies within the unstable mass transfer (MT) regime as recently proposed of M33 X-7. The $f = 1.01$ case differed in stream geometry, accretion disk size, and efficiency, demonstrating non-conservative stable MT through a ballistic uniform-width stream. The non-conservative and stable nature of the $f = 1.01$ case MT also suggests that existing assumptions of semi-detached binaries undergoing RLO may mischaracterize the parameter space of stability for the L1 flow. We also conducted a piecewise evolution of M33 X-7 across nearly the entirety of RLO phase from onset to runaway unstable MT. We present the first model of the efficiency of MT and its associated angular momentum across overfilling factor parameter space. We also present novel relations for binary separation, mass ratio, L1 mass transfer rate, and Roche timescale as they vary with respect to the overfilling factor. These provide constraints on the threshold for the onset of unstable MT which ultimately led in our system to exponentially faster MT evolution. Collectively these parameter constraints, relations, and explorations probe RLO dynamics in HMXBs and their role and distribution as progenitors of binary black holes (BBHs) and common envelopes (CEs).

Quantifying Roche Lobe Overflow in the Formation of Merging Black Hole Binaries

by
David Harper Dickson

A dissertation submitted to the Graduate Faculty of
North Carolina State University
in partial fulfillment of the
requirements for the Degree of
Doctor of Philosophy

Physics

Raleigh, North Carolina
2024

APPROVED BY:

_________________________  _________________________
Rongmon Bordoloi            Carla Fröhlich

_________________________  _________________________
Subhashis Ghoshal           John Blondin
External Member             Chair of Advisory Committee

# DEDICATION

To my wife, Keri, without whom this work would never have been possible, and my cat, Thesis, for whom this work is named.

# BIOGRAPHY

My interest in theoretical astrophysics solidified when an NC State faculty member spoke to my second grade class over twenty years ago. I had always been fascinated by questions about the workings of the universe and its constituent parts. Before their brief speech, which was one among many on that career day, I had never had a word for that curiosity. After, I knew clearly that I wanted to become a theoretical astrophysicist, even if I couldn't pronounce it yet. It is fitting then, that it is in this same institution that I complete the highest degree in that pursuit.

    As I grew, my questions grew in even greater proportion. The complex interconnections of the universe offer me no shortage of questions to explore, and for that I am grateful. Let this stand not only as testament to my work thus far but also my intent to continue asking questions of the universe and sharing what answers I may find.

# ACKNOWLEDGEMENTS

We are grateful for the guidance and mentorship of John Blondin. We acknowledge support from NSF grant AST-2308141. We thank our dissertation committee members Rongmon Bordoloi, Carla Frohlich, and Subhashis Ghoshal as well as the wider research community of North Carolina State University for their support in the creation of this work.

# TABLE OF CONTENTS

# LIST OF TABLES

# LIST OF FIGURES

# CHAPTER 1

# BACKGROUND

This chapter integrates and adapts some work previously published in Dickson (2024) in Sections 1.5.1, 1.5.2, and 1.5.3.

## 1.1 Black Hole Formation and Evolution

Black holes (BHs) are infinitely dense astrophysical objects of potentially varied origins, most often categorized by their masses. Stellar mass BHs are supernova remnants, forming from the core collapse of massive stars at the end of their life cycle with masses of $\lesssim 10 M_\odot$, where $M_\odot$ is the mass of the sun. The second mass category, intermediate mass BHs, is underrepresented in observational data; the unexpected elusiveness of BHs in the range of $100 - 10^5 M_\odot$ may require close constraints of population and distribution to solve (Greene et al. 2020). Supermassive BHs with $\gtrsim 10^6 M_\odot$ are more easily detected, often found in the bright, dynamic centers of galaxies.

Supermassive BHs sometimes drive such intense consumption of nearby matter that they form so-called active galactic nuclei. In the same way, though on a lesser scale, the consumption of nearby matter may allow a stellar mass BH to grow to intermediate mass, and on to the supermassive regime. This evolution pathway may entirely account for the population of BHs we see today, but the observational scarcity of intermediate mass BHs and the drastic range of masses involved may suggest disparate origins (Greene et al. 2020).

Some BHs observed today may have alternative origins from the high-energy early universe before the existence of stars. These primordial BHs and their characteristics and distribution are as-of-yet theoretical. With further research, they may reveal distinct origins of intermediate mass and supermassive BHs from those of stellar mass, as well as BHs of unobserved mass regimes (Carr et al. 2021). The properties, number, and interaction channels of primordial BHs are as of yet unconstrained by observation. Nor are they narrowly constrained by theoretical work, which limits the impact of detailed examinations of primordial BHs (Carr et al. 2021). We instead focus on the best-defined BH formation pathway, that of stellar core collapse.

Not all stars generate BHs; only massive stars that undergo a core collapse can leave a BH remnant (Mapelli 2020). A BH can be formed either by the stalling of the core collapse shockwave, which leaves the star unable to launch its outer envelope, or by significant envelope mass falling back onto the system after a successful supernova following core collapse (Kochanek 2014; Mapelli 2020). While the exact mechanisms of core collapse are the subject of ongoing research, the conditions under which the process begins are more clearly defined. For generating a BH remnant through core collapse to be possible, a star must be of mass $\gtrsim 8 M_\odot$ with an evolved core. Core collapse, whether accompanied by a supernova or a failed nova, can only generate stellar mass BHs (Kochanek 2014). To satisfy the observed range of BHs, we must also examine how they evolve after formation.

Stellar mass BHs are able to grow across astrophysical timescales by consuming stellar material and merging with other BHs. The consumption of stellar material by BHs occurs through accretion, in which the outer layers of the star are peeled away to form an accretion disk around the BH. Gradually, mass is drawn in through the disk to grow the BH. Though the BH itself does not emit light, this accretion process can be sufficiently energetic to generate bright X-ray emission. In some cases, this process may consume the entirety of the host star. BH-BH mergers occur when two BHs become gravitationally-bound into a binary system, then spiral into their common center to form a single larger BH. Despite the high energy of such a collision, these merger events do not produce much radiation; instead, they are most often detected through gravitational waves. These binaries are therefore crucial to understanding the landscape and distribution of gravitational waves and intermediate mass BHs, as well as the validity of primordial origination to the observed supermassive BH population.

## 1.2 Binary Black Holes

Binary black holes (BBHs) are gravitationally-bound systems of two BHs, often distinguished by their masses and orbital separations. The observed population suggests most merging BBHs exhibit total system masses of $4-40 M_\odot$ (Kochanek 2014; Talbot and Thrane 2018). Though higher mass BBHs are more readily detectable, the mass range of stellar BH progenitors suggests this range may be representative of the overall merging population (Talbot and Thrane 2018). Simulated stellar populations suggest, however, that the non-merging population may extend as high as $130 M_\odot$ (Mapelli 2020). The merging population of BBHs arises from so-called close BBHs, which have relatively short initial binary separations at time of formation ($< 100 R_\odot$, where $R_\odot$ is the radius of the sun). BBHs require very long timescales to dissipate sufficient angular momentum to reduce their binary separation and merge. BBHs that form with initial binary separations of $\gtrsim 100 R_\odot$ will not merge within a Hubble time and are not expected to impact merger observations (Mapelli 2020).

BBHs provide a window into the world of gravitational wave astronomy and may be key to solving mysteries as varied and elusive as intermediate mass BHs (Neilsen et al. 2018; Fragione and Bromberg 2019; Greene et al. 2020), BH topology (Bohn et al. 2016; Astorino et al. 2022), and dark matter (Raccanelli et al. 2016; Wang et al. 2018; Scelfo et al. 2018). In isolation of these topics, BBHs alone provide the distinct observational data needed to refine our understanding of gravitational waves and their detection (Abbott et al. 2016, 2019).

In order to best utilize BBH observations to expand our understanding of these varied fields, it is essential to constrain and characterize the population of BBH systems and their distribution. Only by comparison between the observed and theoretical populations can further theoretical quantities (such as the contribution of primordial BHs to the BBH population) be substantially validated (Talbot and Thrane 2018; Carr et al. 2021). To improve our understanding of these systems and the waves they generate, we must therefore critically examine the formation processes and progenitors of BBHs (Marchant et al. 2021; Dorozsmai and Toonen 2022).

However, present day BBH research currently lacks a solid framework with which to define the origins of these systems, or the formation processes involved (Belczynski et al. 2016; Raccanelli et al. 2016; Hainich et al. 2018; Scelfo et al. 2018). The most direct progenitor pathway would be gravitational capture, the method by which a BH catches a nearby star or fellow BH in its cluster. It is unclear how common such a pathway can be, even in very dense clusters. While other progenitor pathways have not been ruled out, the isolated evolution of massive binary star systems forms the most promising progenitor (Marchant et al. 2021; Ivanova et al. 2013; Belczynski et al. 2016; Bavera et al. 2021; Hainich et al. 2018; Scelfo et al. 2018; Dorozsmai and Toonen 2022; Bunzel et al. 2023). Even in gravitational capture of a star, the then binary system must still undergo the same subsequent evolution as a system arising from an isolated stellar binary. We therefore examine binary star systems as the primary progenitors of observed merging BBHs.

## 1.3 Binary Star Systems as BBH Progenitors

Binary star systems, much like BBHs themselves, are gravitationally-bound systems of two stellar mass objects. Binary stars predominantly form in isolation as two stars and exhibit a range of total system masses, mass ratios, and binary separations. Given the variety of binary stars, many such systems cannot form BBHs. The system must be of sufficient mass to induce core collapse on both stars, leaving two BHs as stellar remnants. Of those binaries of sufficient mass, only close binaries may contribute to the population of merging BBHs. Close binaries are not defined by a singular binary separation threshold, as stellar radius varies in the regime of $10-10^4 R_\odot$. Instead, close binaries are defined as those binaries of sufficiently short binary separation for mass exchange to significantly influence the evolution of the individual stars (Mapelli 2020). Some very close binaries have been observed in which one star is massive enough to generate a BH through core collapse and its companion star has already collapsed into a BH visible through its X-ray luminous accretion process (Ramachandran et al. 2022).

These most promising progenitors of merging BBHs in current research are therefore called high mass X-ray binaries (HMXBs) (Bunzel et al. 2023; Dorozsmai and Toonen 2022). HMXBs are close binaries of high mass with an X-ray luminous compact object companion, such as a BH. These systems are so close that they enter the semi-contact regime, in which the binary separation is short enough that massive star's stellar atmosphere begins to escape and transfer onto the accreting companion (Frank et al. 2002; Ivanova et al. 2013).

When a star is in the proximity of another massive object such as a BH, it is no longer the singular endpoint of infalling material. To determine the dynamical effects such a system imposes on nearby material, the space surrounding both objects may be mapped by an effective gravitational potential. As both companions are moving, this potential would vary with time if defined in any inertial frame. It is therefore beneficial to define this potential in the non-inertial frame of reference that rotates with the binary, called the corotating frame. This enables us to consider both the star and its companion to be stationary, and the effective potential of any given point to be constant. We may further improve this effective potential by the use of a relevant offset.

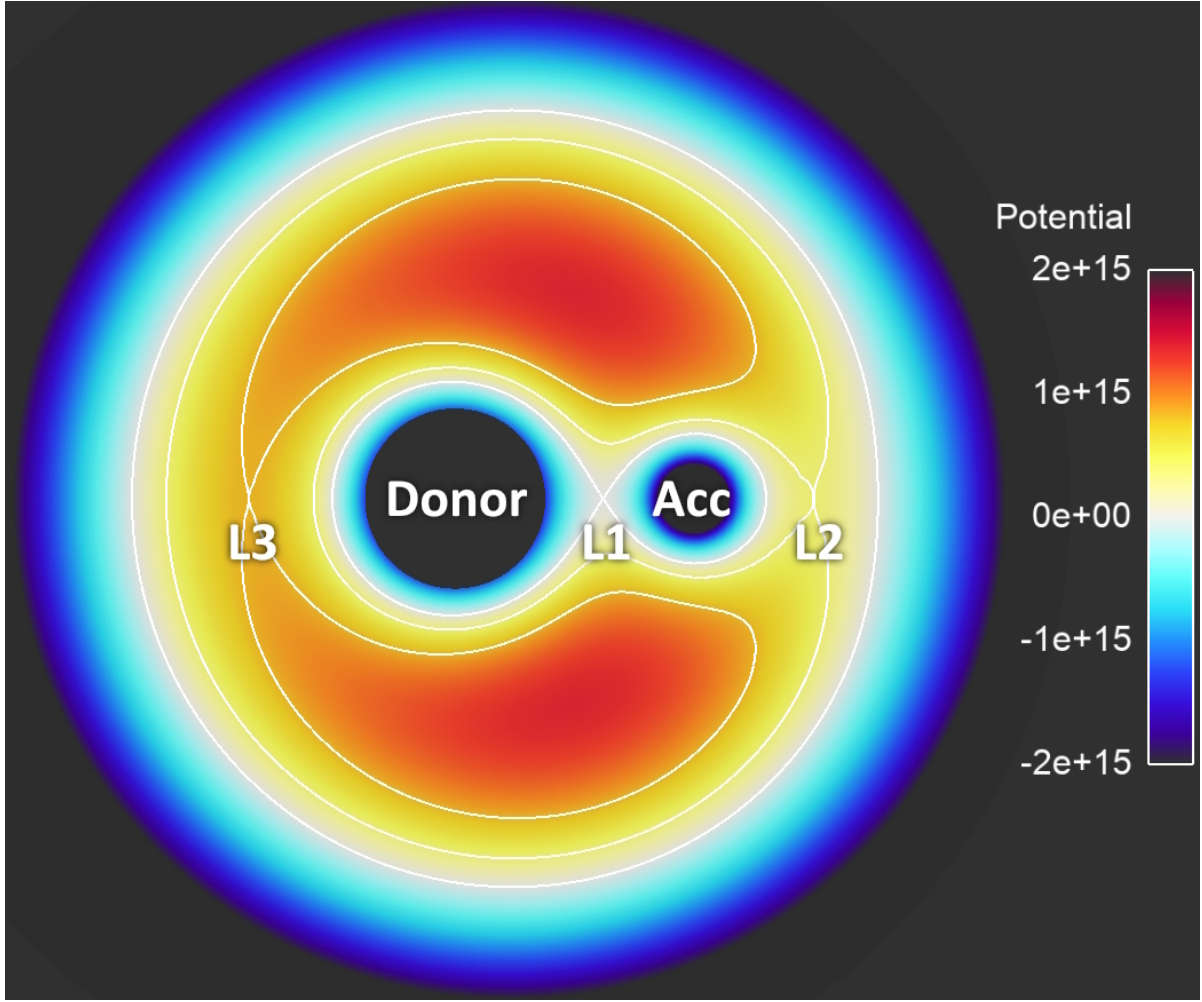

**Figure 1.1** Roche Equipotential Contours. The equatorial contours of effective gravitational potential in the corotating frame for a system of $M_{\text{donor}} = 38.0 M_\odot$ and $M_{\text{acc}} = 11.4 M_\odot$ such as M33 X-7, further described in Section 3.4. These contours are mapped following Equation 1.1, whereby we set $\phi_{\text{L1}} = 0$ as a potential reference. By convention, the Lagrange point beyond the less massive companion is named L2, while L3 is beyond the more massive companion. In systems with more massive accretors than donors, the labeling would be correspondingly reversed.

The natural choice of reference potential is that of the Lagrange point L1. Lagrange points are the inflection points of the effective gravitational potential, shown in Figure 1.2. The Lagrange point L1 denotes the point at which gravitational attraction to two stellar objects is directly balanced against each other and lies on the line connecting their centers of masses. An equipotential contour drawn through the L1 point reveals the bound regions of both companions, called Roche lobes. By use of L1 as a reference potential, we may therefore define the gravitationally-bound region by an effective potential using $M_{\text{donor}}$ and $M_{\text{acc}}$ for the masses of the donor and accretor. This takes the form

$$\phi_{\text{eff}} = - \frac{GM_{\text{donor}}}{R_1} - \frac{GM_{\text{acc}}}{R_2} - \frac{\Omega^2 R_3^2}{2} - \phi_{\text{L1}} \quad , \tag{1.1}$$

where $R_1$, $R_2$, and $R_3$ are the displacements from a specified point to the donor, accretor, and center of mass respectively. The variable $\Omega$ denotes the orbital angular velocity of the rotating system and $G$ is the gravitational constant. The effective potential of the Roche lobes is given by $\phi_{\text{L1}}$, which we take as an offset such that $\phi_{\text{eff}} < 0$ defines the gravitationally-bound region. Within an astrophysical object's Roche lobe, all mass moving at lesser than its escape velocity is bound to the gravity well of that object. Matter in the regime outside either Roche lobe ($\phi_{\text{eff}} > 0$) could variously fall into either object's lobe, co-orbit them both, or escape, depending on its momentum. The stellar atmospheres, and sometimes deeper layers, of massive stars in semi-contact binaries fill their Roche lobes completely and begin to overflow (Hjellming and Webbink 1987; Frank et al. 2002).

## 1.4 Roche Lobe Overflow

The process by which a star in a semi-contact binary exceeds its Roche lobe and overflows onto its companion is called Roche lobe overflow (RLO). Some of the mass lost by the overflowing star falls into the Roche lobe of the companion and is accreted. Therefore, we adopt the terminology of referring to the overflowing star as the donor (or "donor star") and referring to its companion as the accretor. Any companion of a donor star can be considered an accretor, such as a neutron star or smaller main sequence star (Beccari and Boffin 2019). However, the progenation of BBHs from a binary star system requires either that the accretor independently collapse into a BH prior to entering semi-contact, or that the accreted mass accelerates it to that end state. We will primarily discuss the accretor in the context of a BH companion. Regardless, the physics of the system are much the same with any compact object accretor during the RLO of its donor companion.

A star can enter RLO and become a donor either by expansion of its radius or contraction of its Roche lobe. Stellar radii expand based on evolutionary stage, which leads to rapid transitions at the end of each nuclear burning cycle (Kippenhahn et al. 1990; Passy et al. 2012; Quast et al. 2019). Loss of angular momentum through stellar wind can initiate RLO by bringing a system into sufficiently close proximity. This proximity lowers the effective gravitational potential of L1, shrinking the Roche lobes of both companions proportionally to the shrinking binary separation between them (Eggleton 1983).

As a star nears its Roche lobe, it is often quantified in terms of its filling factor, $f$. A stellar object's filling factor represents its radius in comparison to the radius of its Roche lobe. Near a Roche lobe, objects are no longer spherical, and thus the choice of how to define radius becomes critical and non-obvious. For comparison to models of simpler geometries, volume-equivalent radii may be the most informative, and is generally preferred (Eggleton 1983; Marchant et al. 2021). There does not exist an exact analytical solution for the volume of a Roche lobe, but the approximation of Eggleton (1983) is often used. This defines the volume-equivalent radius of a Roche lobe, $R_{\mathrm{RL,vol}}$, as

$$R_{\mathrm{RL,vol}} = d \frac{0.49 q^{2/3}}{0.6 q^{2/3} + \ln(1 + q^{1/3})} \quad . \tag{1.2}$$

The variable $d$ represents the binary separation described by

$$d = \sqrt[3]{\frac{G(M_{\mathrm{donor}} + M_{\mathrm{acc}})}{\Omega^2}} \quad , \tag{1.3}$$

and $q$ is the mass ratio, which can be expressed as

$$q = \frac{M_{\mathrm{donor}}}{M_{\mathrm{acc}}} \quad . \tag{1.4}$$

The Roche lobe radius of the accretor can be similarly obtained by inverting the definition of $q$, though that is rarely needed for compact object accretors.

Filling factors are normalized such that $f(R_{\mathrm{RL}}) = 1$ by definition. Therefore, using volume-equivalent radii, we obtain

$$f = \frac{R_{\mathrm{donor,vol}}}{R_{\mathrm{RL,vol}}} \quad \text{where } f \leq 1 \quad . \tag{1.5}$$

The $f < 1$ regime is defined by wind accretion. Wind accretion is the process by which some of the optically thin stellar wind generated by the donor falls into the accretor Roche lobe. Donor wind is lost in all directions, though this mass loss is not perfectly spherically symmetric, even for donors well inside their Roche lobes. Due to the rotation of the primary, wind density and velocity vary by angle from the ecliptic plane.

The accretor then captures some of the donor wind through the process of Bondi-Hoyle-Lyttelton accretion (Edgar 2004). Bondi-Hoyle-Lyttelton accretion defines a characteristic accretion radius $R_a$ for an accretor in a medium of otherwise uniform density $\rho$ and relative velocity $v_\infty$ by

$$\dot{M}_{\text{acc}} = \pi R_a^2 v_\infty \rho \quad \text{where} \quad R_a \equiv \frac{2GM_{\text{acc}}}{v_\infty^2} \quad . \tag{1.6}$$

This accretion radius is not the Schwarzschild radius of a BH accretor, but rather defines the region within which material of velocity $v_\infty$ becomes captured by the accretor's gravitational field. This method is therefore independent of accretor type and generally applicable to a number of systems, including close binaries. The accretor gains mass through this process at a rate $\sim 10^{-9} M_\odot$/yr, though the exact rate varies by system geometry (Davidson and Ostriker 1973). As $f$ approaches 1, however, wind accretion becomes predominated by a new process, wind RLO.

In the $f \lesssim 1$ regime, the lower boundary of RLO abuts the related process of wind RLO (Mohamed and Podsiadlowski 2007). Wind RLO occurs when only the donor star's optically-thin atmosphere exceeds its Roche lobe, driving low-density stellar wind towards L1 and ultimately onto the accretor. First proposed by Mohamed and Podsiadlowski (2007), this process is generally considered a stable alternative to full RLO, particularly in systems with wide binary separations. Systems undergoing wind RLO are expected to experience mass fluxes of $\dot{M}_{\text{donor}} \leq 10^{-6} M_\odot$/yr across long timescales without inducing as significant an impact on donor evolution as full RLO (El Mellah et al. 2019). By conventional terminology, the wind RLO regime does not extend to $f \geq 1$, where full RLO is considered to take its place (Mohamed and Podsiadlowski 2007).

To describe systems undergoing RLO, we must therefore add to the definition given in Equation 1.5 to span the $f > 1$ regime. In this regime we switch smoothly to mapping the aspherical effective gravitational potential contours of the donor region by their overfilling factor, also denoted by $f$.

We must define characteristic radii carefully in the overfilling regime. While volume-equivalent radii are generally favored for radii at or below the Roche potential, this breaks down for donors in the RLO phase. The region bounded by an overflowing equipotential also includes the accretor Roche lobe, though an overflowing donor does not generally fill the entirety of its companion lobe. A choice of truncation improves the volume-equivalent radius approach only as long as the donor does not exceed its outer Lagrange point, at which point a second truncation in $f$ would be needed (Marchant et al. 2021).

We instead chose to define the donor surface and Roche lobe by their eclipse radii, $R_{\text{donor}} = R_{\text{donor,ecl}}$. We determine eclipse radius with the equipotential surface truncated by the plane passing through the L1 point orthogonal to the line connecting the centers of masses of the system. Without this choice of truncation, any RLO equipotential surface of $R_{\text{donor}} > R_{\text{RL,ecl}}$ fully encloses the accretor, though this equipotential is never completely filled in RLO phase. We thus define overfilling factors by

$$f = \frac{R_{\text{donor,ecl}}}{R_{\text{RL,ecl}}} \quad \text{where } f \geq 1 \quad . \tag{1.7}$$

There exists a curvature discontinuity to address in this definition of overfilling factor. Beyond a certain point, the surface truncated at L1 becomes sufficiently wide that it eclipses the remainder of the donor star, as seen by the accretor sightline. This eclipse inflection, shown in Figure 1.2, represents the point at which an overflowing donor becomes completely eclipsed by the intervening flow of material out through its L1 point. This flow, called the tidal stream, may remain optically thick well beyond L1 at sufficient overfilling factors. If this is the case, the eclipse inflection may occur earlier than the value presented here. The $f$-value of the eclipse inflection depends on system geometry and may mislead observational analyses which take eclipse radius to directly relate to a photometric stellar radius.

Donors undergoing RLO differ substantially from isolated stellar models in more ways than eclipse radius, however. The RLO phase of a donor star is vital to the evolution of close binary systems and can determine the resultant novae, mergers, and/or remnants those systems produce (Lubow and Shu 1975; Pavlovskii et al. 2016; Marchant et al. 2021; Bunzel et al. 2023). Duchêne and Kraus (2013) found at least 80% of O-type stars have one or more stellar companions. Sana and Evans (2010) found most of the surveyed O-type stars in binaries were in short-period orbits ($T_{\text{orbit}} < 100$ days). More specifically, Sana et al. (2012) found at least 60% of O-type stars have at least one companion close enough to interact by RLO. Therefore, the majority of O-type stars may be close enough that RLO or wind RLO (in which only the stellar atmosphere overflows the Roche lobe) could influence their lifespans, novae, and remnants (Ivanova et al. 2013).

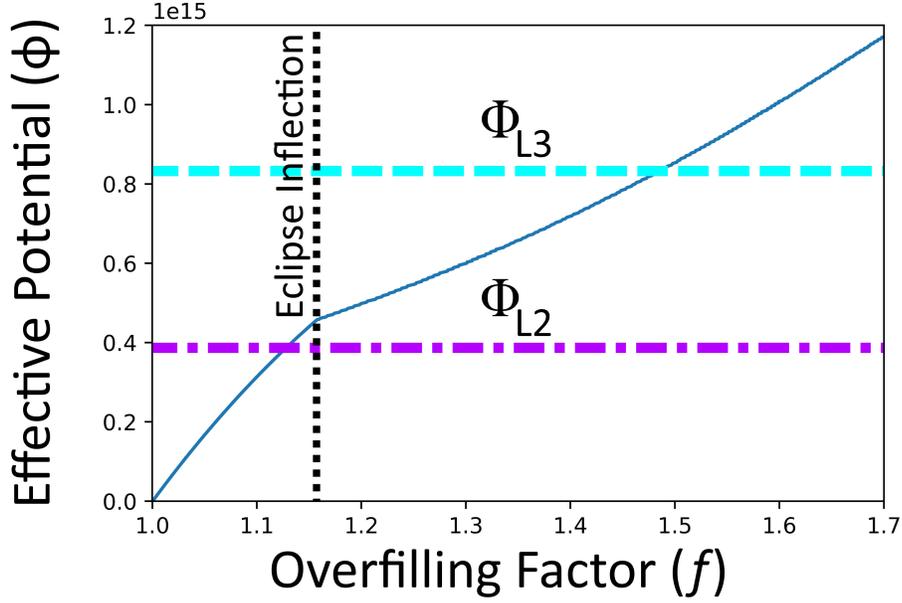

**Figure 1.2** Eclipse Potential vs. Overfilling Factor. Values taken for M33 X-7 with $M_{\mathrm{donor}} = 38.0 M_\odot$, $M_{\mathrm{acc}} = 11.4 M_\odot$, and $d = 35 R_\odot$, further described in Section 3.4. Potentials are derived from Equation 1.1, taking $\phi_{\mathrm{L1}} = 0$. An overflowing donor of $f \geq 1.127$ surpasses the L2 potential, and one of $f \geq 1.482$ overflows its L3 point. Beyond the eclipse inflection ($f = 1.158$), the truncated surface defines the eclipse surface.

The timescale of RLO-induced mass transfer onto BH accretors is expected to be very brief due to their high rate of mass transfer (Lubow and Shu 1975; Marchant et al. 2021; Ramachandran et al. 2022). This brevity limits observational data of this vital progenitor process.

## 1.5 Mass Transfer

Mass transfer (MT) is the loss of material from the donor star onto the accretor, such as during the RLO or wind RLO phases. In binary systems, MT is the primary and most consequential means of material interaction between the donor and accretor outside of merger events.

During true RLO, dense outflow is predicted to be funneled into a narrow columnar emission through the L1 Lagrange point by the tidal effect of the accretor (Lubow and Shu 1975). This dense MT emission is referred to as a tidal stream. The barely overflowing $f \sim 1$ case may see greater MT flux through wind overflow rather than the dense stream; however, the L1 tidal stream predominates MT in the high overflow regime. In extreme overflow, a secondary MT stream may form at the outer Lagrange point as well (Marchant et al. 2021). This secondary stream is presently theoretical but speaks to the complexity of MT dynamics at work in RLO systems.

To address the complex transition of a binary star system into a BBH without computationally intensive multidimensional simulations, assumptions must be made of its MT dynamics. These have often a priori assumed the stability and timescale of the MT process, the MT efficiency from the donor star to its companion, the transfer fluxes of angular momentum in the system, the rate and thus timescale of MT, and the abundance of systems undergoing MT (Bunzel et al. 2023; Lubow and Shu 1975; Marchant et al. 2021; Frank et al. 2002). Reliant upon these assumptions, theoretical and computational models of the overflow MT process have so far been unable to effectively constrain many key parameters.

Simulation-driven physical constraints on the MT process can therefore refine models and aid observation across varied fields. Constraints and validation of theoretical models are needed to address the existing assumptions made of the stability and efficiency of MT as well as the angular momentum carried by MT processes.

### 1.5.1 Mass Transfer Stability

Despite its evolutionary relevance, the RLO MT rate has not yet been constrained as a function of donor radius or the associated overfilling factor $f$. The rate, and therefore timescale, of RLO-induced MT is not only an observational limitation but an unknown parameter of its own (Frank et al. 2002). The MT rate is expected to vary by its stability, which is critical to the subsequent multi-phase evolution of close binary systems (Pavlovskii et al. 2016; Dorozsmai and Toonen 2022; Marchant et al. 2021). Recent work has begun to relate both MT rate and stability to donor radius through spherically symmetric 1D radial simulations and analytical methods (Marchant et al. 2021). Existing work subdivides MT stability into two general regimes, stable and unstable.

Stable MT, characterized by an expanding or constant donor Roche lobe, describes the regime in which net negative feedback occurs in MT. This causes the donor star to be stripped of only its outer envelope, leaving the donor near its Roche lobe throughout a long evolutionary period and potentially forming a BH. Unstable MT describes the opposite case, in which net positive feedback strips ever more of the expanding star away from its diminishing Roche lobe (Frank et al. 2002; Pavlovskii et al. 2016; Marchant et al. 2021).

The stability of MT is determined by the sign and rate of change of the donor's Roche lobe radius. This rate of change is most accurately calculated directly from the generalized analytic formula given in Equation 1.2 (Eggleton 1983; Frank et al. 2002). Without any further assumptions, we obtain a general formula of the proportional rate of change of the form

$$\frac{\dot{R}_{\text{RL,vol}}}{R_{\text{RL,vol}}} = \frac{2\dot{J}_{\text{tot}}}{J_{\text{tot}}} + \frac{\dot{M}_{\text{tot}}}{M_{\text{tot}}} - \frac{2\dot{M}_{\text{acc}}}{M_{\text{acc}}} - \frac{2\dot{M}_{\text{donor}}}{M_{\text{donor}}} + \frac{\dot{R}_{\text{RL,norm}}}{R_{\text{RL,norm}}} \quad , \tag{1.8}$$

where $R_{\text{RL,norm}} \equiv R_{\text{RL,vol}}/d$ and $J_{\text{tot}}$ denotes the total angular momentum of the system. We here use the "donor", "acc", and "tot" subscripts to specify properties of the donor, accretor, and total binary system respectively. If we take the assumption of the system as primarily two co-rotating point masses, we can define $J_{\text{tot}}$ as

$$J_{\text{tot}} = M_{\text{donor}} M_{\text{acc}} \sqrt{\frac{Gd}{M_{\text{tot}}}} \quad . \tag{1.9}$$

In the regime in which $\dot{R}_{\text{RL,vol}}/R_{\text{RL,vol}} \geq 0$, MT is concretely stable. Where it is negative, we must obtain the characteristic timescale of the shrinking Roche lobe to determine stability. This timescale is the inverse of the absolute value of Equation 1.8, given by

$$\tau_{\text{Roche}} = R_{\text{RL,vol}}/|\dot{R}_{\text{RL,vol}}| \quad . \tag{1.10}$$

In the domain where this is of the same order or slower than the nuclear timescale of the star, the MT can also be said to be stable (Frank et al. 2002). If $\tau_{\text{Roche}}$ is significantly shorter than the nuclear timescale of the star, it is undergoing unstable MT.

If we assume fully conservative MT with no angular momentum loss, as is customary, we instead obtain a critical value of the mass ratio. This limit suggests all systems of $q > 5/6$ to be unstable, and those below that threshold to be stable (Frank et al. 2002; Marchant et al. 2021; Ramachandran et al. 2022). By this simplified threshold, a system's MT stability regime is purely defined by $q$. This of course presumes stability to be generally constant throughout the MT phase of any given system, though recent observations suggest systems may transition from one stability regime to another across their evolution (Ramachandran et al. 2022). If, as often presumed in existing literature, MT is strictly stable or unstable, then a system merely entering either phase can be directly informative of that system's subsequent evolution.

Marchant et al. (2021) determined donor stars with radiative envelopes that enter stable MT to be the primary progenitors of BBHs. They arrived at these conclusions assuming

HMXB RLO phase is a 1D hydrostatic system with a steady, transonic, adiabatic, and purely radial tidal stream subject to no stellar rotation or tidal coupling (Marchant et al. 2021). Hydrostatic simulations such as this neglect key elements of the system dynamics that could influence the stability of MT and impact results obtained (Lubow and Shu 1975; Marchant et al. 2021).

Conversely, unstable MT is considered the sole progenitor of common envelopes (CEs), an evolutionary stage in which the lost mass from the donor star forms a co-orbiting distant envelope around both donor and accretor (Marchant et al. 2021; Dorozsmai and Toonen 2022). Recent work suggests that unstable MT-induced CEs can also generate BBHs, but this is unconfirmed due to the violent and brief positive feedback loop that occurs in unstable MT systems, which makes them very difficult to study (Marchant et al. 2021).

Observational verification has proven similarly challenging, largely due to the brevity of RLO-induced MT onto a BH accretor. The MT phase in such systems, especially if unstable, is expected to be very brief, limiting the available observational sample of this vital progenitor process (Lubow and Shu 1975; Marchant et al. 2021; Ramachandran et al. 2022). A simulation-driven characteristic timescale of this phase could better inform future observations.

### 1.5.2 Mass Transfer Efficiency

The efficiency of this binary system MT flux ($\alpha_{\text{MT(tot)}}$) has yet to be critically examined in either the stable or unstable MT regime by hydrodynamic simulations. As shown in Equation 1.11, the efficiency of the binary system is defined by the relative mass gained by the accretor ($\dot{M}_{\text{acc}}$) for each unit of mass lost from the donor ($\dot{M}_{\text{donor}}$), regardless of how that mass is transmitted. Therefore, the binary system MT efficiency may be directly compared to models of other binary phases, such as wind accretion, wind RLO, and CE phases. The efficiency of angular momentum transport in the binary system ($\alpha_{\text{AM(tot)}}$) can be similarly defined as

$$\alpha_{\text{MT(tot)}} = \frac{\dot{M}_{\text{acc}}}{-\dot{M}_{\text{donor}}} \quad , \quad \alpha_{\text{AM(tot)}} = \frac{\dot{L}_{\text{acc}}}{-\dot{L}_{\text{donor}}} \quad , \tag{1.11}$$

where $L$ denotes the angular momentum of a given object. Most existing simulations are not capable of self-consistently transporting angular momentum, limiting its presence in the literature.

Similarly absent from existing study is the relationship between the MT efficiency of the binary system and that of its L1 tidal stream ($\alpha_{\text{MT(L1)}}$ and $\alpha_{\text{AM(L1)}}$), which may predominate accretion. We calculate stream MT efficiency by comparing the L1 mass flux to the optically-thick accretor outflow stream generated by its non-conservative entry onto the accretion disk,

$$\alpha_{\text{MT(L1)}} = 1 - \frac{\dot{M}_{\text{stream,out}}}{\dot{M}_{\text{L1}}} \quad , \quad \alpha_{\text{AM(L1)}} = 1 - \frac{\dot{L}_{\text{stream,out}}}{\dot{L}_{\text{L1}}} \quad , \tag{1.12}$$

by which we can compare the dense inflow and outflows of the accretor region. This definition better characterizes the RLO-induced MT process in isolation of wind accretion and wind RLO. In the high overflow range where RLO predominates by multiple orders of magnitude, these alternate definitions become equivalent such that $\alpha_{\text{MT(tot)}} \approx \alpha_{\text{MT(L1)}}$. However, in the lower limit of RLO where wind accretion and wind RLO make significant contribution to system MT, this distinction is more revelatory. By use of $\alpha_{\text{MT(L1)}}$, models of RLO may directly isolate the impact of the tidal stream without additionally summing across the contributions of other system processes as $\alpha_{\text{MT(tot)}}$ does.

These efficiencies play a pivotal role in subsequent system evolution. Not only does $\alpha_{\text{MT}}$ play an essential role in determining the probability of forming a CE, but if the system does evolve into a BBH, $\alpha_{\text{MT}}$ dictates both the total mass and mass ratio of that system. In the case of BBH formation, greater system mass from conservative MT shortens inspiral and heightens gravitational wave amplitude. Olejak et al. (2021) recently found modern techniques used to synthesize BBH populations to be vulnerable to insufficiently constrained RLO and MT physics. That same vulnerability, born of simplistic assumptions made of the efficiency and rate of RLO-induced MT, limits the efficacy of all models that rely on RLO evolution (Lubow and Shu 1975; Marchant et al. 2021; Bunzel et al. 2023). Additionally, the evolution of the disk around the accretor is deeply linked to the efficiency $\alpha_{\text{MT}}$ by which mass is deposited into and accreted from that disk (Frank et al. 2002).

$\alpha_{\text{MT}}$ comes into play outside of single system modeling as a relevant parameter in determining galactic metal enrichment as well. The metal-enriched feedback of the system depends on its end state; core collapse that fails to drive a supernova and yields a BBH releases far less stellar material back into the interstellar medium than a successful nova or CE. Additionally, any impact made on the evolutionary timescale of a system alters how long it takes for that system to feedback supernova-enriched material.

The dominant nucleosynthesis processes required for heavy element enrichment, as they are currently understood, rely on a handful of high energy events, including binary mergers and supernovae. The astrophysical dynamics of supernovae and high energy binary mergers are therefore critical to the metallicity timescale of the universe and in particular the enrichment regimes across cosmological timescales. A relatively small variation in the rates and efficiencies of these dynamics can result in drastically different nucleosynthesis yields across a statistically large sample (O'Shaughnessy et al. 2008, e.g. ).

In all of these applications, $\alpha_{\mathrm{MT}}$ has been thus far assumed a priori rather than directly determined, as has the angular momentum MT carries (Lubow and Shu 1975; Frank et al. 2002; Marchant et al. 2021).

### 1.5.3 Angular Momentum Transfer

Angular momentum, as well as mass, is lost from the system by inefficient MT. However, simple models and prescribed MT simulations often neglect this possibility. CE formation and initial BBH separation are dictated by angular momentum loss during the RLO phase (Ivanova et al. 2013). In the seminal work of Lubow and Shu (1975), the angular momentum carried by MT has been assumed equivalent to that of a stable orbit of the center of mass that passes through the L1 point. They also assume fully conservative MT and neglect angular momentum transfer within the accretion disk, eliminating the possibility of MT-induced angular momentum loss (Lubow and Shu 1975). With these assumptions, they concluded the MT tidal stream to launch at an angle of 19.5°-28.4° into a uniform-width column for much of its length before tapering on its curved approach to the disk (Lubow and Shu 1975). They also provide prescriptions of stream width, density, and accretion disk radius.

Despite the underlying assumptions of Lubow and Shu (1975) that neglect angular momentum loss, their conclusions have been widely used as prescriptions of L1 MT. Their results in characterizing the angle and width of MT tidal streams have been widely cited in more granular works where such assumptions may no longer hold, even if they do in the general case. Their prescriptions do not allow for the transfer of angular momentum from mass to mass inside the accretion disk, despite the essential nature of angular momentum transport in mass accretion (Frank et al. 2002).

The loss of angular momentum from the system reduces its orbital period and binary separation. As BBHs cannot lose angular momentum through the same means as RLO binaries, their merger timescales are directly dependent upon their initial binary separation at BBH formation (Mapelli 2020). The merging BBH population is therefore determined by the efficiency of angular momentum transfer during the RLO phase of their progenitor systems. Similarly, the donor star's evolution timescale is also dependent on the impact of MT.

## 1.5.4 Evolutionary Response to Mass Transfer

Stellar evolution and MT exist in constant feedback, each influencing the course of the other. On the appropriate timescales, donor stars may vary in radius. Even slight radial changes generate substantial changes to the overfilling factor $f$. These changes do not occur instantly and should only be considered in context of the evolution timescales of a given donor's stellar classification.

Massive stars evolve on multiple relevant timescales. The thermal response timescale of a star defines the time required for the star to re-establish thermal equilibrium in response to perturbation. This response is defined by the Kelvin-Helmholtz timescale given by

$$\tau_{\text{KH}} \approx \frac{GM_{\text{donor}}^2}{2R_{\text{donor}}L} \tag{1.13}$$

with $L$ denoting the donor's luminosity. For the high masses of BH progenitor stars, $\tau_{\text{KH}}$ may be on the order of 10 kyrs, or $10^4$ years.

The nuclear timescale of a star describes the chemical evolution of the core as it varies across time. Due to changes in radiative output, this timescale also describes changes in stellar radius that occur outside of evolutionary responses to MT. For high-mass BH progenitor stars, the nuclear timescale is on the order of 1 Myr or $10^6$ years (Quast et al. 2019; Klencki et al. 2022).

Existing research by e.g. Quast et al. (2019); Klencki et al. (2022) suggests that main sequence donor stars do not exhibit a radial response to MT occurring on thermal timescales. Stars undergoing MT on nuclear timescales respond by evolution of their radii which, depending on the evolutionary state of the star, may increase or decrease the star's radius. These variations naturally impact the evolution of RLO in binaries undergoing prolonged MT across nuclear timescales. Unstable MT cannot occur steadily on nuclear timescales by our definition given here, but stable MT may become unstable due to a radial expansion of the donor.

Klencki et al. (2022) found that donors can maintain thermal equilibrium with partially stripped envelopes throughout their main sequence evolution. However, these systems may have their subsequent progenitor pathway determined by expansion that occurs after the main sequence in their transition to become, often, giants.

The outer envelopes of stars in the asymptotic giant branch show similar levels of response. These stars exhibit thermal response timescales on the order of one year and are do not exhibit a radial response to changes occurring on comparable or shorter timescales (Passy et al. 2012). Simulations suggest that the radii of giant stars do not significantly change in response to extreme MT rates exceeding $10^{-3} M_\odot$/yr, but may respond to slower MT rates (Passy et al. 2012). Stellar expansion after the main sequence introduces key uncertainties into the progenitor pathway and may substantially impact the relative rates of CE and BH formation (Klencki et al. 2022).

Not only does stellar evolution alter the course of MT, but MT can alter the course of stellar evolution. Vartanyan et al. (2021) suggest massive donor stars whose envelopes have been stripped by MT are more likely to achieve core collapse supernovae. With the gravitational impact of their envelope reduced, MT-stripped stars experience expansion of their cores. This reduction in core compactness increases the likelihood of a core collapse supernova occurring; additionally, these supernovae are more likely to generate low-mass BHs in the range of $2-5 M_\odot$ (Vartanyan et al. 2021).

The model of Schneider et al. (2021) suggests the opposite may be true; they found complete envelope stripping by MT to reduce the formation rate of BH remnants when compared to isolated stars. It is unclear whether this result is also applicable to donor stars with partial or incomplete envelope stripping; recent research suggests BBHs may be most likely to form from systems still undergoing active MT at time of core collapse (Klencki et al. 2022). Those RLO systems that do not form BBHs, particularly those in which the donor is completely stripped of its envelope, are likely to instead follow the alternative path of CE formation.

## 1.6  Common Envelopes

While RLO in semi-detached binaries with a BH accretor forms the most promising progenitor of BBHs, it also serves as the sole progenitor of common envelope (CE) systems and the exotic remnants they can produce (Paczynski 1976; Frank et al. 2002; Ivanova et al. 2013; Abbott et al. 2016; Bavera et al. 2021; Dorozsmai and Toonen 2022; Bunzel et al. 2023). CEs are formed when unstable MT strips away nearly all of a donor's mass on a short timescale, filling the co-orbiting space around the binary with an optically thick shared envelope. The donor star so far exceeds its Roche lobe that its outer envelope expands and is shed through the outer Lagrange point, removing angular momentum in the process. This lost mass is no longer bound to the donor and is free to distribute along the equipotential, forming an enclosed shell around the binary. This enables the binary to reduce its orbital period drastically and transfer sufficient mass to initiate high energy events such as a supernova or merger or strip a giant donor star to form a smaller remnant (Paczynski 1976; Tylenda et al. 2011; Ivanova et al. 2013).

The CE phase was proposed as a solution to unexplained discrepancies in angular momentum transport (Sparks and Stecher 1974; Refsdal et al. 1974; Paczynski 1976). Binary systems that form with intermediate or long orbital periods experience orbital decay through the transfer of angular momentum by stellar winds (Hjellming and Webbink 1987; Podsiadlowski et al. 1992; Bate et al. 2002). Orbital decay occurs on very long timescales, accelerating somewhat as the binary separation shortens.

As the stars reduce to a sufficiently short binary separation, they enter the contact phase and begin RLO-induced MT. In simulations, even unstable MT does not remove sufficient angular momentum from the system to explain the formation of the observed population of extremely close binaries with very short orbital periods (Paczynski 1976; Ivanova et al. 2013). To explain these observations, we must also consider CE evolution (Paczynski 1976).

During CE evolution, the ejection of the donor's envelope strips it of nearly all of its angular momentum. Traditional models suggest this would lead to a merger event before the donor can evolve to a BH. However, recent research suggests very close BBHs may be a possible result of CEs (Marchant et al. 2021). Therefore, we must consider both stable and unstable MT phases, and the subsequent CE phase, as possible pathways in the progenation of BBHs.

BBHs are not the only possible outcomes of CE systems. Another natural result of common envelope evolution (CEE) may be the formation of Helium-rich white dwarf stars (He-WDs). Helium enrichment consistent with observational data would require conditions similar to the core of a red giant or supergiant star, which cannot form in a close binary system. This suggests a physical process which can transition an intermediate or long period binary to a very short period binary in which the giant donor has been able to shed its envelope entirely, leaving the bare core as a He-WD (Paczynski 1976). It therefore seems likely that CEE serves a key role in the formation of He-WDs, as does the process by which a system transitions from intermediate period binary to CE binary. This is further substantiated by the discontinuity in the statistical distribution of white dwarf binary separations, which suggests a sudden transition event rather than a gradual gradient (Korol et al. 2022).

Some dynamical phases in binary evolution are internally well-defined and show promising similarities between simulated and observed data. However, the transition between phases, and specifically from rapid MT in RLO to the formation of a CE, is relatively untouched by both observational and simulated data (Ivanova et al. 2013). This is owed to the logistical challenges of CE simulation, as well as ongoing debate as to which physical parameters and approximations are valid in CEE.

The range of timescales and length scales that may be critically relevant to the dynamics of CEE are many orders of magnitude beyond what can be practically simulated intensively with modern computing. These issues do not apply equally to the transition phase in which the CE is initially formed, as the parameters of the RLO phase can be expected to be continuous with the initialization of the transition phase. Furthermore, any distinct physical observables of the RLO-CE transition identified could serve as advanced warning of a CE phase for future observations, significantly increasing the chance of observing a form of stellar evolution that is thought to occur on very brief timescales (Izzard et al. 2011).

A similar technique has been successful in the study of stellar-mass BHs by examining the processes by which they could form, which in the case of BHs refers to supernovae and high-energy binary mergers. Simulations of supernovae and binary mergers have yielded practical constraint parameters on the initial formation of BHs (Fryer 1999). By considering the elusive CE binary system through the context of an exotic object rather than a developmental stage, the same principle arises. Once further practical constraints can be applied to CEE, the range of CE formation rates and yield efficiencies can carry this constraint into statistical models to better determine the nucleosynthesis pathways and timescales through which the observable universe's heavy element enrichment has occurred.

Without computationally detailed models to constrain the progenitor path of CEs, they have so far been constrained primarily by observation and simulation. These each present their own challenges that require substantial assumptions to be made.

### 1.6.1 Detection and Classification

Beginning with Paczynski (1976), there have been several attempts to identify CE systems or CEE remnants from observational data. The mention of remnants is significant, as current analysis suggests that the CEE phase occurs over short timescales, perhaps on the order of years (Ivanova et al. 2013). This leaves the observable population of contemporary CE systems very low compared to the observable population of remnant binaries. There is one particular candidate that presents a strong possibility for observed CEE, V1309 Sco, originally reported by Nakano et al. (2008) and further analyzed in Tylenda et al. (2011). Between 2002 and 2008, the orbital period of V1309 Sco decreased by 1.2%. In 2007, the light curve showed a transition from two maxima per orbital period to a single maximum, which may be an indication of a CE binary no longer resolvable as distinct objects. This phase lasted until a red nova in 2008 which likely indicates a merger event (Nakano et al. 2008).

There are observational concerns that limit the conclusions. Most critically, the observed features such as the light curve transition cannot be taken as evidence uniquely of CEE as an alternative explanation exists. The two maxima in the light curve of 1309 Sco could correspond to two surface spots on a single star until the fluid dynamics changed to feature a single spot instead, as is suggested by Tylenda et al. (2011). It is worth noting that Tylenda et al. (2011) and Ivanova et al. (2013), among others, conclude that this sunspot hypothesis is significantly less likely than a binary as the observational data would correspond to an extremely unusual single star. V1309 Sco has catalyzed a re-examination of similar transient outbursts as signs of CE binaries. This work is ongoing but suggests that red novae or V838 Mon-type outbursts are a promising field of potential CE observations (e.g. Ivanova et al. 2013).

Observational data of CE remnants such as SN 1987A have been analyzed in great detail and found to be very consistent with the theoretical outcomes of CEE, though these observations still come with difficulties in verification (Podsiadlowski 1992). The first difficulty in this area of research is false negatives and false positives. Any determined constraints made on CE dynamics based on proposed CE remnants is at risk from the misidentification of a particular remnant. This risk is mitigated by a large population of identified CE remnants, but a sufficient sample is still being accumulated (Ivanova et al. 2013). As it is, another issue to be mindful of is observational bias. There may exist CE systems which do not rapidly lead to a burst or SN that go unobserved. This constraint of observational data limits the applicability of CE remnant observations to determinations of the rate or remnant-production efficiency of CEE.

### 1.6.2 Simulations and Logistical Limitations

CE system simulations have encountered arduous logistical limitations. A number of dynamical processes are at work in CEE that may include a supergiant donor and a compact accretor both engaged in non-symmetric evolution as well as the envelope and wind acting at length scales beyond the scale of the supergiant. Making some assumptions about the accretion onto a compact object, one might still need to simulate across a range in length scale from the Schwarzschild radius of a BH to several times the diameter of a giant donor, as much as $\sim 10^8$ times longer. This is then compounded by a possible range in timescale between a neutron star's dynamical timescale and the thermal timescale of CE, as much as $\sim 10^{10}$ longer (Ivanova et al. 2013).

Despite these challenges, CEE simulations date back to two years after the development of the theory (Taam et al. 1978). More sophisticated models have recently begun to show significant similarity with observational constraints, particularly in 3D hydrodynamic simulations with fine spatial resolution. These simulations are still very computationally intensive, taking months to process (Reichardt et al. 2019). The advancement of computational tools will inevitably improve these matters, but presently only a few detailed simulations can be performed through the full CEE process each year. This is particularly limiting as the types of binaries which may undergo CEE are widely varied, with the properties of both donor and accretor likely to present as significant variables in the outcome of the evolution (Ivanova et al. 2013).

Whether directly modeling the CE or RLO phase or the transition between them, the methods used can be critical to the value of a simulation. The assumptions made in crafting simulations limit the power of those simulations to independently validate or constrain physical processes and observations.

# CHAPTER

# 2

# INTRODUCTION

This chapter integrates and adapts some work previously published in Dickson (2024) in Sections 2.1.1 and 2.1.2.

## 2.1 Roche Lobe Overflow Simulations

RLO is both a critical and complex process in the progenation of BBHs, CEs, and binary mergers. This evolutionary phase is vital in determining the subsequent evolution of a binary; the abundance, distribution, and characteristics of BBHs and CEs directly depend on the rate, efficiency, stability, and dynamics of MT during RLO (Lubow and Shu 1975; Pavlovskii et al. 2016; Marchant et al. 2021; Bunzel et al. 2023).

Many of these factors have been as of yet unconstrained by observation due to prevalent use of spectral analysis models that do not allow them to freely vary (Ramachandran et al. 2022). Theoretical calculations cannot eliminate these degrees of freedom a priori, and thus offer no direct solution to constrain them either (Lubow and Shu 1975; Marchant et al. 2021). Computational approaches, the remaining method available, have been forced to simplify and prescribe the dynamics present for computational efficiency.

Much like CEE, RLO has been under-examined by existing simulations due to its computational complexity. A holistic model of RLO-driven MT onto a black hole accretor would require simulating a thin disk of angular resolution $\lesssim 1°$ that transports angular momentum to accrete infalling mass on scales as small as the event horizon of a stellar mass BH on timescales of less than a second. Simultaneously, the same holistic model would need to model distances as large as the binary separation, which may exceed $1000 R_\odot$ across timescales as long as the donor's evolutionary response (Ivanova et al. 2013).

### 2.1.1 RLO Dimensionality

Such a model would also require tracking the impacts of hydrodynamics, radiation, and chemical enrichment in 3D, due to the aspherical deformation of the donor and the geometry of the MT tidal stream. 3D modeling compounds the computational demands of resolution and scale, yet it is essential to realistically modeling a system with so few symmetries. A complete model of such a make would also need to simulate across the entire span of the RLO phase, then be repeated for a number of systems across the parameter space of initial conditions.

Simulations relying on prescribed L1 MT, rather than a fully realized donor star, have become standard in the study of RLO-fed accretion (Whitehurst 1988; Meglicki et al. 1993; Kunze et al. 2001; Zhilkin et al. 2019, 2022). Donor envelopes have primarily been modeled separately using the hydrostatic stellar evolution code Modules for Experiments in Stellar Astrophysics (MESA) (Paxton et al. 2010; Pavlovskii and Ivanova 2015; Valsecchi et al. 2015; Marchant et al. 2021; Renzo et al. 2023; Gossage et al. 2023). The spherical symmetry of MESA precludes the asymmetric deformation of an overflowing envelope; MESA donor simulations therefore rely on prescriptive MT rates or indirect inferences rather than self-consistent hydrodynamical tidal stream modeling (Paxton et al. 2010; Marchant et al. 2021).

Naturally, the accretion of one mass onto the other is not a spherically symmetric process, nor is the accretion disk surrounding the secondary. Spherically symmetric simulations thus neglect key elements of the system dynamics that could impact the results obtained (Lubow and Shu 1975; Marchant et al. 2021). To constrain RLO mass and angular momentum flux, and their associated transfer efficiencies, multidimensional hydrodynamics are needed in modeling the complete binary system (Olejak et al. 2021; Marchant et al. 2021).

### 2.1.2   Resolution in 3D RLO Simulations

When multidimensional hydrodynamic simulations of complete RLO binaries have been attempted through grid solutions or smoothed particle hydrodynamics (SPH), they have been thus far constrained to low resolution.

SPH tracks individual point mass particles flowing through system contours, allowing a wide range of length and time to be examined with minimal computational power, in exchange for relatively low resolution. SPH methods, for their computational efficiency, cannot simulate multiple mass scales effectively, so hydrodynamic grids are preferred in higher computational load simulations. So called "oil and water" SPH solutions split their resolution between two non-interacting mass scales, but even these solutions cannot span the range of relevant mass scales in RLO-induced MT. Even at comparable resolutions to a grid solution, SPH provides lesser insight into system physics due to the largely ballistic trajectory of SPH particles. SPH cannot model the deposition of a tidal stream onto an accretion disk without first applying assumptions of disk diameter, density, and temperature, among other quantities. SPH models have thus far been limited to $10^4 - 10^6$ particle resolutions for 3D RLO (Nazarenko and Glazunova 2003; Lajoie and Sills 2010; de Vries et al. 2014; Reichardt et al. 2019).

Grid solutions must reconcile varying length scales, limiting the effectiveness of uniformly-distributed grid zoning to constrain small-scale flow. For 3D RLO, even well-designed non-uniform grids have been almost exclusively low-resolution, spanning $10^5$ zones (Bisikalo et al. 1998a,b,c,d,e, 2000; Oka et al. 2002). Nazarenko and Glazunova (2006) achieved a landmark non-uniform Cartesian grid resolution of $10^6$ zones. Rather than resolving the accretion process, they still assume 50% of incident mass entering the accretion disk ultimately accretes onto the compact object. These simulations allow detailed study of accretion disk dynamics, but do not truly simulate the process of overflow and the dynamics it entails of the donor envelope.

### 2.1.3   Limitations of RLO Simulation

The breadth of systems that are thought to undergo RLO limits the broader applicability of any given simulation. While it is not expected that all such systems generate BBHs, their relevance to other subsequent binary phases and mergers merits investigation. RLO systems may combine any of a wide array of donors and accretors that span a range of mass and length scales.

Stars can become binary donors in RLO systems at any stage, with main sequence stars and giant stars both becoming donors with some regularity (Beccari and Boffin 2019). As their name suggests, giant stars can be substantially larger and more massive than main sequence stars. The evolutionary timescale and phases of giant stars are distinct from those on the main sequence as well (Passy et al. 2012; Quast et al. 2019; Klencki et al. 2022). Considering both giants and the main sequence, donor masses range between the orders $1-10^3 M_\odot$ and donor radii range between the orders $1-10^3 R_\odot$ (Dumm and Schild 1998). In order for RLO to occur, the orbital separation must be greater than but of the same order as the donor radius. This range makes choice of system all the more important, as there is exponentially greater computational expense required to simulate a larger system to the same resolution. In the extreme case of giant donors, MT tidal streams may need to be modeled across $10^3 R_\odot$ before reaching the accretor.

As the total length scale of the simulation is determined by the donor, the resolution required is similarly determined by the accretor. The accretor may be any of a range of stellar objects or remnants, including BHs, neutron stars, and main sequence stars (Eggleton 2012). Among a number of differences these imply of the system, the resolution required to model down to the event horizon of a BH accretor may be a factor $\sim 10^3$ shorter than the photospheric scale height of a main sequence accretor. Smaller accretors also compound the resolution concern with higher velocity accretion disks that may require exponentially smaller steps in time evolution to resolve. Semi-contact binaries may feature any combination of these donor and accretor species, though some may also appear as contact binaries.

In contact binaries, both accretor and donor overflow their respective Roche lobes and MT occurs through a shared envelope similar to CEE, unlike the tidal stream of semi-contact binaries. As BHs cannot exceed their Roche lobes, these systems are not considered likely progenitors of BBHs. In order for a contact binary to evolve into a BBH, both stars would need to separately achieve core collapse without destroying or merging with the other. Conversely, semi-contact binaries with BH accretors need only reach core collapse in the donor. This is made easier by the evolutionary impact of MT, detailed in Section 1.5.4.

Currently all such systems are subject to semi-analytical methods that rely on a number of key assumptions detailed above. As such, any high-resolution 3D simulation of a binary in the RLO phase has the potential to begin to validate or refute the applicability of these models to a range of drastically different binary systems undergoing RLO. Therefore, despite the limited nature of a single RLO simulation, one or a set thereof may still be of substantial impact if well-chosen.

While many other binary systems undergo RLO, HMXBs are considered the most likely progenitor of BBHs (Bunzel et al. 2023; Dorozsmai and Toonen 2022). Given the rise of BBH observations through gravitational waves, simulation of HMXBs is perhaps the most needed. These systems are at the center of intersectional research interests, directly impacting the BBH and CE populations as well as informing the active fields of intermediate mass BHs, massive stars, and accretion disks.

A single simulation is not only limited by the system it models, but also the duration across which it spans. While simple models that rely heavily on time-saving approximations can span much longer timescales, more complex models must be more brief. Even with access to the supercomputer resources available today, modeling an accretion disk's evolution in steps of seconds in 3D across the entire span of the RLO phase is prohibitive to the point of impossibility. Instead, a single simulation is limited to a brief window within the longer, more complex process. There may be emergent variabilities in real RLO systems that such short duration simulations miss entirely. It is therefore imperative for a complete model of RLO to span as great a duration as possible.

Practical simulations are therefore also limited by their choice of initial conditions. Beginning a simulation at the onset of RLO will not be informative of the dynamics of its final transition into CEE, merger, or supernova. Studying the final transition state relies on assumptions made of the system's RLO process prior to the simulation's start.

This decision is made all the more impactful by the necessity of settling time. RLO simulations begin from artificial values generated based on the expected system state. These artificial values could bias the data obtained if not properly accounted for. In hydrodynamic simulations, running for a sufficient duration will enable the simulated physics to govern a steady state without being substantially biased by variations in initialization. The simulation runtime required to achieve this steady state is referred to as its settling time. Less understood and more dynamically variable system states, such as the final stages of RLO, require longer settling times. Especially in this regime, the computational time required to reach a steady state may substantially shorten the effective duration of a single simulation.

Energy transport presents an additional limitation to stellar simulations, including those of RLO binaries. In reality, radiation emitted from a single point distributes energy in all sightlines to the limit of their respective optical depths. These optical depths are determined by the density, elemental abundances, and ionization states along a given line of sight. As this radiation is absorbed, ionization states, energies, and entropies change across the affected area. At the intensities generated in HMXBs, radiation also carries significant pressure and force, altering the trajectory of the absorptive material.

Neither SPH nor grid-based hydrodynamics can self-consistently model radiative energy transport without substantial additional prescriptions. Even at incredibly high resolutions, SPH does not inform column densities and optical depths in any line of sight. Hydrodynamic grids are theoretically able to approximate radiative transport freely, though optical depths unaligned with the grid must be taken as estimates. Practically, however, a full radiative transport prescription would require each grid zone to depend on each other grid zone at every computational step. For example, a grid across $10^5$ zones would require $(10^5)^2 = 10^{10}$ calculations for radiative transport alone in order to advance one computational step. Sundqvist et al. (2018) computed such a grid in VH-1 at great computational expense even in a small 2D grid comprised only of donor wind.

While some regimes in HMXBs are dominated by dynamical modes of energy transport such as convection, radiation transport cannot be neglected entirely. Recent research by Marchant et al. (2021) suggests that the mode of energy transport in the outermost layers of the donor envelope is a key predictor of system end state.

## 2.2 Radiative versus Convective Envelopes

Stellar envelopes are broadly classified into two energy transport modes, radiative and convective. This classification depends on the stability of the entropy gradient against convection. An envelope that is stable against convection generates negative feedback against convective motion, diminishing that motion over time. Such envelopes are therefore considered radiative as they predominantly rely on radiative transport over convection. Envelopes that are unstable against convection generate positive feedback that propagates further convective motion across the now-convective envelope (Kippenhahn et al. 1990; Gabriel et al. 2014).

This convection transports energy by the (generally cyclical) radial displacement of mass; in order for this process to transport substantial energy, it must occur on a timescale shorter than the rate at which the displaced mass exchanges heat with its surroundings. In the regime in which flow occurs more slowly than local heat exchange, the displaced mass remains close to equilibrium with its surrounding environment throughout its displacement and very little energy is transported. That is to say that an envelope cannot develop convective regions, localized or general, without near-adiabatic flow rates (Gabriel et al. 2014). This may be modeled by the relative timescales of radiative energy transport ($\tau_{\text{rad}}$) versus that of adiabatic dynamical flow ($\tau_{\text{dyn}}$).

These characteristic timescales are quantified

$$\tau_{\text{dyn}} = \Delta r / |\vec{v}| \quad , \quad \tau_{\text{rad}} = \kappa \Delta r^2 \rho / c \quad , \tag{2.1}$$

where $|\vec{v}|$ gives the speed of flow through a region of density $\rho$. The characteristic length is denoted $\Delta r$ and $\kappa$ represents the opacity of that region. The speed of light is given by $c$. In the case of grid-based simulations, $\Delta r$ may be taken to be the width of a particular grid zone in order to resolve timescales for individual zones or taken collectively across the simulated envelope for a bulk solution.

Convection, once generated, is typically quantified by its location, scale, and rate. The average local convective velocity $v_{\text{conv}}$ and local convective turnover time $\tau_{\text{conv}}$ define the speed of displaced material and the rate at which it completes a convective cycle. These cycles, called turnover, may be identified in simulations by the complete inversion of the velocity direction across the convective region. The height of the convective zone is given by the mixing length parameter $\alpha_{\text{MLT}}$ normalized to units of the local pressure scale height $H_P$, which is the radial distance required for pressures to differ by a factor of $e$. These combine to define a convective turnover time of

$$\tau_{\text{conv}} = \frac{\alpha_{\text{MLT}} H_P}{v_{\text{conv}}} \quad . \tag{2.2}$$

These factors are related to the possibility of dynamo-generated magnetic activity by the Rossby number

$$\mathfrak{R}_{\text{o}} = \frac{T_{\text{rot}}}{\tau_{\text{conv}}} \sqrt{\frac{\alpha_{\text{MLT}} H_P}{R_b}} \quad , \tag{2.3}$$

where $T_{\text{rot}}$ is the period of rotation of the donor star, which is identical to the period of the binary system for a corotating donor (Gilliland 1985). The position of convection within the star is defined by the radius at the base of the convective zone $R_b$; in the case of aspherical overflowing donor envelopes, this value is approximate (Gilliland 1985; Charbonnel et al. 2017). While we may obtain a Rossby number for the dynamics in a given simulation, the modeling of magnetic effects substantially increases the computational load of a simulation and is therefore outside the scope of this work.

The dominance of radiative versus convective energy transport modes in donor envelopes has been found to impact post-RLO evolution. Marchant et al. (2021) found their simulated systems undergoing RLO with radiative envelopes to have a narrower set of end states than otherwise. Their simulated population never saw a radiative envelope success-fully ejected during CE phase. They concluded that each of their simulated radiative RLO systems that became CEs would merge during CEE and therefore could not generate BBHs.

In contrast, Marchant et al. (2021) concluded that CEs can generate BBHs if a deep convective envelope forms prior to CEE. They note however that most BBHs produced from CEs may have sufficiently wide binary separations to not contribute to the observed BBH merger population.

## 2.3   Sets of Simulations

While the scope of any single simulation is limited, sets of simulations can enable more broad conclusions to be drawn. By spanning a range of values, simulation sets are able to identify trends and differences between systems subject to specific variations. Sets of simulations are generally assembled in one of two modes, parameterized or sequential. Parameterized simulations sweep a range of parameter space for a specific variable or set of variables, while sequential simulations span a range of time evolution for a single set of initial conditions. We here present two sets of simulations to examine a test HMXB system undergoing RLO. Firstly a set of parameterized models across overfilling factor $f$ space and secondly a sequential set of models initialized from the onset of RLO to characterize its duration and changes with respect to time.

Parameterized simulations span a set parameter space and run independently with a variety of initial conditions operating under the same physics. These simulations enable a range of donor and accretor species to be tested across differing masses, mass ratios, and binary separations. Any individual study performing a set of parameterized simulations chooses which initial conditions to vary and across what range to vary them, as well as the density of simulations with which to span that range. Parameterized simulations therefore have a broad application to test simple models but suffer proportionally more for the computational cost of each simulation. This requires parameterized simulations to simplify their models as efficiently as possible. In the case of RLO simulations, this has often meant prescribed MT applied to 1D radial simulations in order to examine a range of systems in the same phase (Pavlovskii and Ivanova 2015; Valsecchi et al. 2015; Marchant et al. 2021; Renzo et al. 2023). In contrast, sequential simulations examine a single system across its evolution.

Though a single simulation must trade off resolution for duration to speak to more gradual transitions, a series of simulations may collectively inform the entirety of the RLO phase more effectively. Sequential simulations enable high-resolution modeling to span a greater timescale by utilizing the steady state of one to inform the state of the next simulation.

We implement a series of limited-duration sequential simulations to sweep overfilling factor $f$ parameter space to simulate the entirety of RLO phase. This approach enables high-resolution precise imaging of a test HMXB across the duration of RLO independent of the characteristic timescale of that phase. A model in steady state may be used to set the subsequent simulation incrementally from $f = 1$ until the expected onset of a CE, supernova, or merger. By extracting the mass and angular momentum fluxes throughout the system from each simulation, a characteristic timescale may be obtained. We therefore use this series to quantitatively model the timescale of RLO for the first time.

By use of this timescale and the observed fluxes, the next simulation of the sequential set may be generated by piecewise evolution. Piecewise evolution is the process of subdividing evolution into a series of sufficiently short pieces that each can be approximated by a single stable state. This enables an advancement of system time to be taken by only the stable state of the previous simulation. With this time advancement performed, then another simulation in the series is required to achieve a new steady state. Under the assumption that the quasi-steady state evolves smoothly, the fractional error this method induces can be minimized with sufficiently short evolutionary steps and sufficiently long settling times.

While the evolutionary trend of an HMXB is widely informative, we also apply a direct comparison of a small set of parameterized models to observations of a test system to offer an additional axis of cross-verification. This allows us to not only verify the efficacy of our methodology, but also directly compare the interpolated results of our method to the generated values of more simplified spectral fitting and semi-analytical methods and identify points of failure in the methods of e.g. Lubow and Shu (1975), Marchant et al. (2021), and Ramachandran et al. (2022). In combination, our parameterized simulations enable us to achieve the first model of the efficiency of MT and angular momentum transfer as they vary across overfilling factor parameter space.

We describe the construction by which we achieve a holistic RLO model of unprecedented resolution in Chapter 3. The specific construction of our simulation sets is defined in Sections 3.5 and 3.6. We detail the direct results of our two simulation sets in Chapter 4 and contextualize the implications of those findings in Chapter 5. A concise summary of our most consequential findings is provided in Chapter 6.

# CHAPTER 3

# METHODS

This chapter integrates and adapts some work previously published in Dickson (2024) in Sections 3.1.3, 3.2, 3.3, and 3.4 as well as Figures 3.4 and 3.5.

We employ 3D hydrodynamic simulations to constrain the mass and angular momentum loss from the donor and associated efficiencies of accretion onto the compact companion in HMXBs undergoing RLO (Section 3.1). In order to precisely model the efficiencies and rates of MT, we utilize a hydrodynamic grid of unprecedented resolution using the VH-1 code (Section 3.1.1). To optimize our computational efficiency, we subdivide the binary system into a set of spherical non-uniform overset grids (Sections 3.1.2 and 3.1.3). This solution maps effectively to system symmetries while balancing a range of length scales.

We map a 1D radiative envelope solution onto the effective gravitational potential contours of the Roche geometry (Section 3.2). This enables us to easily fill a radiative envelope in stable state into the aspherical geometry of RLO at any choice of eclipse radius.

We pair this solution with a thermodynamical transition into the isothermal wind regime, as well as the dynamical MT tidal stream (Section 3.3). While previous models have generally limited themselves to a universal isothermal or adiabatic approximation, this approach ensures a smooth transition between energy transport regimes.

This solution is applied to the test HMXB system M33 X-7, chosen for its recent obser-vations (Section 3.4). These observations have differed from previous findings of M33 X-7 regarding its MT stability and overfilling factor $f$ (Ramachandran et al. 2022).

We present two sets of simulations for the same system. In the first set of simulations, we compare the system at overfilling factors $f = 1.01$ and $f = 1.1$ with otherwise identical initial conditions (Section 3.5). This set serves to directly compare to observations and theoretical predictions of MT. The second set performs a piecewise evolution from a barely overflowing $f = 1.001$ initial state to examine the system across RLO phase (Section 3.6). While this set includes simulations with similar overfilling factors to the first set, the system has been allowed to evolve in masses and binary separation such that it is no longer directly comparable to observations beyond the initial state.

## 3.1   3D Hydrodynamics

The inherent structure of a non-SPH hydrodynamic simulation requires a grid of zones collectively evolving across time steps, in which the evolution of an individual zone is affected by the state of the surrounding zones. To minimize calculation required at each time step, the time between each step is selected such that no significant hydrodynamic change can propagate by more than one grid zone. This is achieved through Courant time evolution limiting based in each zone's dimensions and contained velocities and then standardized across the grid. This enables any zone to be evolved purely based on its own state parameters and those of immediately adjacent zones.

In order to simultaneously evolve a complete 3D grid, we have integrated the Message Pathing Interface (MPI) to separate the grid into smaller subsections and communicate across the boundaries between them effectively (Walker and Dongarra 1996). While these subsection boundaries require the passing of hydrodynamic variables at regular intervals, the more time-consumptive process of evolving individual grid zones can be tasked effectively to hundreds or thousands of processors and therefore distribute the computational burden of a 3D grid. Given the efficient hydrodynamical calculation implementation and substantial size of datasets, the communication time between processors can outweigh their relative advantage to processing speed, especially when operating on servers without modern and standardized multi-node threading. This practical limitation confines the scale and duration of simulations our code can achieve.

We have maximized the power of our simulation through use of a highly-efficient simulation code applied to a carefully distributed non-uniform grid.

### 3.1.1 Virginia Hydrodynamics One

Our simulation code relies on the framework of Virginia Hydrodynamics One (VH-1), a hydrodynamical simulation system capable of highly-optimized system evolution (Blondin and Pope 2009). VH-1 is a Fortran-based code, enabling it to more rapidly perform repeated calculations than simulations structured through higher level languages. VH-1 is capable of conserving angular momentum and enacting radiative cooling in 3D non-uniform grids, making it ideal for our purposes (Blondin and Taylor 2024). These radiative transfers of heat use the methodology defined in Blondin et al. (1990a).

VH-1 uses the piecewise parabolic method of Colella and Woodward (1984) to solve Euler's equations for an ideal gas with an adiabatic index $\gamma = \frac{5}{3}$ (Blondin and Taylor 2024). This yields a total energy and equation of state of

$$E = e_i + \frac{1}{2}v^2 \quad \text{and} \quad P = (\gamma - 1)\rho e_i \tag{3.1}$$

respectively, where $e_i$ denotes the internal potential energy per unit mass enclosed by hydrodynamic region of uniform density $\rho$, pressure $P$, energy $E$, and velocity $v$. We then apply conservation of mass, which may be expressed as

$$\frac{\partial \rho}{\partial t} + \nabla \cdot \rho \vec{v} = 0 \quad . \tag{3.2}$$

Next, we must apply momentum conservation in order to effectively track angular momentum transport across the system. This takes the form

$$\frac{\partial \rho \vec{v}}{\partial t} + \nabla \cdot \rho \vec{v}\vec{v} + \nabla P = \rho \vec{a}_{\text{eff}} \quad , \tag{3.3}$$

where the effective acceleration $\vec{a}_{\text{eff}}$ is defined

$$\vec{a}_{\text{eff}} = -\frac{GM_{\text{donor}}}{R_1^2}\hat{R}_1 - \frac{GM_{\text{acc}}}{R_2^2}\hat{R}_2 - \vec{\Omega} \times \left(\vec{\Omega} \times \vec{R}_3\right) - 2\vec{\Omega} \times \vec{v} \quad , \tag{3.4}$$

where $R_1$, $R_2$, and $R_3$ respectively denote the displacement from the donor, accretor, and center of mass. This acceleration represents the sum of gravitational forces acting on a grid zone from both donor and accretor. We treat both as point masses for ease of calculation. The distributed mass on-grid varies between simulations but is generally sufficiently small as to contribute negligible self-gravity. We next regard conservation of energy. The acceleration $\vec{a}_{\text{eff}}$ also accounts for the Coriolis and centrifugal forces generated by simulating the system in the corotating frame. The Coriolis and centrifugal forces in our system are defined by the third and fourth terms in the right-hand side of Equation 3.4, respectively.

The third Euler equation regards conservation of energy. In addition to dynamical mixing and the forces described above, VH-1 also enables us to account for energy change due to radiative cooling, $\dot{q}$. We therefore maintain our hydrodynamic evolution with the additional constraint

$$\frac{\partial \rho E}{\partial t} + \nabla \cdot \left((E+P)\vec{v}\right) = \rho \dot{q} + \rho \, \vec{v} \cdot \vec{a}_{\text{eff}} \quad . \tag{3.5}$$

Euler's equations, given in Equations 3.2, 3.3, and 3.5, collectively provide the basis of hydrodynamic time evolution in the piecewise parabolic method. This method treats the contents of the grid as ideal compressible gases (Colella and Woodward 1984). Observation suggests all elements of HMXBs, except the BH, behave as ideal compressible gases so this assumption is consistent with verified observations.

The VH-1 code has previously been utilized to model the HMXB Vela X-1 consistent with observations in the steady state (Blondin and Taylor 2024). This version of VH-1 was only implemented for the wind and accretion disk, and did not feature a donor envelope undergoing full RLO. The version of VH-1 used by Blondin and Taylor (2024) assumes an isothermal equation of state everywhere on-grid; this assumption breaks down in the stellar envelope, limiting the depth to which it can model full RLO. In order to achieve a holistic simulation of full RLO, our method differs with a number of additional prescriptions described in Sections 3.2 and 3.3.

### 3.1.2 The Yin-Yang Grid

We choose to apply a 3D spherical coordinate system to best resolve system symmetries. Due to the radially-driven donor wind and circular accretion disk, circular and spherical symmetries are substantially present in system dynamics, despite the aspherical nature of the overflowing donor. As further described in Section 3.2, we also extend our grid deep into the donor Roche lobe, where the inner equipotentials are nearly circular as well.

Traditional spherical grids encounter a computational issue resolving the grid poles, where a coordinate singularity occurs; zones across the entire azimuthal coordinate space converge on the grid pole with side lengths approaching zero (Figure 3.1). Even exempting the zones directly affected by this singularity, the remaining zones in the polar region become substantially smaller than those near the ecliptic. In Courant limited timesteps, unnecessarily brief timesteps would be required of the ecliptic zone in order to smoothly resolve the time evolution of the polar regions. As Courant time evolution depends on the smallest dimension of each zone, a hydrodynamic grid can evolve more computationally efficiently with zones of directionally invariant diameter, which is not achievable on a traditional spherical grid.

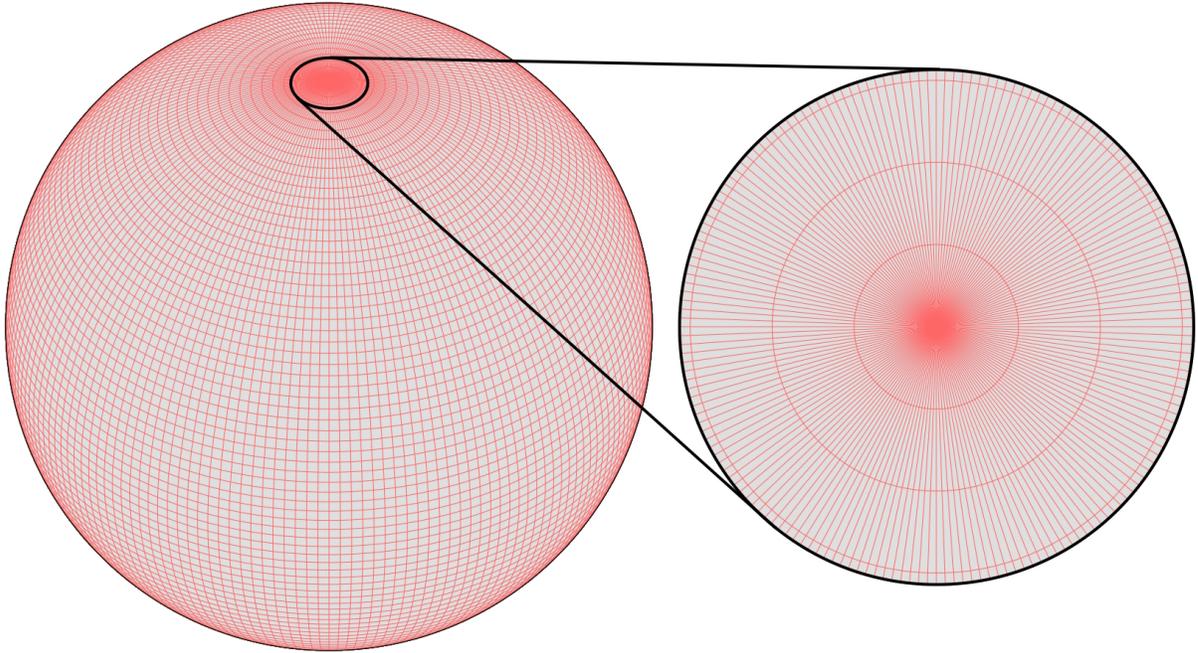

**Figure 3.1** Traditional Spherical Grid. Spherical coordinates demonstratively gradated for 2° zones. The polar region is shown in greater detail on the right. The polar zones differ in shape from the square-zone ideal for Courant time evolution.

In order to resolve the coordinate singularity, we implement a set of two partial spherical "yin-yang" grids following the method of Kageyama and Sato (2004). This system oversets two partial grids with a coordinate transform $x \rightarrow -x; y \rightarrow z, z \rightarrow y$ such that the grid poles are orthogonal to each other, shown in Figure 3.2. This system positions any point on the sphere $\leq 45°$ from the ecliptic of its respective partial grid, thereby almost entirely eliminating the polar variation in zone diameter. Despite the overlap of fringe zones, the elimination of the polar zone density enables our yin-yang grid to achieve 1° resolution universally with 15% fewer zones than a traditional spherical grid requires.

In order to pair the grids effectively, some region of overlap is required (Kageyama and Sato 2004). This overlap region, shown in the right-hand side of Figure 3.2, may span multiple zones at the high grid resolutions we intend to use. This presents the issue that, taken simply, the surface area of a unit sphere on a yin-yang grid is greater than $4\pi$. For calculation of mass and angular momentum fluxes in the system, we require a self-consistent method of comparison of simulated zones to physical area.

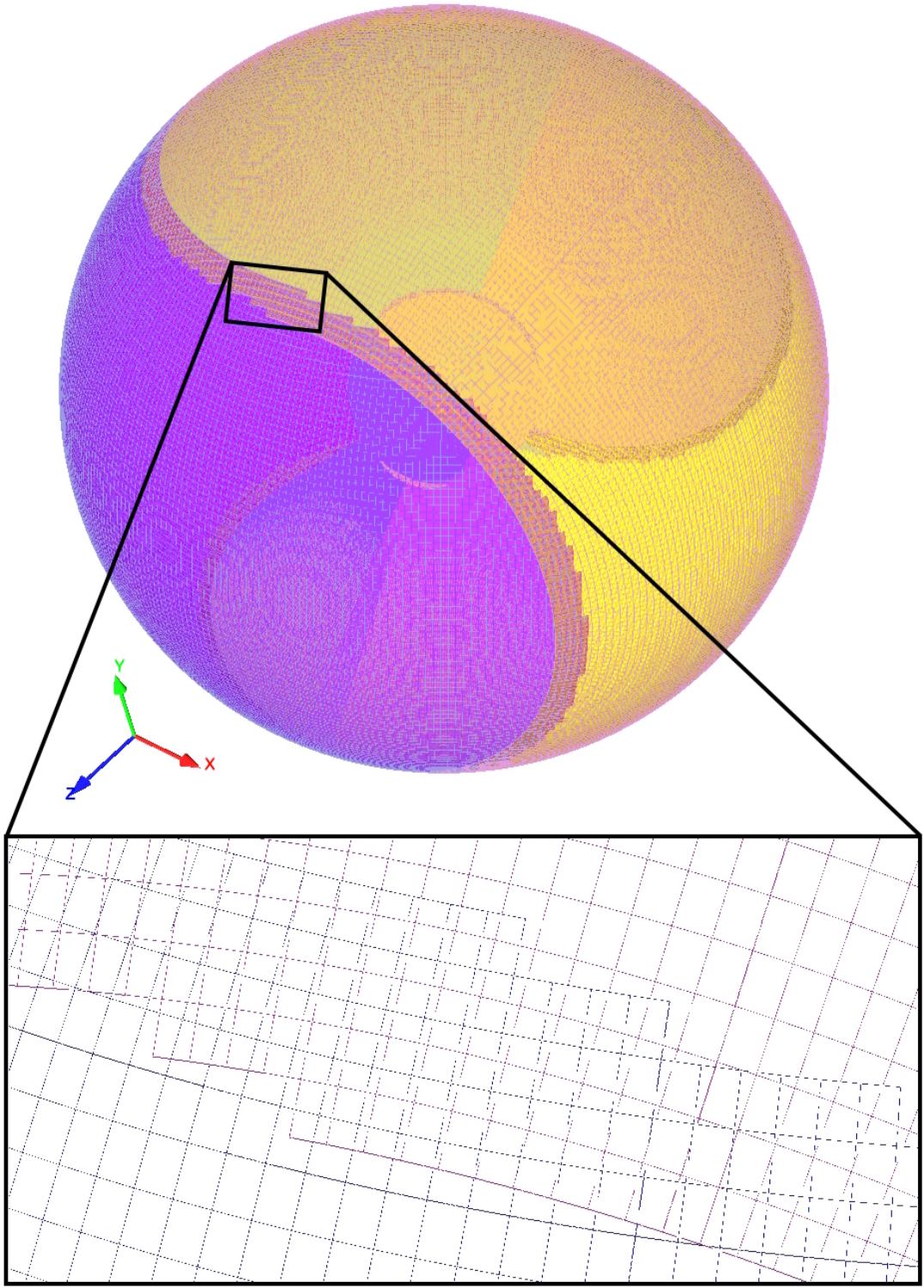

**Figure 3.2** The Yin-Yang Grid. Depicted gradated for 1° zones. The macro grid is shown with transparency and coloration for better 3D visualization. The fringe region is shown in greater detail on the bottom, without transparency or coloration.

We therefore apply a weighting solution to the fringe region. The fringe region is defined by sweeping the entirety of the yin (or yang) grid independently to determine the smallest angle $\Theta_{\text{crit}}$ that occurs between two zones represented on-grid. This naturally excludes the portion of the traditional sphere truncated to form the yin-yang grid. We then apply a sweep of the opposite (yang) grid. All zones that are an angle less than $\Theta_{\text{crit}}$ away from the nearest zone on the opposite grid are considered fringe zones. The total surface area of fringe and non-fringe zones are summed separately, and the fringe is weighted by a constant $0 < C_{\text{fringe}} \leq 1$ such that

$$A_{\text{non-fringe}} + C_{\text{fringe}} A_{\text{fringe}} = 4\pi R^2 \quad \text{at a given radius } R. \tag{3.6}$$

This method allows fluxes to be extracted across any surface without biasing results by non-physical dimensions. However, another modification is needed to the method of Kageyama and Sato (2004). Binary systems engage with spherical and circular symmetries both about the donor and accretor, limiting the efficacy of a single choice of origin.

### 3.1.3 The Double Yin-Yang Grid

We elect to represent our system by a set of two overlapping yin-yang pairs as shown in Figure 3.3. We utilize one set of spherical grids centered on the donor and likewise on the accretor, computed in the frame corotating with the binary system. This grid-based code enables us to simulate the donor envelope, photosphere, and wind simultaneously with a model of the accretion flow onto the compact companion.

An additional advantage of this double yin-yang solution is that VH-1 is well-suited to evolving grids on different timescales. Due to the highly dynamical evolution of the disk, we evolve the accretor grids on more rapid subcycles than we utilize to evolve the donor yin-yang. The relative number of subcycles is adaptive to compensate for any degree of timescale mismatch. For our chosen grid boundaries, we observe between $2-300$ subcycles on the accretor grid for each time step required by the donor grid. The range of subcycles required is due to the accretor grid's sensitivity to choice of inner boundary; the closer the inner boundary is drawn to the BH, the faster orbital motion occurs in the innermost zones on-grid.

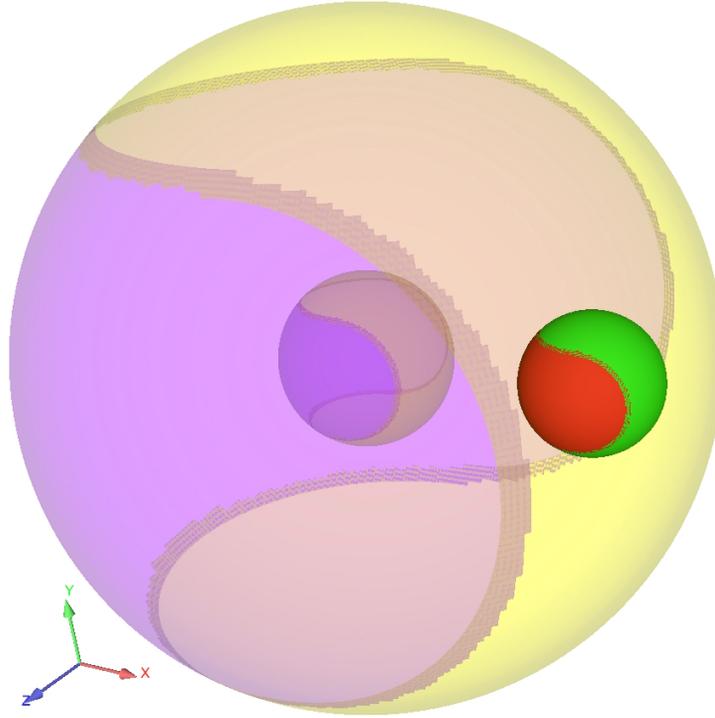

**Figure 3.3** The Double Yin-Yang Grid. The inner and outer boundaries of the donor grids are shown transparently in yellow and purple. The outer boundaries of the accretor grids (right) are shown opaquely in green and red.

We therefore perform an initial evolution with an arbitrarily high accretor inner boundary followed by incremental reductions to make best use of our computational resources. We choose to end the model well outside of the event horizon as VH-1 is not equipped for relativistic calculations; the characteristic timescale of zones in the relativistic region would be prohibitively rapid even if additional prescriptions were made. Our approach differs in our piecewise evolution series of sequential runs, described in Section 3.6, by maintaining a single inner boundary limit on the accretor grid from initialization.

We holistically resolve the entire interacting region of the system at unprecedented resolution. This was made possible by use of our well-optimized non-uniform set of paired grids. We employ four grids of 512 radial zones, 96 polar increments, and 288 azimuthal increments, combined into two overset spherical yin-yang configurations to achieve an angular resolution of 1° across the entire simulation window (Kageyama and Sato 2004; Blondin and Taylor 2024).

Our grid set is subsequently scaled to the geometry of a test system further described in Section 3.4. The scaling of our grid varies slightly between simulations in our sequential series of runs (Section 3.6) but remains of the same order in all respects. This variation is made to improve resolution as the system contracts, with the outer bound of both grids defined proportionally to the binary separation. The parameterized simulations detailed in Section 3.5 operate on identical grids to the quantities described here.

For fine-scale detail in the disk, the accretor is mapped by a set of higher-resolution grids housed inside the larger donor grids and paired across boundary zones. The accretor grids' innermost zones are as small as $\Delta R = 1210$ km. Accretor zones smoothly scale larger with radius, maintaining a constant $\Delta R/R$ for approximately cubic zones with respect to the angular resolution, $\Delta R = R\Delta\theta$. To fully span the accretor Roche lobe, we chose an accretor-centered grid outer limit $\approx 10 R_\odot$. We model the accretor grids down to the inner range of the accretion disk, limited by our code's slow angular momentum transfer and lack of X-ray feedback from the accretor. The accretor grids therefore span radially from 0.12 $R_\odot$ to 10.6 $R_\odot$.

Within the donor envelope and photosphere, the donor grid resolution is set to maintain a steady well-resolved flow. We define this resolution by half of the photospheric scale height $SH$ which yields

$$\frac{SH}{2} \equiv \frac{c_s^2 R_{\text{donor}}^2}{2GM_{\text{donor}}} = 7,114 \text{ km} \quad , \tag{3.7}$$

where $c_s$ denotes the sound speed. The donor grids extend to larger zones in the wind regime where scale lengths are significantly longer; in total, they span radially from 13.3 $R_\odot$ to 52.8 $R_\odot$ to encapsulate the accretor grids while accounting for the binary separation.

Our grid dimensions of $512 \times 96 \times 288 \times 2 \times 2$ span a total of $5.7 \times 10^7$ zones. The donor grids are then evolved on timesteps of $\sim 10$s while the accretor grids evolve on timescales as brief as 0.01s, for a total simulated span on the order of days post-settling. With these improvements to resolution, our complete binary grid enables us to examine the MT rate, its associated efficiency ($\alpha_{\text{MT}}$), and angular momentum loss, each insufficiently constrained by previous simulations. In order to holistically model MT, however, we must first model the donor envelope from which the tidal stream is generated.

## 3.2 Envelope Solution

In order to model RLO in cases like our test system M33 X-7, in which the donor radius significantly exceeds the Roche surface, we have added an outer envelope to the numerical method of Blondin and Taylor (2024). The initial outer envelope is constructed as a spherically-symmetric radiative envelope and then mapped as a function of gravitational potential to the effective potential of the binary system. This solution is computed assuming a constant Kramer's opacity and the stellar parameters mass, radius, and luminosity as given in Table 3.1 for M33 X-7 (Kippenhahn et al. 1990).

We begin with the assumption of an ideal gas of power law opacity as defined by

$$\kappa \propto \rho^a T^{-b} \quad , \tag{3.8}$$

and take the envelope to be initially at hydrostatic equilibrium. We next convert the system of equations for hydrostatic equilibrium of a spherically-symmetric radiative envelope into the Schwarzschild variables of Schwarzschild (1946). We therefore utilize $y$, $t$, $q$, and $x$ as dimensionless replacements for pressure $P$, temperature $T$, mass $m$ enclosed, and radius $r$ enclosed respectively. This yields

$$P = \frac{GM_{\text{donor}}^2}{4\pi R_{\text{donor}}^4} y \quad , \quad T = \frac{\mu G M_{\text{donor}}}{\mathfrak{R} R_{\text{donor}}} t \quad , \quad m = M_{\text{donor}} q \quad , \quad r = R_{\text{donor}} x \quad , \quad \rho = \frac{M_{\text{donor}}}{4\pi R_{\text{donor}}^3} \frac{y}{t} \quad , \tag{3.9}$$

with $\mathfrak{R}$ denoting the gas constant and $\mu$ denoting the mean molecular weight of the stellar material. These Schwarzschild variables are defined at the stellar surface by the known boundary conditions,

$$y = 0 \quad , \quad q = 1 \quad , \quad x = 1 \quad , \quad \frac{y}{t} = 0 \quad . \tag{3.10}$$

We then derive the coupled ordinary differential equations with respect to $q$ from the equation of state and hydrostatic equilibrium condition, obtaining

$$\frac{dy}{dq} = \frac{-q}{x^4} \quad , \quad \frac{dt}{dq} = -C_{\text{schw}} \frac{y^a}{t^{a+b+3} x^4} \quad , \quad \frac{dx}{dq} = \frac{t}{x^2 y} \tag{3.11}$$

where the factor $C_{\text{schw}}$ is a constant described by

$$C_{\text{schw}} = \frac{3\kappa_0}{16\sigma(4\pi)^{a+2}} \left(\frac{\mathfrak{R}}{\mu G}\right)^{b+4} L R_{\text{donor}}^{b-3a} M_{\text{donor}}^{a-b-3} \quad . \tag{3.12}$$

For a radiative envelope, we may apply $a = 1$ and $b = \frac{7}{2}$ to Equations 3.8, 3.11, and 3.12 befitting Kramer's opacity law.

We numerically integrate the system of coupled ordinary differential equations across $10^6$ numerical steps using the computational approach of Shampine (1975). These steps are exponentially distributed to provide fine resolution to the near-surface region in which $\frac{dq}{dx} \sim 0$ due to the region's relatively low densities. At each numerical step, a series of numerical integrations across statistically varied steps are averaged together to provide a smooth solution.

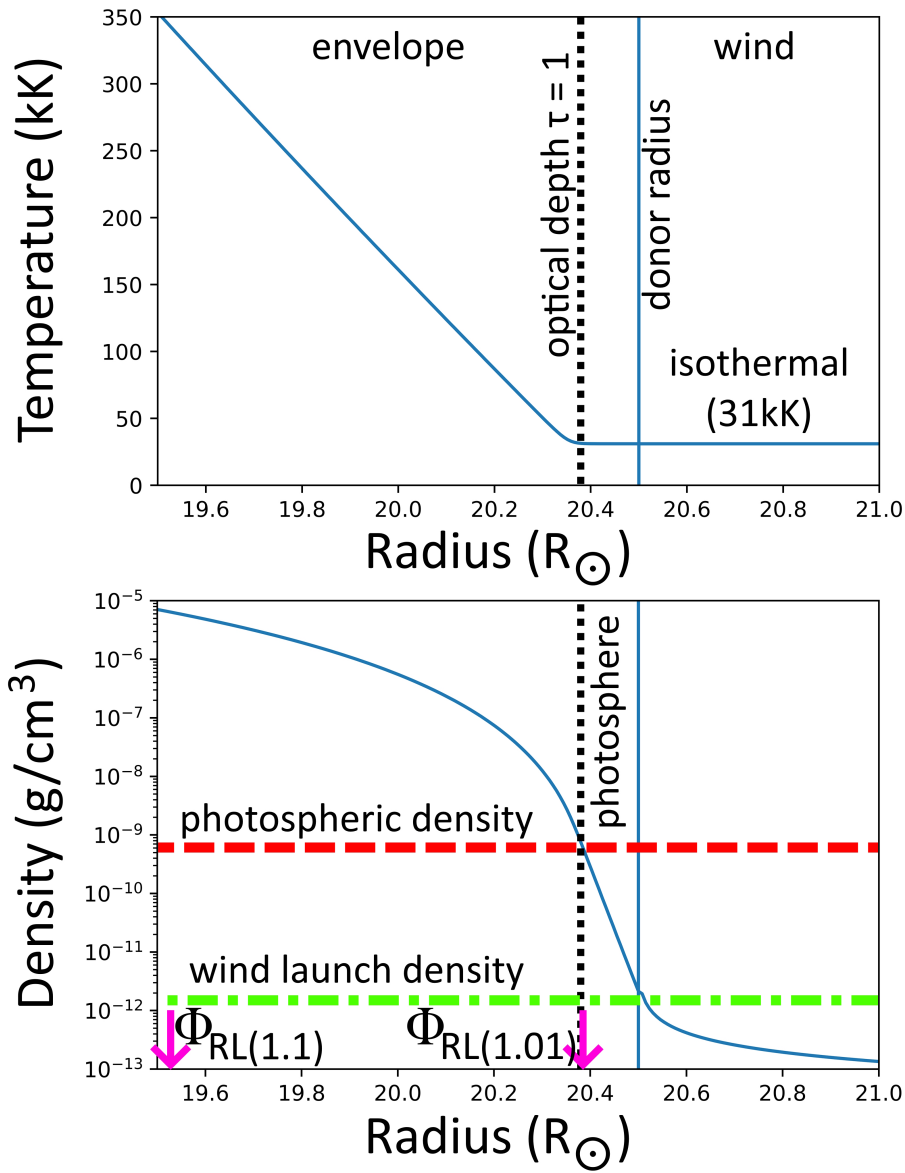

**Figure 3.4** 1D Donor Envelope Profile. The envelope to wind transition is plotted in temperature and density. Photospheric density corresponds to an optical depth $\tau = 1$, in keeping with the model O-II atmosphere of Hainich et al. (2019). Wind launch density indicates where velocities become supersonic. Pink arrows labeled by $\Phi_{\mathrm{RL}(1.01)}$ and $\Phi_{\mathrm{RL}(1.1)}$ demarcate the effective potential offset of the Roche lobe relative to the donor radius for our parameterized simulations, both the 1.01- and 1.1-cases.

We then convert the numerically integrated dataset back into physical coordinates for $P$ and $\rho$ using Equation 3.9 and derive the corresponding spherical gravitational potential values. This 1D solution, depicted in Figure 3.4, generates both the isothermal photosphere and the radiatively-dominated envelope in keeping with predictions of the test system (Hainich et al. 2019; Ramachandran et al. 2022). This methodology applies to any donor star with a radiative envelope (Kippenhahn et al. 1990). Figure 3.4, as it applies to the parameters of M33 X-7 specifically, is discussed in greater detail in Section 3.4.

In order to map this solution onto an overflowing star with $f > 1$, we must first calculate the effective gravitational potential of each zone on our grid. This follows the calculation given in Equation 1.1.

To obtain the eclipse radius and therefore overfilling factor of each equipotential contour, we numerically sweep an ecliptic quarter-plane. This quarter-plane is additionally truncated at L1, as described in Section 1.4. At each point in the sweep, we calculate the potential and eclipse radius of that point. We then sort by the local maxima of eclipse radius along each potential and compare to the L1 equipotential surface to obtain overfilling factors $f$. While this method is fully general, the results obtained are specific to the geometry of the binary system. See Figure 1.2 for the results of this method applied to the test system M33 X-7. With the equipotential of the desired eclipse radius ascertained, we offset our set of 1D potential values such that the stellar surface occurs at the chosen overfilling factor.

For each zone on the simulation grid within $r \leq R_{L1}$ of the overflowing primary with effective potential less than the eclipse radius equipotential, we initialize with the corresponding pressure and density of the potential profile. The choice of $r \leq R_{L1}$ prevents the inverted density gradients that would be generated by the potential contours of the accretor lobe or those beyond the outer Lagrange point. This necessitates a brief settling time for the L1 flow to reach a steady state.

We model the donor envelope to a depth sufficient to be characterized by quasi-static, highly subsonic motion with negligible radial flux; this ensures the envelope mass flows that feed the tidal stream are steady and physical. We therefore simulate the full extent of relevant mass during the MT phase and enable mass upwelling into the outer regime stripped by MT. This inner boundary gradually supplies mass to the grid to maintain constant pressure in the innermost grid zones. Due to the negligible radial flux near the boundary, this flow does not substantially vary the system dynamics other than to refill the donor envelope as mass overflows and is lost through L1; system dynamics were insensitive to a complete cutoff of the inner boundary flow on timescales briefer than days, though simulations of greater duration saw gradual reduction in overflow as the envelope diminished.

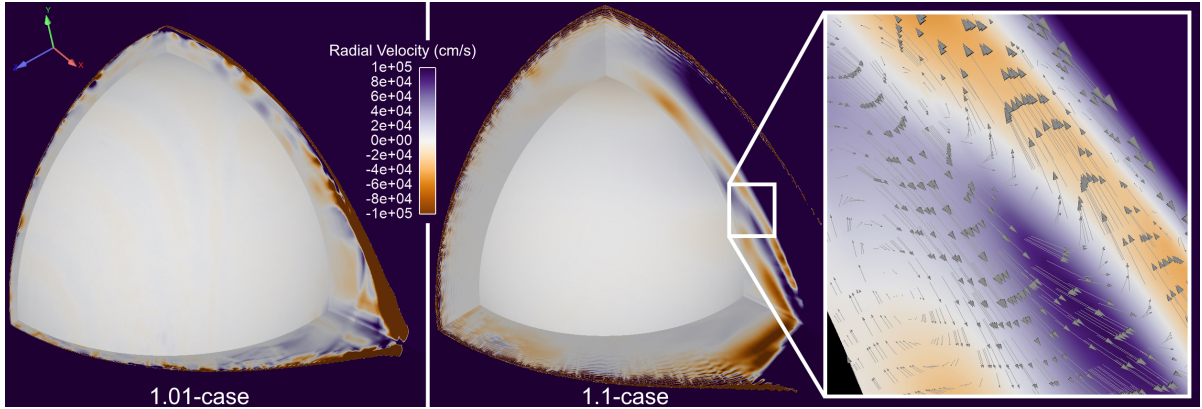

**Figure 3.5** Interior Velocities of Donor Envelope. The stellar interior is mapped by radial velocities. Images taken from our parameterized simulations detailed in Section 3.5. For both cases, the inner boundary of the grid is mapped by the central region and the ecliptic is set in the upper right quadrant. In the ecliptic plane, the deep tidal-feeding flow structure forms in contrast to the relatively quiet poles. Deep eddy flows that emerge from the 1.1-case are further detailed by velocity vectors in the ecliptic plane. Velocities are all significantly sub-sonic, with sound speeds $c_s \geq 6 \times 10^6$ cm/s everywhere in the stellar interior.

Figure 3.5 shows the radial velocity component within the envelope and particularly at the inner boundary. Perturbed surface flows develop in the photosphere, unable to propagate deeper into the envelope due to the pressure and density gradient. The $f = 1.1$-case (detailed in Section 3.5) also exhibited deep eddies with complete turnover near the ecliptic, mapped by velocity vectors on the right-hand side of Figure 3.5.

Our radiative envelope solution does not account for stellar evolution since the timescale of stellar evolution has previously been well beyond the scope of individual simulations. However, across our piecewise evolution sequence, we span sufficient time that evolutionary response may become relevant. We discuss the timescale of evolutionary response in greater detail in Section 1.5.4. For future research, we would like to incorporate MESA into our system initialization to model varied stellar evolutionary stages (Paxton et al. 2010).

For continuity, we then pair the envelope to the wind profile modeled to the limits of our grid resolution. Our model initializes with spherically-symmetric winds corresponding to a Sobolev-driven $\beta$-law velocity profile and both donor photosphere and envelope corotating with the system (Hummer and Rybicki 1985; Kudritzki and Puls 2000). The tidal stream and disk are both absent from the initialization, instead allowed to form freely during the settling phase. The time evolution required for model settling occurs computationally quickly in all components other than the accretion disk. To avoid propagating artificial instabilities during settling, we artificially restrict the system thermodynamics until a quasi-steady state is achieved. Once initial settling is complete, we begin to self-consistently model the transitions between thermodynamical regimes in our steady state system.

## 3.3  Simulating Thermodynamics

Holistically modeling RLO requires careful mediation between disparate thermodynamical extremes. Deep within the stellar envelope, radiative transport should dominate the flow of energy that dictates the entropy profile with respect to effective potential (Kippenhahn et al. 1990). In contrast, the surface flows entering the overflowing tidal stream are sufficiently dynamic to transport energy adiabatically, and recent works have taken the stream as adiabatic as well (Marchant et al. 2021).

In preliminary simulations, we noted the emergence of tidal-feeding flows upwelling from deep within the envelope. These suggest the invalidity of an adiabatic assumption such as used in e.g. Marchant et al. (2021) and require us to take a more careful approach in the simulations we present here. These deep flows lead to a steady rise in the entropy of the tidal stream across simulation time as gas is drawn up from ever deeper in the envelope (Figure 3.5). As we are unable to compute radiative energy transport in a 3D grid solution, we instead maintain the deep envelope in a radiative steady state and implement a smooth numerical transition between the steady-state radiative profile and adiabatic dynamical flow.

Our numerical method begins by calculating the radiative and dynamical timescales to cross each individual zone within the envelope at each computational step. This calculation is shown in Equation 2.1, using $\Delta r$ to denote to the radial dimension of a particular grid zone and $\kappa$ to denote opacity.

We then apply a weighted average within each zone, weighting the pressure derived by unconstrained adiabatic flow against that of the initial radiative thermal gradient as

$$P = \left( \frac{P_{\text{adiabatic}}}{\tau_{\text{dyn}}} + \frac{P_{\text{radiative}}}{\tau_{\text{rad}}} \frac{\text{dt}}{\Delta t_{\text{settling}}} \right) \times \left( \tau_{\text{dyn}}^{-1} + \tau_{\text{rad}}^{-1} \frac{\text{dt}}{\Delta t_{\text{settling}}} \right)^{-1} . \qquad (3.13)$$

The variable $P_{\text{radiative}}$ corresponds to the pressure required to maintain equal temperature across the appropriate equipotential surface, while $P_{\text{adiabatic}}$ is calculated by adiabatic dynamical evolution at each step. This allows a smooth transition from the radiative region to the dynamical flows induced by the tidal stream.

Our weighted average is taken at each iterative step, which must be mediated by a factor of $dt/\Delta t_{\text{settling}}$. This factor is selected based on the settling timescale of the tidal stream, so as not to bias dynamical evolution by the varying frequency of iterative steps. The system is insensitive to changes to $\Delta t_{\text{settling}}$ of less than an order of magnitude such that the settling timescale need not be known to any further precision than its order. Systems are visually distinct across changes of this timescale by one or two orders, but mean system fluxes post-settling require changes of several orders of magnitude to differ significantly.

Radiative energy loss dominates over both of these considerations in the regime of the thin disk and optically-thin wind. We define the wind regime as the regime with density below the photospheric density of the donor star, which inherently defines it as optically thin. The thin disk has far more surface area per unit volume than stream or envelope, enabling it to more effectively dissipate heat despite densities greater than the photospheric density. We set a floor temperature on the near-accretor region such that the disk cannot thin to less than one grid zone ($\sim 1°$) in scale height; the high-density regime within this near-accretor region we define as the disk regime. The outer reaches of disk and stellar wind are therefore best characterized by an assumption of highly efficient radiative cooling that results in an isotherm (Blondin et al. 1990b). The temperature of this isotherm is dependent upon the choice of system to be modeled.

## 3.4  Selection of M33 X-7

As with any simulated real-world system, we must select a physical system with which to validate our findings in comparison to verifiable observed data. For system validation, we select to simulate the observed system parameters of M33 X-7, an HMXB thought to be observed during the RLO-induced MT process (Ramachandran et al. 2022). The accretor of M33 X-7 is the oldest confirmed stellar mass BH by an order of magnitude and occupies a close binary with a massive stellar companion. Despite the binary's short orbital period, observations have confirmed that no CE phase has occurred (Orosz et al. 2007). Due to M33 X-7's age, high mass, short period, and lack of CE phase, it is the best candidate system in existing observation to be a future progenitor of a BBH system.

**Table 3.1** M33 X-7 System Parameters.

| Parameter | Symbol | Value |
|---|---|---|
| Mass of accretor black hole | $M_{\text{acc}}$ | 11.4 $M_\odot$ |
| Mass of donor star | $M_{\text{donor}}$ | 38 $M_\odot$ |
| Orbital period | $T_{\text{orbit}}$ | 3.45 days |
| Binary separation | $d$ | 35 $R_\odot$ |
| Temperature of donor star | $T_{\text{eff}}$ | 31 kK |
| Terminal velocity of donor wind | $v_\infty$ | 1500 km/s |
| Mass loss rate of donor wind | $\dot{M}_{\text{wind}}$ | $4.0 \times 10^{19}$ g/s |
| Stellar radius of donor star | $R_{\text{donor}}$ | 20.5 $R_\odot$ |
| Hydrogen mass fraction of donor star | $X_{\text{H}}$ | 0.60 |
| Helium mass fraction of donor star | $X_{\text{He}}$ | 0.395 |

The accreting BH binary M33 X-7 has provided a rare observational window into the RLO phase in HMXBs. The spectral analysis of Ramachandran et al. (2022) concludes M33 X-7's donor star to be significantly overfilling its Roche lobe, supplying the accretor with the excess mass required to generate the observed high X-ray luminosity (Table 3.1). This differs from the previous analysis of Orosz et al. (2007), which reported the donor star fueled the BH's X-ray luminosity with wind accretion alone as it did not fill its Roche lobe. They derived a filling factor $f \sim 0.77$ (Orosz et al. 2007). Ramachandran et al. (2022) predict M33 X-7's MT phase is in process of becoming unstable and subsequently may form a CE, rather than a BBH as its currently-stable MT phase would suggest (Pavlovskii et al. 2016). They also propose a much higher filling factor $f \sim 1.2$.

By simulating the system dynamics of M33 X-7, we intend to capitalize on this rare glimpse of RLO onto a BH accretor to improve parameter-space constraints on this elusive process. Detailed examination of M33 X-7 may reveal broader implications about three parameters as of yet unconstrained by existing research: MT rate, MT efficiency, and tidally-fed angular momentum loss. We also intend to utilize our analysis to compare the implications of the parameters given by Ramachandran et al. (2022) as they differ from those of Orosz et al. (2007).

We set the 1D profile to a radius $R_{\text{donor}} = 20.5 \ R_\odot$ following observations of Ramachandran et al. (2022) (Figure 3.4) . This is consistent with a photospheric density (optical depth $\tau = 1$) of $5.72 \times 10^{-10}$ g/cm$^3$ in keeping with the model O-II atmosphere of surface gravity $\log g = 3.4$ and $T_{\text{eff}} = 31$ kK presented by Hainich et al. (2019). We observed wind launch-ing to supersonic velocities at densities of $1.4 \times 10^{-12}$ g/cm$^3$ from the exponential-density isothermal atmosphere. For each 3D simulation, we map the overflowing donor surface at an equipotential corresponding to our choice of overfilling factor, further detailed in Sections 3.5 and 3.6. In all of our 3D models, the envelope is initially identical on any radial slice to the contours given in Figure 3.4, except mapped by aspherical effective potential rather than radius.

We set isothermal region (wind and disk) to a temperature $T_{\text{eff}} = 31$ kK in keeping with observations of the test system M33 X-7 (Ramachandran et al. 2022). Recent observations also suggest M33 X-7 may be near its Eddington luminosity. If this is the case, the feedback ionization and radiation pressure could substantially differ from this prescription. The outer envelope may also have significant Helium abundance that could impact the system's response to X-ray feedback from BH accretion (Ramachandran et al. 2022). X-ray heating and astrochemistry are outside the scope of this paper but may improve the scope of applications of this method if incorporated in future work.

We choose to examine M33 X-7 through two sets of simulations. In the first, we examine overflow in two parameterized simulations which differ only by overfilling factor. Through this set we provide a baseline of comparison to both the extreme overflow observations of Ramachandran et al. (2022) and the less overflowing earlier conclusions of the system. This set also enables direct comparison to theoretical work that examines the geometry of overflow as it varies with overfilling factor. Namely, we compare to the widely-used model of Lubow and Shu (1975) as well as the most current update to that model given by Marchant et al. (2021). For the second set, we begin from $f \sim 1$ and perform a piecewise overflow through a set of sequential simulations. This set engages more directly with the evolution of the rate, efficiency, and timescale of MT and angular momentum transfer across $f$ parameter space, which has previously been only minimally constrained.

## 3.5 Parameterized Simulations

We chose to simulate the system at two thresholds of overflow. In the more extreme overflow case, we chose the envelope surface to match an eclipse radius of $R_{\text{donor}} = 20.5 R_\odot$ in agreement with Ramachandran et al. (2022). This yields an overfilling factor $f = 1.10$ and thus we call it the 1.1-case. The 1.1-case most closely corresponds to the observations of Ramachandran et al. (2022) for direct comparison. While the work of Ramachandran et al. (2022) infers an overflow of $f \sim 1.2$, their method of calculation differs sufficiently that we anticipate a model of $f = 1.10$ by our method will best fit the data they present.

A second, less overflowing case is needed to compare with broad RLO models, such as the work of Lubow and Shu (1975) and Marchant et al. (2021). We therefore examine a second model at an eclipse radius $R_{\text{donor}} = 18.7\ R_\odot$. We call this slightly overflowing model the 1.01-case, with an overfilling factor $f = 1.01$.

To fit varied overflows, we set both case's envelope solution to an eclipse radius corresponding to its overfilling factor. By use of Equations 1.1 and 1.7 and the quarter-plane sweep method described in Section 3.2, we obtain an effective potential curve given in Figure 1.2. We therefore fit our envelopes by differing eclipse radii with only the 1.1-case corresponding to the $R_{\text{donor}} = 20.5 R_\odot$ observation of Ramachandran et al. (2022).

We otherwise maintain every other initial system parameter constant between the 1.01- and 1.1-cases. We assign both models identical system geometries given in Table 3.1. By holding the mass ratio constant, the traditional stability model would suggest all cases experience identically unstable MT (Frank et al. 2002). By examining high and low overflow in parallel on a well-observed system, we can test the traditional stability model in comparison to the computed timescale of overflow we obtain through Equation 1.8 for each case.

Parameterized simulations also empower us to examine the boundaries of validity of semi-analytical models of RLO. These models are widely-used in lieu of multidimensional hydrodynamic modeling across a diverse range of systems for which the implicit assumptions of those models may not apply. Particularly of note is the efficacy of existing semi-analytical models in fitting high overflow that lacks significant observational constraints (Lubow and Shu 1975; Marchant et al. 2021).

In keeping with the method of Blondin and Taylor (2024), we maintain the treatment of both donor and accretor as point sources for gravitational calculations for both cases. This prescription is worth noting as the 1.1-case envelope contains a distributed mass of $\sim 0.01 M_\odot$ on-grid, while the 1.01-case envelope maps only $\sim 0.0002 M_\odot$ on-grid. This difference may generate small but non-zero self-gravity in the 1.1-case unaccounted for by our present method.

$\Delta t_{\text{settling}}$ was chosen to be $10^6$ seconds ($\sim 12$ days) for both models. The 1.01- and 1.1-case models each ran for a simulated duration of $\sim 50$ days in order to resolve the settling timescale of the accretion disk. This settling timescale is highly dependent upon choice of inner boundary, though the accretion disk outside $1 R_\odot$ settles on the order of days.

## 3.6 Sequential Simulations

While our computational resources are insufficient to continuously model the millennial timescale of the thermal envelope, our simulated systems reach quasi-steady states that can be utilized for a stepwise approach through sequential simulations. From that state, the initialization of the next model can be derived by utilizing the dynamical timescale and prevailing fluxes of the system. Incrementally, this stepwise approach can span a much longer timescale than any single 3D hydrodynamic simulation could achieve, allowing us to characterize the evolution of MT during RLO.

This method also allows us to account for variations in the grid structure, which occur as mass and angular momentum are transferred away from the primary, by modifying structural parameters between sequential simulations. As this applies to M33 X-7, we relocate the accretor-centered grids closer to the donor as the binary separation diminishes to maintain spherical symmetry about the accretor. We also reduce the outer limits of the grid proportionally to the change in binary separation in order to obtain higher resolution on the shrinking system.

We elect to begin with a simulation set with as small of an overflow we can achieve on our simulation resolution, which equates to an overfilling factor $f = 1.0005$. We choose to denote this simulation as model A, such that the subsequent models may be each denoted in alphabetical order as models B through T. Other than its overfilling factor and the associated eclipse radius, model A is maintained with the system parameters observed of M33 X-7 given in Table 3.1.

The overflow mapping utilized in model A is summarized in Section 3.2 with the results of the method applied to the geometry of model A defined in Figure 1.2. This distribution of effective gravitational potentials across eclipse radii varies by binary separation and mass ratio. As we elect to treat binary separation and the masses of both components as variable in this evolution, the distribution shown in Figure 1.2 applies only to model A. The same method is used to obtain the effective potentials of each subsequent model.

As the purpose of our sequential simulations is to span a wide range of overflows across a smooth progression, we elect to simplify our disk model. By more than an order of magnitude, the most computationally intensive aspect of our simulation method is the rapid dynamical evolution of the inner region of the accretion disk. While the parameterized models described in Section 3.5 are suited to holistically modeling the majority of the disk, this simulation set is better served by a greater number of models with only the outer region of the disk simulated on-grid. In order to validate that our results would not be impacted by this change, we studied the parameterized set of models at a variety of inner boundaries. We found the observed fluxes of the system, including in the outer disk, to be invariant with choice of inner boundary between $0.028 - 1.4 R_\odot$ on our grid.

As the parameterized series of simulations identified a natural disk band gap to occur in the region $\sim 1 - 1.6 R_\odot$, further detailed in Section 4.1, we selected an inner boundary of $1.4 R_\odot$ for the accretor grids of model A, with an outer limit set at $10.2 R_\odot$ as noted previously. For subsequent models, we scale these limits with changes in the binary separation $d$ such that the accretor grids range from $0.039 d - 0.3 d$. As the binary separation changes by less than an order across the piecewise evolution of our sequential simulations, the accretor grids are each comparably sized.

The initial models in this sequential series either developed no disk outside of the raised inner boundary or formed a low-mass disk that stabilized within the settling phase of the tidal stream dynamics. However, beginning with model J, we initialize each model with the accretor grid filled proportionally to the mass and temperature distribution of the end state of the previous model in the series. This differs from our previous method of initializing the accretor grids with the stellar wind profile of the donor taken in isolation. The presence of highly dynamic accretion during the settling phase increased the runtime required to achieve the same model duration by $\sim 30\%$ but substantially reduced lingering noise from prolonged accretion disk settling.

With these changes, model settling is reduced to $\sim 1$ day in simulated time, and we set a simulation duration of $\sim 3$ days per model. This provides a substantial steady state sample from which to extract the mean mass and angular momentum fluxes between system regimes.

As we intend this series of simulations not for visualization but for efficient runtime, we examined the impact of $\Delta t_{\text{settling}}$ on system fluxes and settling time. After the initial dynamics resolve, the activation of the method described in Equation 2.1 forms a non-zero contribution to the duration of settling required to achieve a steady state. The most physical choice of $\Delta t_{\text{settling}} \sim 1$ day would contribute substantially to the computational time required to perform each model, so we tested a range of values. While the choice of a lower $\Delta t_{\text{settling}}$ value makes the transition between radiative and adiabatic regimes less smooth, we found system fluxes of the steady state to be invariant to changes in $\Delta t_{\text{settling}}$ in the range from 1 second to 1 day. We therefore choose $\Delta t_{\text{settling}} = 1$ second to optimize the efficiency by which models are generated without compromising the physicality of the results obtained.

Each time a model in the sequential series completes, its steady-state system fluxes may be used to precisely calculate $\dot{R}_{\text{RL,vol}}/R_{\text{RL,vol}}$ by means of Equation 1.8. While other timescales, fluxes, and efficiencies may be observed from each model, the timescale $\tau_{\text{Roche}}$ most directly determines the subsequent model. In order to maintain a reasonable time resolution, we follow each model with a piecewise evolution of duration $0.01 \times \tau_{\text{Roche}}$, 1% of the characteristic Roche lobe timescale. We assume constant mass and angular momentum fluxes during this evolution. We therefore directly calculate the change in $M_{\text{donor}}$, $M_{\text{acc}}$, and $J_{\text{tot}}$. This change in $J_{\text{tot}}$ is also translated into a change of binary separation $d$ and period $T_{\text{orbit}}$.

Our grid solution requires longer timescales to dissipate sufficient angular momentum to accrete material through the inner boundary of the disk. Therefore, while we note the rate of accretion at this inner boundary, we calculate the mass gained by the secondary differently. The net inflow rate of mass through the spherical surface surrounding the disk region is used in place of the inner bound accretion rate. We find this value to be consistent with the deficit between the mass loss rate of the donor and the mass loss rate through the outer boundary of our grid.

With these changes to mass and angular momentum computed, we then derive a new eclipse radius and overfilling factor. These are drawn from mapping the previous model's donor surface overlaid onto the potential contours of the new system geometry. This process changes the donor eclipse radius only slightly as the aspherical donor strikes slightly different sightlines from the now closer accretor. The overfilling factor varies more consequentially as the changes in mass ratio and binary separation slightly reduce the eclipse radius of the Roche lobe.

# CHAPTER 4

# RESULTS

This chapter integrates and adapts some work previously published in Dickson (2024) in Section 4.1 as well as Figures 4.1 and 4.2.

Our simulation methodology, applied to the parameters observed of M33 X-7 described in Ramachandran et al. (2022), has achieved a quasi-steady hydrodynamic equilibrium within donor wind, donor envelope, and tidal flow in two sets of simulations of unprecedented resolution. Angular momentum transfer in RLO binaries, previously limited to analytical assumptions, is simulated here for the first time. We subdivide our results into those of our two parameterized simulations and those of our twenty sequential simulations.

The parameterized set examines the system parameters of M33 X-7 given in Table 3.1 at two overfilling factors. We selected to study the system at $f = 1.01$ for direct comparison to existing theoretical models of RLO, which we refer to as the 1.01-case. For comparison to recent observations of M33 X-7 by Ramachandran et al. (2022), we also studied the system dynamics at $f = 1.1$, which we similarly name the 1.1-case. We here empirically describe the resultant systems for each dynamical regime: the donor wind and envelope, the dense outflows through L1 and exiting the accretion disk, and the disk itself. We particularly detail the upwelling flows and eddies induced in the envelope and the banded disk structure which formed in our simulations, as both represent novel findings.

To study the efficiency, rate, and timescale of RLO processes across their evolution, we performed a sequential series spanning from the onset of full RLO to the final decades of unstable MT. We denote these simulations alphabetically, as models A through T. As quantitative modeling of RLO timescale and efficiency is here conducted for the first time, we provide particular detail to these findings in Sections 4.2.1 and 4.2.3. This discussion is enriched by additional context in regard to the evolution of the system's orbital geometry, provided in Section 4.2.2.

The efficiency of RLO is a vital parameter that our method is well-suited to constraining more substantially than prior research. We therefore study RLO mass and angular momentum efficiency across four variables, individually defined in Section 1.5.2.

## 4.1  Parameterized Simulations

Our two parameterized simulations at $f = 1.01$ and $f = 1.1$ differed substantially in both flux and geometry in the envelope, tidal stream, and disk regimes despite maintaining comparable donor winds. Both cases saw the launch of an optically thick MT tidal stream through the L1 point, which formed a disk around the accretor. We compare the scales, geometry, and dynamics of the 1.01- and 1.1-cases in each of these regimes.

Our 3D profile demonstrates a steady-state donor wind which averages $v_\infty = 1500$ km/s and $\dot{M}_{\text{wind}} \sim 5$ to $6 \times 10^{19}$ g/s across all sightlines not perturbed by the tidal flow, in keeping with Ramachandran et al. (2022). To resolve the launched donor wind, we examine its flux at the L1 radius. We exempt the tidal stream to be summed separately as the L1 mass flux, shown in Table 4.1. The Sobolev-driven wind launches from the aspherical surface of the donor envelope, rapidly accelerating to higher velocities than occur within the stellar interior (Figure 3.5). The donor wind launches with angular momentum and mass fluxes almost identical to the quantities observed exiting our simulation's external boundary. These were constant across simulation time in both cases (Figure 4.1). This corresponds to a system angular momentum loss timescale of $\tau_{\dot{L}} \equiv L/\dot{L} \sim 20$ Myrs in both cases.

The donor envelope solutions we present in Figure 3.5 are subsonic and radiatively-dominated throughout, with the near-exception of the substantial 1.1-case ecliptic flow. The 1.01-case donor envelope developed significantly sub-sonic flows which die off quickly towards the interior. The deep envelope flows induced were sufficiently insignificant that we elected to alter the donor boundary conditions; by raising the inner boundary of the envelope, we were able to increase the resolution of the donor wind regime near the external boundary. We found the L1 and wind mass fluxes to vary by less than 1% with an increase in the donor grid inner radius to $15.4 R_\odot$. This elevated inner boundary is shown in Figure 3.5. This change in resolution did not result in a significant change in wind flux, though the lesser envelope depth of the 1.01-case model increased its computational efficiency.

**Table 4.1** Simulated Mass and Angular Momentum Fluxes. Values expressed in cgs units where appropriate. System properties listed by AM refer to angular momentum.

| Region | Property | Symbol | 1.01-Case | 1.1-Case |
|---|---|---|---|---|
| Donor Wind | Mass Loss Rate | $\dot{M}_{\text{wind}}$ | $5.36 \times 10^{19}$ | $6.03 \times 10^{19}$ |
| | AM Loss Rate | $\dot{L}_{\text{wind}}$ | $1.67 \times 10^{39}$ | $2.03 \times 10^{39}$ |
| Donor L1 | Mass Loss Rate | $\dot{M}_{\text{L1}}$ | $1.7 \times 10^{18}$ | $8.80 \times 10^{22}$ |
| | AM Loss Rate | $\dot{L}_{\text{L1}}$ | $3.1 \times 10^{37}$ | $1.18 \times 10^{42}$ |
| External | Mass Loss Rate | $\dot{M}_{\text{ext}}$ | $5.14 \times 10^{19}$ | $5.84 \times 10^{19}$ |
| | AM Loss Rate | $\dot{L}_{\text{ext}}$ | $1.64 \times 10^{39}$ | $1.96 \times 10^{39}$ |
| L1 Tidal Stream | Mass Transfer Efficiency | $\alpha_{\text{MT(L1)}}$ | 0.86 | 1.00 |
| | AM Transfer Efficiency | $\alpha_{\text{AM(L1)}}$ | 0.20 | 1.00 |
| Binary System | Mass Transfer Efficiency | $\alpha_{\text{MT(tot)}}$ | 0.07 | 1.00 |
| | AM Transfer Efficiency | $\alpha_{\text{AM(tot)}}$ | 0.04 | 1.00 |

The 1.1-case donor developed more dynamic flows in the ecliptic. A significant deep outflow emerged which feeds the higher tidal stream density we observed accompanying the greater overflow. The dynamical nature of this deep outflow also generated eddy currents exhibiting complete turnover deep within the radiative envelope (Figure 3.5). The 1.1-case donor envelope was simulated to a greater depth sufficient to model the entirety of the tidal-feeding outflow. This flow exceeds Mach 0.6 near the surface deposition of the tidal-feeding stream and has dynamical timescales as brief as $\tau_{\text{dyn}} \sim 10 \tau_{\text{rad}}$ but remains radiative. In both cases, the stellar interior motion is entirely subsonic, unlike the tidal stream.

The complete turnover of stellar material by dynamical perturbation of a radiative envelope, as shown in Figure 3.5, presents an unexpected result. Though the 1.1-case envelope was distinctly radiative everywhere outside of the near-L1 ecliptic region, complete turnover is generally considered to be a defining and exclusive trait of convective envelopes (Gilliland 1985; Charbonnel et al. 2017). This is especially consequential as convection is "an essentially non-local process" which by the contemporary model of convective envelopes cannot be isolated to a localized region; our result in the 1.1-case better fits the now-disfavored tra-ditional model of turbulent eddies in a localized adiabatic gradient (Spruit 1996). Following the model of Gilliland (1985), we define this convective region by the parameters described in Table 4.2. We cannot precisely constrain the observed turnover parameters due to the local complexity of the deep outflow, but all turnover parameters obtained are within the expected range for convective envelopes; this suggests the possibility of a magnetic dynamo effect unaccounted for in existing RLO hydrodynamics (Gilliland 1985; Landin et al. 2010; Charbonnel et al. 2017).

**Table 4.2** Turnover Parameters of 1.1-Case Convective Region. Values expressed in cgs units except where otherwise specified. Due to the localized and dynamic convection simulated, all values should be taken as estimates.

| Property | Symbol | 1.1-Case |
| --- | --- | --- |
| Average Local Convective Velocity | $v_{\text{conv}}$ | $1.5 \times 10^5$ |
| Local Convective Turnover Time | $\tau_{\text{conv}}$ | 8 days |
| Local Pressure Scale Height | $H_P$ | $5 \times 10^{10}$ |
| Radius at the Base of the Convection Zone | $R_b$ | $9.3 \times 10^{11}$ |
| Mixing Length Parameter | $\alpha_{\text{MLT}}$ | 2 |
| Rossby Number | $\mathfrak{R}_o$ | 0.14 |

The L1 tidal stream differed significantly between the 1.01- and 1.1-cases, in both structure and scale. In both cases, we saw the tidal stream accelerate through L1 onto the outer edge of the accretion disk. That mass slowly accreted through our disk to the BH grid boundary (Figure 4.2). The accretion of the L1 tidal stream onto the outer edge of the disk forms an impact shock front that leads to a dense outflow stream beyond the disk. The accretor outflow stream is markedly similar in mass flux across both cases. However, the mass flux of the L1 stream itself differed greatly from $\dot{M}_{\text{L1(1.01)}} = 1.7 \times 10^{18}$ g/s in the barely overflowing 1.01-case to $\dot{M}_{\text{L1(1.1)}} = 8.80 \times 10^{22}$ g/s in the extreme overflow 1.1-case (Table 4.1). The L1 tidal stream in the 1.01-case exceeds the donor wind in density but not in total mass flux, unlike the 1.1-case which substantially exceeds the associated wind mass loss rate (Figure 4.1). This MT is conservative, with an efficiency $\alpha_{\text{MT(1.1)}} = 1.00$. The barely overflowing 1.01-case differed from the conservative model, with MT efficiency $\alpha_{\text{MT(1.01)}} = 0.86$ on average across the simulated time window. This discrepancy is a result of the relative invariance of external flows; much like the donor wind, the accretor outflow stream remains of the same order in mass and angular momentum flux in spite of the fifty thousand fold increase in $\dot{M}_{\text{L1}}$ between the two cases. In the 1.01-case, this accretor outflow therefore represents a much more substantial proportion of the overflowing material than it does in the 1.1-case.

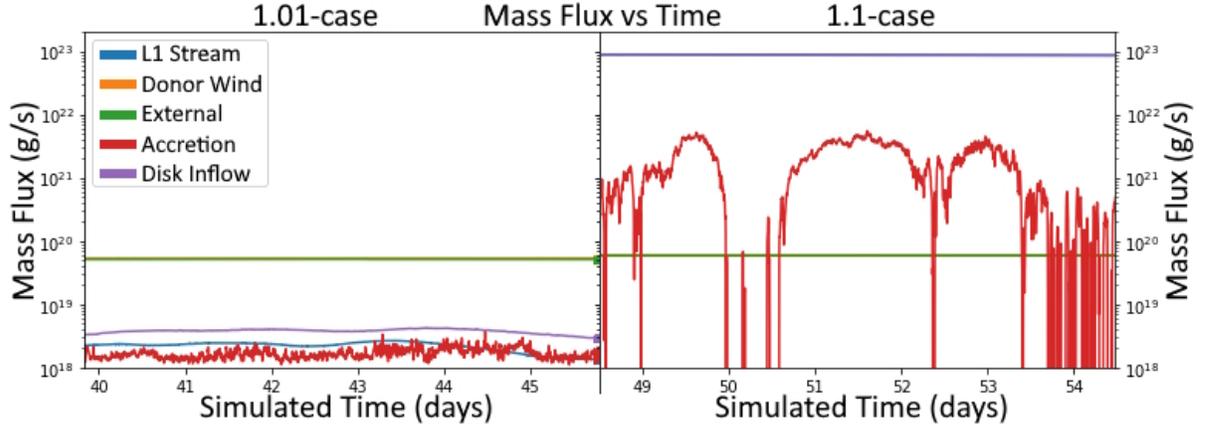

**Figure 4.1** Mass Fluxes in Parameterized Simulations. L1 Stream sums zones exceeding the maximal density of the supersonic wind at the L1 radius, $3 \times 10^{-13}$ g/cm$^3$. The lower mass zones at the same radius are summed for the Donor Wind. Disk Inflow corresponds to the net inflow through an accretor-centered spherical surface fully enclosing the disk. Accretion demarcates the net inflow through a spherical surface in the disk gap. L1 Stream and Disk Inflow of the 1.1-case are nearly identical across simulated time. Donor Wind and External mass fluxes are nearly identical across simulation time in both cases. We exclude the settling time required to achieve a quasi-steady state from initial conditions, in which fluxes were not generally observed nor physically meaningful.

As this outflow feature persists throughout both sets of simulations, we must ensure we refer to it consistently. This dense outflow from the accretor region exits the system via the outer Lagrange point; however, in systems of different mass ratios, the term "L2 stream" is otherwise defined. We therefore refer to this optically-thick outflow that exits the system from the accretor region due to inefficient accretion as the "accretor outflow stream".

In both the 1.01- and 1.1-cases, we observed the separation of the accretion disk into distinct concentric bands. In Figure 5.1, a visible separation forms between the tidal stream influx and the inner band of the accretion disk in both cases. Across a spherical shell separating the two bands, the mass flux contained to within ±5° of the ecliptic outweighs the off-ecliptic mass flux across both models. Despite the gap, this indicates that mass accretion is primarily driven by the tidal stream, not the donor wind. While our code can develop this banded disk structure, the computation time required to simulate the inner deposition onto the black hole is prohibitive. Additionally, observation of M33 X-7 suggests the innermost regime would be dominated by radiative feedback our model does not account for (Ramachandran et al. 2022). Therefore, we simulate only to an inner bound of $0.12 R_\odot$ and assume all mass that crosses the disk gap ($R_{\text{gap}} \sim R_\odot$) is eventually accreted, which is to be taken as an upper limit of accretion onto an X-ray source BH.

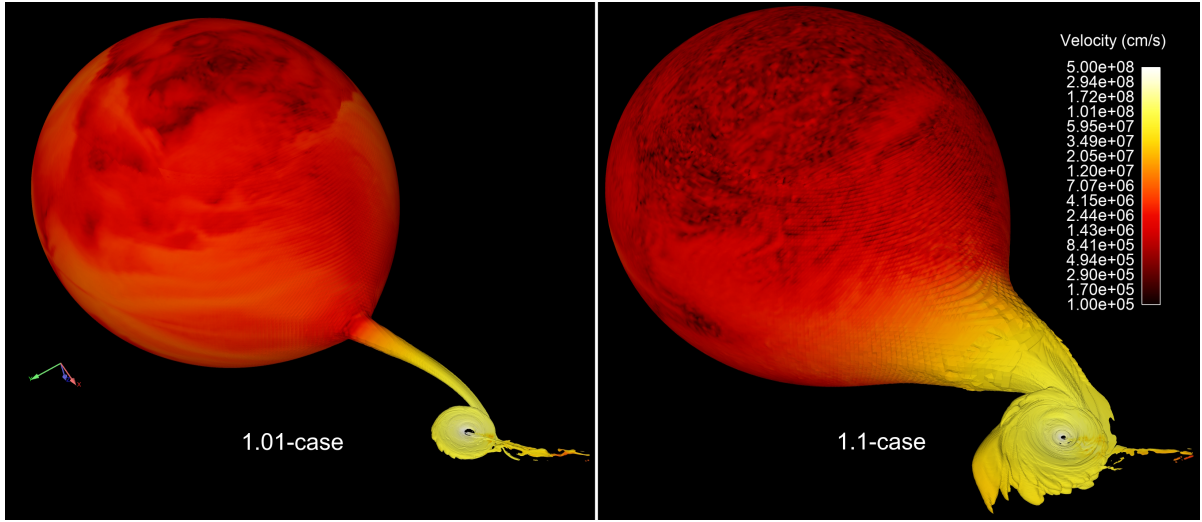

**Figure 4.2** Binary System Visualized by Parameterized Simulations. The tidal stream is angled with respect to the line connecting centers of mass and is centered on the ecliptic. View given from below the ecliptic to better image the stream impact shock front and the non-conservative accretor outflow stream formed where the tidal stream collides with the disk.

## 4.2 Sequential Simulations

Our series of twenty sequential simulations spanned a total of $616,307$ years of system evolution, comparable to the nuclear timescale of the donor star (Quast et al. 2019; Klencki et al. 2022). During that span, the system was evolved piecewise from $f = 1.0005$ to $f = 1.1779$ by increments equivalent to 1% of the characteristic timescale of the donor Roche lobe. We therefore map an 18% reduction in Roche lobe radius, reducing the volume of the Roche lobe by a factor of two. This simulation set is summarized in Table 4.3 and visualized in Figure 4.6, with the remaining models visualized in Appendix B. Our simulations begin with the unexpected absence of full RLO in model A.

Model A does not launch an optically thick tidal stream as its overfilling factor would indicate. We find model A remains in the wind RLO phase with an enhanced wind transferred through the L1 region onto the BH accretor. Due to this wind RLO, the accretor captures 3% of the donor wind as an average taken over simulation steady state. This rate corresponds to seventeen times the rate of Bondi-Hoyle-Lyttelton wind accretion for this system geometry, as calculated in Equation 1.6 (Edgar 2004; Davidson and Ostriker 1973). As wind RLO is expected to span up to a hundred times the wind accretion rate while lacking an optically thick tidal stream, we can confidently define model A as occupying the wind RLO phase (Mohamed and Podsiadlowski 2007). We discuss this further in context of prior research in Section 5.2. Despite occupying the traditionally stable wind RLO phase, the piecewise evolution induced by model A progressed model B into the onset of full RLO our series of simulations was intended to examine.

**Table 4.3** Overfilling Factors and Orbital Geometries of Sequential Runs. The symbol $f$ denotes the overfilling factor. "End State" describes the system that results from the piecewise evolution of our final simulation, model T.

| Model | $f$ | Mass Ratio | Binary Separation ($R_\odot$) | Period (Days) |
|---|---|---|---|---|
| A | 1.0005 | 3.33 | 35.3 | 3.45 |
| B | 1.0083 | 3.31 | 35.0 | 3.43 |
| C | 1.0163 | 3.29 | 34.8 | 3.40 |
| D | 1.0246 | 3.26 | 34.6 | 3.37 |
| E | 1.0329 | 3.24 | 34.3 | 3.33 |
| F | 1.0416 | 3.22 | 34.1 | 3.30 |
| G | 1.0501 | 3.20 | 33.9 | 3.27 |
| H | 1.0587 | 3.18 | 33.6 | 3.23 |
| I | 1.0673 | 3.16 | 33.4 | 3.20 |
| J | 1.0762 | 3.14 | 33.2 | 3.17 |
| K | 1.0849 | 3.12 | 33.0 | 3.14 |
| L | 1.0941 | 3.10 | 32.8 | 3.11 |
| M | 1.1030 | 3.08 | 32.5 | 3.07 |
| N | 1.1120 | 3.06 | 32.3 | 3.04 |
| O | 1.1210 | 3.04 | 32.1 | 3.01 |
| P | 1.1303 | 3.02 | 31.9 | 2.98 |
| Q | 1.1394 | 3.00 | 31.7 | 2.95 |
| R | 1.1487 | 2.98 | 31.5 | 2.92 |
| S | 1.1581 | 2.96 | 31.3 | 2.89 |
| T | 1.1681 | 2.94 | 31.0 | 2.87 |
| End State | 1.1779 | 2.92 | 30.8 | 2.84 |

We implemented the sequential series of simulations to accomplish multiple objectives. Firstly, we seek to map the relevant parameter space of overfilling factor by the associated rates, timescales, and efficiencies of the MT process. For both mass and angular momentum transport, these values are largely unconstrained by theory and observation; multidimensional simulations have also insufficiently explored these MT process parameters, as discussed in Sections 1.5.1, 1.5.2, and 2.1. To map as much of the relevant parameter space as possible, we endeavor to span the entirety of the RLO phase or nearly so. In that span, we also endeavor to simulate multiple RLO regimes, including crossing the eclipse inflection and exceeding the L2 potential. While each simulation provides additional insight, we chose to end our series after twenty runs due to exponentially briefer timescales, grid limitations, and the span of dynamical thresholds covered.

### 4.2.1 Span of Simulations

We intended our sequential runs to span as much of the RLO phase as possible. In that regard, we were successful. If we assume the trends obtained across our set of models continue smoothly to the end of the RLO phase, the system cannot maintain RLO for more than an additional 15 years. We therefore conclude that we have modeled $\geq 99.997\%$ of the RLO phase.

While the duration remaining is quite short, the timescale of each model is exponentially briefer than its predecessor, as illustrated in Figure 4.3. 88.5% of the 616,307-year duration of our series transpired in the first two piecewise steps, and 97.6% transpired in the first three. The piecewise evolution step calculated from the end of model T spans less than 4 years. Assuming the trends of our data persist, an additional $\sim 100$ simulations would be required to model the remaining 15 years required to reach a definitive end state to RLO. At that point, a merger would occur if CE onset or core collapse do not alter the rate of angular momentum loss in the system. Well before this merger occurs, the system would diverge from the RLO phase as constrained here. Even as early as model T, the system dynamics strain the grid geometry at our current resolution.

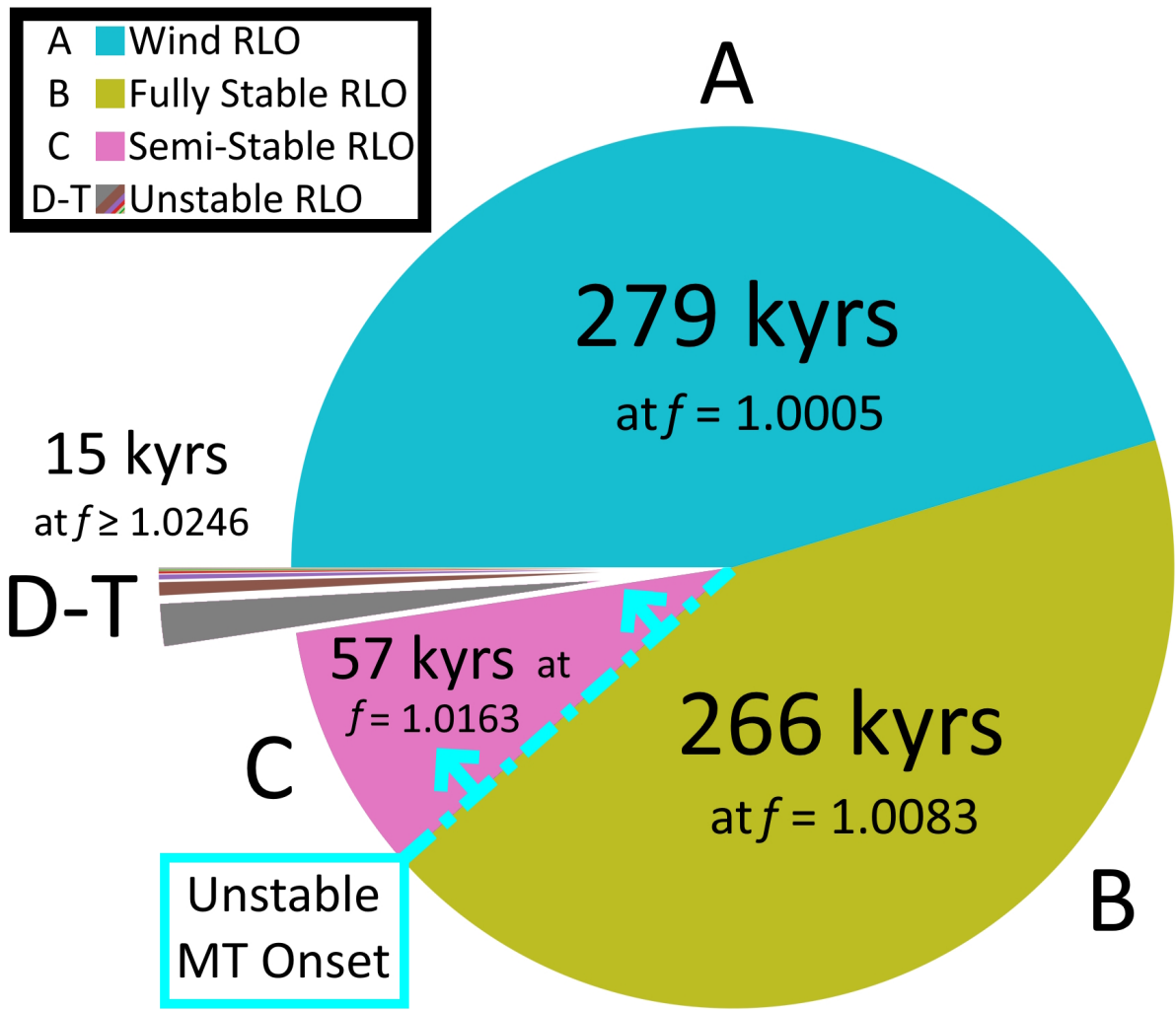

**Figure 4.3** Time Distribution of Sequential Simulations. We subdivide the models by their RLO regime (top left) and group models D through T due to their brief durations. The cyan threshold indicates the inferred unstable RLO threshold around $f \sim 1.012$ derived in Section 5.3.

This choice of endpoint is in part due to the limits of our resolution. Increasingly high overflows induce increasingly deep outflows in the envelope in order to feed the mass flux of the tidal stream. While our present grid allows for the donor to be well-resolved to a depth of $0.65 R_{\text{donor}}$, extending deeper with the same resolution would extend zones to larger than half of the photospheric scale height defined in Section 3.1.3. We elected instead to maintain this resolution across the series. As of model T, the inner boundary of the envelope experiences significant radial outflow in the ecliptic region, contributing to our decision to end the series with that model.

Finally, we are satisfied with our set of simulations as they span several key thresholds in RLO. Model M surpasses the overfilling factor of our 1.1-case simulation, guaranteeing our sequential series provides an even greater range of overfilling factors. Model S surpasses the eclipse inflection point of its system geometry, defined in Section 1.4. This ensures our series spans fluidly across the eclipse inflection threshold. Model T surpasses the effective potential of L2 for its system geometry, which impacts the escape of donor material from the accretor lobe after MT.

The next significant RLO threshold our simulation set could surpass would be the effective potential of L3. Based on the potential profile of the initial system geometry in Figure 1.2, this may require an overfilling factor as high as $f \sim 1.5$. As the system geometry has shifted to lower mass ratios, this may occur sooner, but remains well beyond the overflow of model T. We expect an additional $\sim 25$ simulations would be needed for our series to reach L3, by which point the deep tidal-feeding outflows in the envelope would be well beyond the scale our current grid resolution can fit smoothly.

### 4.2.2 Evolution of Orbital Geometry

Our sequential set of simulations were spaced by Roche timesteps equivalent to 1% of the Roche lobe characteristic timescale, but naturally situated with nearly even spacing in other system variables as well. In Table 4.3, we demonstrate the smooth evolution of overfilling factor, mass ratio, binary separation, and orbital period across our simulation set. These simulations were spaced drastically unevenly in time evolution, with the timestep between models A and B that is $\sim 10^5 \times$ longer than the timestep between models S and T. However, we found the variables listed in Table 4.3 to vary nearly-linearly across Roche timesteps.

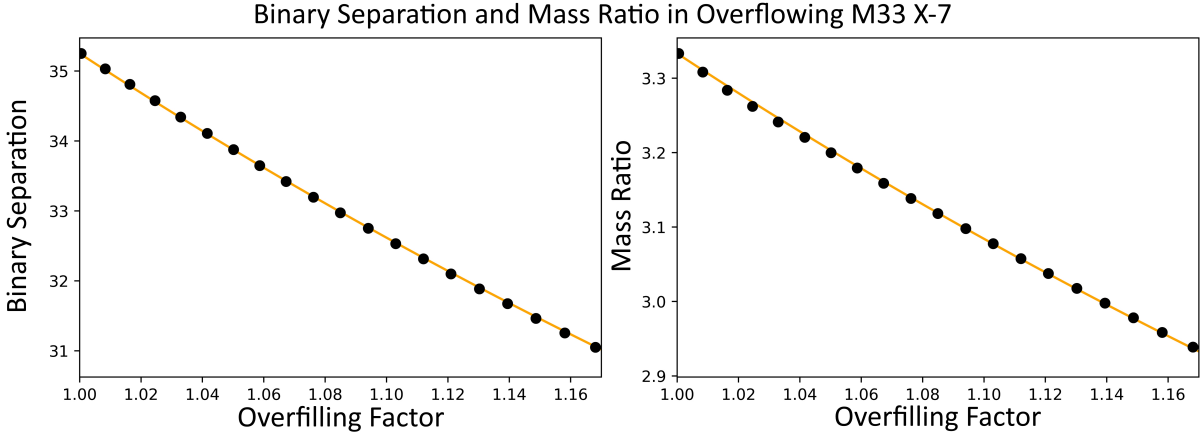

**Figure 4.4** Binary Separation and Mass Ratio vs. Overfilling Factor. The orange trendline corresponds to our power law solution described in Equation 4.1.

Increments of mass ratio and binary separation remained approximately constant across our sample, best fit by a power-law dependence on the overfilling factor. This implies, at least for the system geometry of M33 X-7,

$$q \quad \propto \quad d \quad \propto \quad f^{D_1} \tag{4.1}$$

where $D_1 = -0.815$ in the test system, though that value may not be general. We here use $q$ and $d$ to respectively denote the mass ratio and binary separation, and $f$ to denote the overfilling factor defined by Equation 1.7. We demonstrate this fit in Figure 4.4 with a coefficient of determination of $R^2 = 0.9999$ across the two datasets. The scope of applicability of this proportionality cannot be determined without the future examination of additional sequential simulations. Even limited to HMXBs of comparable mass ratios, Equation 4.1 may prove beneficial to future models, as will our measurements of MT efficiency.

### 4.2.3 MT Efficiency versus Overfilling Factor

We define four parameters of efficiency, given in Equations 1.11 and 1.12. In the combination of these parameters, we capture the total efficiency of mass and angular momentum transfer across the binary system as a whole as well those efficiencies specific to the L1 tidal stream. We have tracked these variables across overfilling factor parameter space and obtained the trends depicted in Figure 4.5. There are several trends present in Figure 4.5, both physical and artificial, which we here subdivide into four regimes.

The first regime is that of wind RLO prior to the onset of RLO. Model A, denoted by the first marker in each trendline and the first slice of Figure 4.3, did not experience sufficient overflow to drive an optically thick tidal stream. We observed this regime to be consistent with previous observations of wind RLO, with a binary system efficiency of 3%. This MT occurred almost exclusively through the enhancement of the wind near the L1 point. We discuss this exploration of the wind RLO regime further in Section 5.2.

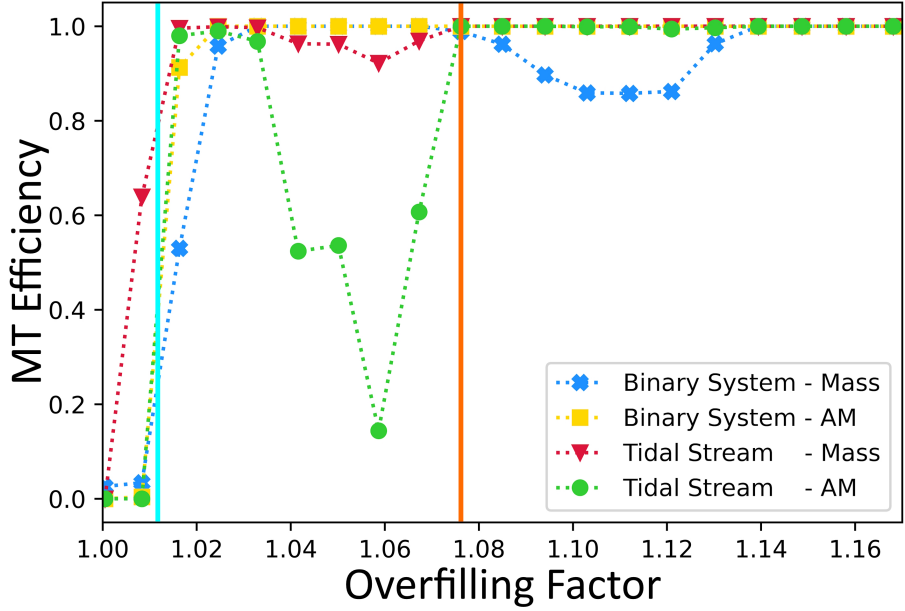

**Figure 4.5** Mass Transfer Efficiency vs. Overfilling Factor. Depicted are the efficiencies of mass and angular momentum (AM) fluxes through the L1 tidal stream and overall system across overfilling factor parameter space. The cyan bar indicates the inferred unstable RLO threshold $f \sim 1.012$. The orange bar indicates the modification made to disk retention with model J; those models at or right of the bar retain the disk solution of the prior model.

Notably, we found no significant accretion of angular momentum onto the BH on the timescale of our simulation. As model A lacked a tidal stream, its associated efficiencies are given as zero. This arises from our definition in Equation 1.12, as model A did feature a very slight optically-thick accretor outflow stream. The overflowing wind converged on the accretor, commingling its angular momentum from differing trajectories deflected towards the BH. We suspect the same grid with a deeper inner boundary would have allowed this wind to form a bow shock, as is traditionally expected in wind RLO. In the absence of a bow shock, we saw the formation of a dense stream outflowing from the accretor region. This accretor outflow stream carried away nearly the totality of the angular momentum which had entered the accretor region via the overflowing wind. The effect of this stream offers a probable explanation for the lack of angular momentum gained by the accretor in model A.

The second regime describes stable MT. This spans the next few models $f \lesssim 1.02$. These models lie near the threshold we discuss in greater detail in Section 5.3, in which we examine the rate and timescale of MT. This threshold approximately defines the beginning of the transition between stable and unstable MT. We depict the unstable MT threshold with the cyan vertical bar in Figure 4.5, but our third and fourth models remain somewhat stable beyond that threshold. In this regime, we see increasing rates of mass flux in the growing tidal stream that quickly surpass the total mass flux of the donor wind between our third and fourth models (C and D).

In this regime, the tidal stream is not immediately fully conservative. In model B, we see the dense tidal stream outweigh its corresponding accretor outflow stream in mass by a factor of 3, leading to an efficiency of 64%. However, the accretor outflow stream still matches the L1 stream in angular momentum, removing nearly all the angular momentum the L1 stream would deposit onto the accretor. This rapidly changes with model C, in which the tidal stream carries 92% as much mass as the donor wind. Throughout the remainder of the stable MT regime, mass and angular momentum transfer through the tidal stream are nearly conservative.

The binary system efficiencies in the stable MT regime provide a fascinating contrast. In our second model, the relatively small tidal stream increases the overall efficiency of the system by less than 1% on the scale of Figure 4.5 in both mass and angular momentum metrics. However, our third model shows a drastic rise. As expected for a fully conservative tidal stream that drives approximately half of the donor's mass loss, the tidal stream of the third model raises the total system efficiency to $\sim 50\%$. The more consequential change is the shift of angular momentum efficiency from $< 1\%$ to 91%. This indicates that the tidal stream is at least an order of magnitude more effective at transmitting angular momentum onto the accretor than wind accretion or wind RLO per unit mass. We see system efficiencies in both mass and angular momentum approach conservative in the following model.

The third regime defines unstable MT without disk inheritance. This spans from approximately model D to model I, positioned just before the orange vertical bar on Figure 4.5. In this regime, unstable MT rapidly expands the disk as the overfilling factor and tidal mass flux increase. The accretion disk extends substantially beyond the inner disk band we consider as an inner boundary for this set of simulations and forms an outer band on-grid. However, prior to model J, we did not sufficiently account for the rate at which this expansion was occurring. As such, models in this regime did not fully evolve the formation of their outer disk bands in the duration of the model. This problem was accentuated by initializing without any disk present. As successive models differed more from this initialization, the efficiency of the tidal stream was disrupted. This is primarily the product of erratic dense outflows that occur during disk settling, which contribute to tidal stream efficiency but whose effects are muted in the total system efficiency.

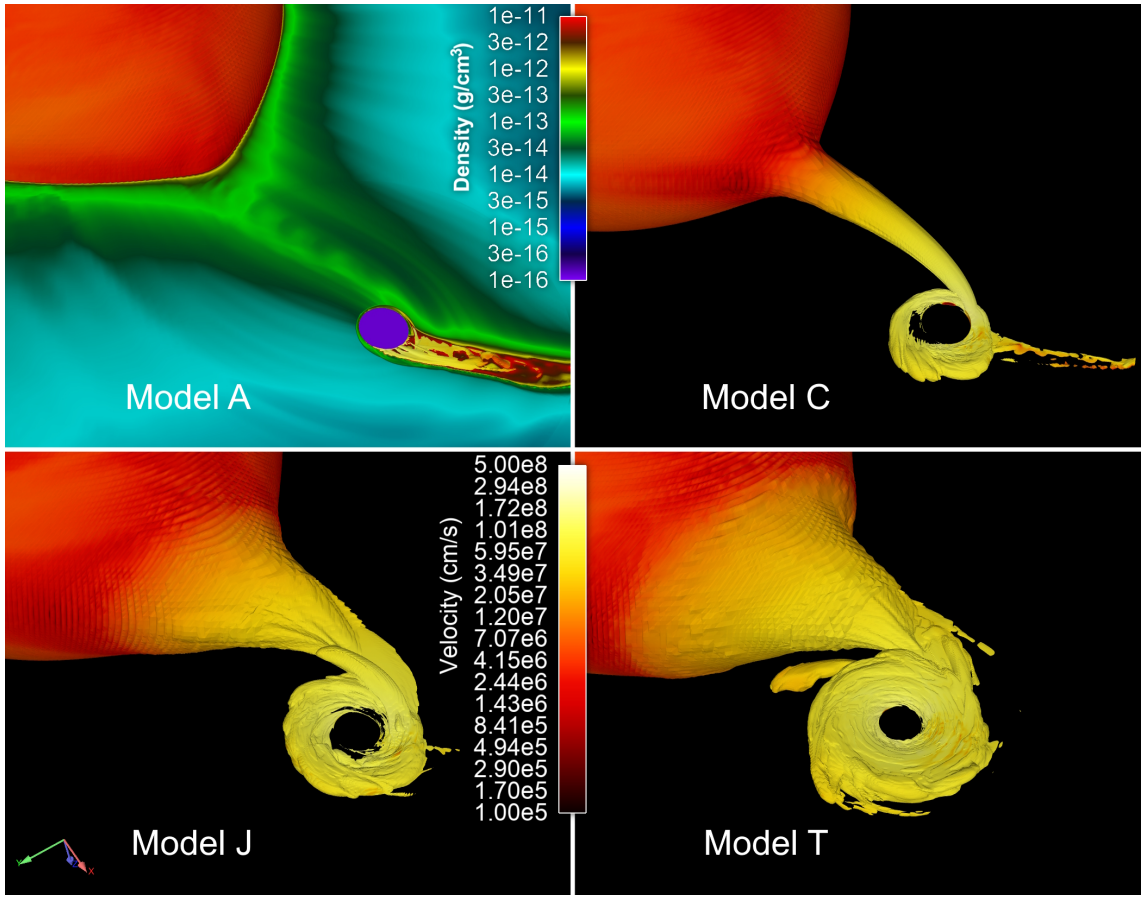

**Figure 4.6** Binary System Visualized by Sequential Simulations. The donor, tidal stream, and outer band of the accretion disk are visualized by an isodensity surface for sequential models A, C, J and T. The equatorial plane is additionally mapped for model A, as no dense stream is generated. For the corresponding visualizations of all twenty sequential models, see Appendix B.

As validated by the total binary system efficiencies, we can take this regime to be fully conservative in both mass and angular momentum transport. In physical systems, this regime spans thousands of years. Therefore, the growth of the disk should be gradual enough as to almost entirely mitigate the issue of erratic dense outflows due to the perturbation of increased L1 mass flux. We take deviation from the fully conservative case in this regime to be fully due to non-physical constraints of our method.

Starting with model J, we enter the fourth regime, denoted by the orange vertical bar in Figure 4.5. This regime spans a century in which more and more rapid unstable MT occurs, rendering it effectively invisible on Figure 4.3. This regime differs substantially from the third regime due to a change in method; subsequent models inherit the evolved disk solution of the prior model. By maintaining a settled disk across models, we no longer see the erratic dense outflows of the third regime. As such, the tidal stream is fully conservative in both mass and angular momentum across this regime. The binary system mass flux differs due to the growth of the disk, though the external boundary of our grid validates that no additional mass is escaping the accretor lobe. Across the span of several simulations, the disk grows beyond the radius anticipated. This deposition of mass in outer radii is unaccounted for in the calculation of binary system mass flux, leading to an apparent deficit. While this has more physical merit than the erroneous inefficiency of the tidal stream in the third regime, this regime still spans sufficient time that such growth will not noticeably reduce the efficiency of MT.

We may therefore conclude unstable MT to be fully conservative across a large range of overfilling factors in our test system. Similarly, stable MT is highly efficient, particularly in the transport of angular momentum. Stable MT rapidly becomes conservative as the tidal stream surpasses the donor wind in mass flux. In our simulation, this transition occurred around $f = 1.016$.

These efficiencies are presented without consideration to the efficiency of mass and angular momentum transport after entry into the accretion disk. While our model is capable of transporting angular momentum to allow for accretion, we lack a prescription for radiative feedback in the near-BH region. Especially in the high $f$ regime, the system is likely to enter super Eddington accretion rates in which radiative feedback is essential to determining the complete efficiency of mass gained by the accretor (Ramachandran et al. 2022).

# CHAPTER

# 5

# DISCUSSION

This chapter integrates and adapts some work previously published in Dickson (2024) in Section 5.1 and Figure 5.1.

In two sets of simulations, we have modeled RLO in the HMXB M33 X-7 in examination of the rate, efficiency, and timescale of RLO-induced MT and angular momentum transfer. We here interpret the results obtained in comparison to the existing state of RLO research.

Our simulations are divided into a parameterized set and a sequential set. The parameterized set spans two models differing across overflow parameter space at $f = 1.01$ and $f = 1.1$, which we refer to by their overfilling factors as the 1.01- and 1.1-cases. These models are otherwise identical in initial conditions. Our sequential set, by contrast, begins with one simulation at $f = 1.0005$ and proceeds through a piecewise evolution across the duration of the RLO phase. This set spans a total of twenty simulations named alphabetically for their order as models A through T.

We subdivide our detailed discussion of our results across three realms of analysis. We first compare our parameterized simulations on the basis of the RLO-induced geometry of both disk and tidal stream. We also examine model A as an evident wind RLO system in the $f > 1$ regime traditionally modeled by full RLO. Finally, we provide a detailed discussion of MT properties as they vary with overfilling factor, particularly in models C-T.

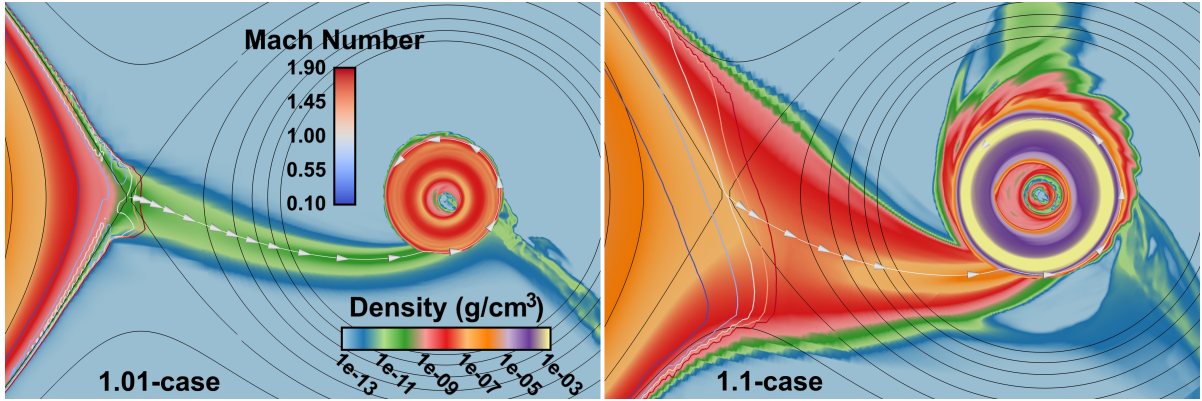

**Figure 5.1** Tidal Stream Geometry in Ecliptic Plane. The system ecliptic is mapped with equipotential lines in black. Streamline arrows indicate flow pattern extending from the L1 point. Colored contours have been provided to map the transonic region. Accretion disk bands separated by the lower density gap are shown by density gradient.

## 5.1 Geometry of RLO

Theoretical models of RLO exist in the legacy of the predictions of Lubow and Shu (1975) and have advanced with more recent models thus far culminating in the work of Marchant et al. (2021). We therefore use these two models as our primary points of comparison to theoretical prescriptions of RLO. These prescriptions center the geometry and density of the L1 stream as well as the way other system fluxes vary with respect to changes in L1 mass flux. We also compare to the widely-used mass ratio stability prescription well-summarized by Frank et al. (2002).

Our model was built from the values observed by Ramachandran et al. (2022). As such, we can comment only on the validity of their particular observational model of M33 X-7 without significant comment to be made on earlier observations of the same system. In particular, they characterized the system as undergoing a transition from stable to unstable MT at an overfilling factor of $f \sim 1.2$. Our simulations can speak to the validity of that model as well as to the mass fluxes of donor wind and BH accretion their model obtained.

The tidal stream geometry set out by Lubow and Shu (1975) is largely validated by our 1.01-case model, though the extreme overflow of the 1.1-case differs. In keeping with their semi-analytical model, both our 1.01-case and 1.1-case tidal streams achieve Mach 1 precisely at the L1 point (Figure 5.1). We also found the angular momentum carried by the tidal stream to be on the order of the circular orbit through the L1 point about the system's center of mass as predicted by Lubow and Shu (1975) (Table 5.1). They also prescribe the angle of the tidal stream from the most direct line between the two bodies as between 19.5° and 28.4° across all mass ratios. For the mass ratio of M33 X-7, the predictions of Lubow and Shu (1975) differ from our observed angle by 2.5°. Shown in Figure 5.1, both models follow equivalent angles of deflection despite disparate overfilling factors, as predicted.

**Table 5.1** Comparison to Quantitative Predictions of Lubow and Shu. Values expressed in cgs units except where otherwise specified. Vertically centered values indicate predictions independent of overfilling factors or their associated L1 mass fluxes.

| Property | Case | Our Data | Marchant et al. |
|---|---|---|---|
| Tidal Stream Angle | Both | 17.5° | 20° |
| Tidal Stream Width | 1.01 | $4.2 R_\odot$ | $1.3 R_\odot$ |
|  | 1.1 | $12 R_\odot$ |  |
| Donor L1 Angular Momentum Loss Rate | 1.01 | $3.1 \times 10^{37}$ | $3.2 \times 10^{37}$ |
|  | 1.1 | $1.18 \times 10^{42}$ | $1.67 \times 10^{42}$ |
| L1 Lagrange Point Density | 1.01 | $4.0 \times 10^{-10}$ | $1.12 \times 10^{-10}$ |
|  | 1.1 | $1.1 \times 10^{-7}$ | $5.94 \times 10^{-6}$ |
| Disk Outer Radius | 1.01 | $2.47 R_\odot$ | $4.9 R_\odot$ |
|  | 1.1 | $3.83 R_\odot$ |  |

Unlike the predictions of Lubow and Shu (1975), the 1.1-case tidal stream forms a conical funnel, rather than remaining of nearly uniform width across the nearly straight segment preceding its inspiral. This discrepancy is likely the product of their assumption that overflow occurs only in a small region about the L1 point; in the 1.1-case, the tidal stream width at L1 was $12 R_\odot$, of the same order as the binary separation. Lubow and Shu (1975) predicted stream width to be invariant with respect to L1 mass flux, which arises from their small overflow region assumption. For the parameters of M33 X-7, they prescribe the tidal stream to be of width $1.3 R_\odot$. Both of our cases substantially exceeded this value, and did so to varying degrees that suggest a dependence upon overfilling factor.

In both cases, a cross-section at L1 revealed the streams to exhibit circular symmetry with heights identical to their widths to the limits of our grid resolution. We observe donor envelope flows to be primarily constrained to the ecliptic except in the solid angle beneath the tidal stream in both cases. Significant outflows occur outside the ecliptic in this region, which is much larger in the high overflow case. These flows generate circular symmetry in the stream well away from the axis connecting centers of masses, where stream densities are substantially lower than at its center. These outflows remain shallower and slower than those observed in the ecliptic. This suggests the majority but not entirety of mass entering the tidal stream is first driven into the equatorial region prior to overflowing as predicted by Lubow and Shu (1975).

**Table 5.2** Comparison to Quantitative Predictions of Marchant et al. Values expressed in cgs units except where otherwise specified. Mass capture rates indicate net mass entering the disk regime.

| Property | Case | Our Data | Marchant et al. |
|---|---|---|---|
| Tidal Stream Width | 1.01 | $4.2 R_\odot$ | $1.4 R_\odot$ |
|  | 1.1 | $12 R_\odot$ | $10 R_\odot$ |
| L1 Lagrange Point Density | 1.01 | $4.0 \times 10^{-10}$ | $2.29 \times 10^{-10}$ |
|  | 1.1 | $1.1 \times 10^{-7}$ | $1.60 \times 10^{-7}$ |
| Donor Wind Mass Capture Rate | 1.01 | $2.3 \times 10^{18}$ | $3.49 \times 10^{18}$ |
|  | 1.1 | $2 \times 10^{18}$ | $4.38 \times 10^{18}$ |
| Donor L1 Mass Capture Rate | 1.01 | $1.5 \times 10^{18}$ | $2.27 \times 10^{18}$ |
|  | 1.1 | $8.80 \times 10^{22}$ | $3.61 \times 10^{22}$ |

The discrepancy of tidal stream width correlates closely with a discrepancy in L1 density. While the 1.01-case had comparable density to their predicted value at the L1 point, our 1.1-case was significantly broader and less dense than their model would predict for the same mass flux. They also predict the outer radius of the accretion disk to be independent of tidal stream mass flux, prescribing a disk radius of $4.9 R_\odot$. This arose from the radius at which they predicted a ballistic stream would collide with itself after one orbit of the accretor. Both our stream widths and deflection angles differed from this geometry, which may account for the difference between their predicted value and the more compact disks we observed (Lubow and Shu 1975). As we do not examine angular momentum transport all the way in to the BH, there exists a possibility that this transport would broaden the disk radii. However, this broadening is limited by the stream-disk interaction, which primarily expels angular momentum through the accretor outflow stream, which carries a specific angular momentum of $1.1 \times 10^{20}$ cm$^2$/s.

Recent work by Marchant et al. (2021) provides an updated semi-analytical method which generalizes beyond the small overflow approximations of Lubow and Shu (1975) and presumes adiabatic tidal flow rather than isothermal; their method more closely fit our results of tidal stream mass flux and geometry. Their method was accurate to within an order of magnitude in predicting the wind and L1 stream mass capture rates for both the 1.01- and 1.1-cases, where the mass capture rate is defined by the donor mass flux captured by the accretor (Table 5.2). We similarly validated their method with respect to the density at the L1 Lagrange point. Their methodology relies on the assumption of the L1 point being solely fed by surface flows along the Roche equipotential surface, which differed from our observations; this did not substantially impact the density obtained, though it may play a more substantive role at greater overflow than we studied here. Their method predicts tidal stream widths from the effective potential difference in the L1 plane. This method underestimates our findings but provides a substantially better fit than the Lubow and Shu (1975) prescription (Marchant et al. 2021). Our final point of comparison to the work of Marchant et al. (2021) is in their calculation and examination of MT stability.

MT stability is often reduced to a simple threshold of mass ratio, but we here substantiate that non-conservative dynamics can generate stable MT well beyond the customary mass ratio limit (Frank et al. 2002). MT stability in semi-contact binaries is defined by the proportional rate of change in the volume-equivalent Roche lobe radius $R_{\rm RL,vol}$. If we assume fully conservative MT with no angular momentum loss modeled in isolation of the donor wind, as is customary, we may instead use a critical value of mass ratio (Frank et al. 2002). M33 X-7 has a mass ratio $q = M_{\rm donor}/M_{\rm acc} = 3.3 \gg 0.83$, which would place this system well within the customary unstable MT regime (Frank et al. 2002; Marchant et al. 2021; Ramachandran et al. 2022). However, instead of using the $q \sim 0.83$ stability limit, we may more accurately calculate $\dot{R}_{\rm RL,vol}$ from the generalized analytic formula given in Eggleton (1983). This yields a Roche lobe shrinking timescale of 24 Myrs in the 1.01-case, greatly differing from the 1.1-case timescale of 4200 yrs (Equation 1.10). The 1.1-case timescale may be sufficiently short for corotating synchronicity to break down; if that is the case, use of the Roche geometry may not fully capture the system dynamics.

In the 1.01-case, the change in Roche lobe radius is driven primarily by the wind overflow, and its Roche lobe shrinks on a timescale of the same order as the total angular momentum loss and total mass loss of M33 X-7. This case therefore corresponds to stable MT occurring well outside of the customary mass ratio limit $q \sim 0.83$ (Frank et al. 2002).

With a Roche timescale of 24 Myrs, the 1.01-case occurs significantly slower than the nuclear timescale of M33 X-7. Furthermore, this timescale exceeds the entirety of M33 X-7's main sequence evolution (Weidner and Vink 2010). Therefore, the 1.01-case is subject to evolutionary response from the donor which may vary its radius substantially. In contrast, the 1.1-case overflow occurs on the donor's thermal timescale, which may preclude radial response from the donor star (Quast et al. 2019; Klencki et al. 2022).

The 1.1-case's rapid Roche lobe evolution, driven by the high-overflow L1 tidal stream, places it in the unstable MT regime. The discrepancy of stability regime between the 1.01- and 1.1-cases supports the conclusion of Ramachandran et al. (2022) that the system is transitioning from stable to unstable MT. The stability of the 1.01-case, in contrast to the 1.1-case of identical mass ratio, suggests the mass ratio's invalidity as a predictor of system stability and progenitor pathway. Stability regime may be more effectively constrained by regime of overfilling factor $f$.

In our sequential simulations, we extended to near the proposed $f = 1.2$ overflow of Ramachandran et al. (2022) with a final system state of $f = 1.1779$. In this high overflow regime, we found overflow to evolve on the timescale of years, much shorter than its Roche timescale (Equation 1.10). Our model set suggests that M33 X-7 cannot sustain overflow of $f \geq 1.1$ for a period exceeding 100 years due to the high rate of mass loss in the extreme overflow regime. Observing a system in this regime is rendered implausible by the brevity of the phase.

In the rapid overflow regime of $f \geq 1.0246$ visualized in Figure 4.3, our assumption of initial thermal equilibrium breaks down. Both the 1.1-case of our parameterized set and the simulations D through T of our sequential set exist in a regime of overflow in which dynamical system changes occur more rapidly than the thermal timescale of the donor star. This prevents the star from maintaining thermal equilibrium in its outer envelope, which would drive hotter, denser gas into the L1 stream than we have modeled here. We may therefore take the timescale and stability prescriptions we provide for the rapid overflow regime as upper limits, with physical systems expected to be briefer and less stable than the results of our analyses.

MT stability in this extreme regime may be further altered by a secondary tidal stream at the outer Lagrange point of the donor. The formation of a secondary MT stream at L2/L3 that could significantly modify the stability regime has been suggested by Marchant et al. (2021); our simulations were unable to launch such a stream except under extreme overflow significantly beyond the 1.1-case and simulation T. The overfilling factor required to generate a secondary stream in M33 X-7 is $f = 1.482$, as visualized in Figure 1.2.

Our simulation provides novel constraints on the efficiency of mass and angular momentum delivery in RLO-fed MT. In keeping with the predictions of Marchant et al. (2021), we found the 1.1-case to have significant loss of angular momentum from the donor, which would result in orbital hardening without requiring a CE phase (Table 5.2). The results of both cases validated the prediction of Lubow and Shu (1975) that the L1 tidal stream carries angular momentum equivalent to that of a circular orbit of the center of mass that passes through the L1 point. While our simulation cannot resolve all the way to the BH accretor surface, we provide upper limits of mass accretion taken from the mass flux inflowing through the disk gap without X-ray feedback. Even at this upper limit, we found the 1.01-case did not exceed the Eddington luminosity. Our 1.01-case accretion averaged $1.7 \times 10^{18}$ g/s, or $\sim 73\%$ of the Eddington value; this exceeds but is of the same order as the $\sim 12\%$ finding of Ramachandran et al. (2022). Our 1.1-case accretion was orders of magnitude higher, $\sim 1000\times$ the Eddington value, suggesting an extreme rate of X-ray feedback would be required to reach equilibrium.

**Table 5.3** Comparison to Quantitative Predictions of Ramachandran et al. Values expressed in cgs units. Vertically centered values indicate predictions independent of overfilling factors or their associated L1 mass fluxes.

| Property | Case | Our Data | Ramachandran et al. |
|---|---|---|---|
| Donor Wind Mass Loss Rate | 1.01 | $5.36 \times 10^{19}$ | $4.0 \times 10^{19}$ |
|  | 1.1 | $6.03 \times 10^{19}$ |  |
| Accretion Rate | 1.01 | $1.71 \times 10^{18}$ | $2.8 \times 10^{17}$ |
|  | 1.1 | $2 \times 10^{21}$ |  |

Beyond stability, another side-effect of RLO MT in our simulation was the enhancement of the wind. Though the wind remained of the same order, we saw a significant discrepancy in $\dot{M}_{\text{wind}}$ between the 1D case and our two parameterized 3D cases, 1.01 and 1.1. We fit our 1D model to $\dot{M}_{\text{wind(1D)}} = 4.0 \times 10^{19}$ g/s as observed by Ramachandran et al. (2022). The same simulation parameters defined our two 3D cases, however they differed unavoidably. Due to the rotation of the primary, we saw slower, denser wind near the ecliptic and faster, less dense wind driven from the poles. We therefore discuss wind mass flux as a sum across the spherical surface that excludes the solid angle represented by the optically-thick L1 tidal stream. With this method, our 1.01-case saw a $\sim 35\%$ increase over the 1D wind mass flux. The 1.1-case saw an even greater $\sim 52\%$ increase over the 1D case (Table 5.3). This is made all the more notable for the 1.1-case's wide L1 tidal stream excluding an even greater solid angle from wind contribution. We also observed this wind enhancement effect in our sequential set of simulations, in a range comparable to the parameterized set. Model A saw a $\sim 34\%$ enhancement while later models saw as much as $\sim 41\%$. Though recent work by Blondin and Taylor (2024) has examined an analogous effect in the wind RLO regime, further research is required to separate the impact of 3D modeling wind asymmetries, the result of greater donor surface area in RLO systems, and the direct consequence of the L1 tidal stream on wind enhancement.

## 5.2 Wind RLO of Model A

Wind RLO has primarily been proposed as a mechanism of wide binaries, and it has therefore rarely been simulated for HMXBs (Mohamed and Podsiadlowski 2007). In our system at $f = 1.0005$, we found the BH of model A to capture 3% of the donor wind without the formation of an optically thick tidal stream. As our sequential models do not simulate close enough to the BH to resolve a bow shock as is generally expected, we take our results obtained as an upper limit of BH accretion. We highlight two previous works with similar systems for comparison.

In El Mellah et al. (2019), a system was simulated with similar wind RLO to that of our A model. They modeled an HMXB with a BH accretor at a mass ratio of 2 and filling factor $f = 0.9$, for which they estimated the BH to capture 7% of the donor wind at our wind velocities, approximately an order of magnitude higher than the value predicted by wind accretion alone. While their system differs on several key parameters, the wind capture rate they obtained is comparable to our result of 3%.

Blondin and Taylor (2024) used 3D hydrodynamics to model wind RLO in the HMXB Vela X-1. For donor of $v_\infty = 1450$ and $f = 0.99$ with a mass ratio of 12.8, they present an L1 mass flux of 5% of the donor wind. While their mass ratio differs significantly from the 3.33 mass ratio of M33 X-7, the similarities in terminal wind velocity and filling factor provide a promising point of comparison. The 5% yield they report is not directly a statement of mass captured by the accretor, and such a measurement may not be directly informative given the presence of a neutron star accretor rather than a BH in Vela X-1. However, their L1 mass flux is analogous to first order with our method of calculating accretor wind capture rates.

Model A is therefore in keeping with the range of values set out in comparable models and identifies that wind RLO may occur even in the $f \gtrsim 1$ regime, rather than strictly defined by the limit $f < 1$ as previously suggested (Mohamed and Podsiadlowski 2007; El Mellah et al. 2019). This further supports recent research examining RLO in close binaries and HMXBs specifically.

Wind RLO in high mass ratio binaries has also been discussed in the context of stability. El Mellah et al. (2019) suggest wind RLO provides a consistently stable alternative to RLO-induced MT in high mass ratio binaries. This arises from their conclusion that wind RLO is always stable due to its low rate of mass flux. As has become customary in such works, they take the criterion $\dot{M}_{\text{donor}} \leq 10^{-6} M_\odot/\text{yr}$ to define the stable regime (El Mellah et al. 2019; Sun et al. 2023). Our model A was slightly below that limit with a mean value of $\dot{M}_{\text{donor}} = 8.5 \times 10^{-7} M_\odot/\text{yr}$, of which only $2.2 \times 10^{-8} M_\odot/\text{yr}$ was captured by the accretor. By our method of Roche lobe timescale calculation, model A was definitively stable as well; it exhibited a characteristic timescale on the same order as the entire lifespan of an O star, in the tens of millions of years.

Despite both of these stability regimes being satisfied, our wind RLO model led directly to the onset of unstable RLO. The possibility of wind RLO leading to unstable RLO-induced MT has been recently proposed by Sun et al. (2023). Their parameterized set of simulations examined low mass binaries at similar mass ratios to M33 X-7. They found that systems modeled with prescriptions for both wind RLO and full RLO were more likely to result in unstable RLO end states than those that neglected the possibility of wind RLO.

Our results validate this discrepancy. Our sequential set of simulations found the Roche lobe shrinking induced by wind RLO to directly result in a transition to full RLO with stable MT. This stable MT continued the same process of increasing overfilling factor and transitioned into rapid unstable MT well beyond the $10^{-6} M_\odot/\text{yr}$ limit. This series of transitions occurred across a duration two orders of magnitude shorter than the characteristic timescale of the wind RLO exhibited by model A.

## 5.3   MT as a Function of Overfilling Factor

For the first time, we map the related trends of mass flux through the L1 region and the Roche timescale defined by Equation 1.10 across variation in overfilling factor. Through this examination, we characterize the evolution of these parameters and identify a threshold of overfilling factor across which system dynamics shift. These conclusions are made possible by our sequential simulations, which allow the system geometry to evolve piecewise as described in Section 3.6.

Our sequential series of models achieved a wide range of mass fluxes across the overfilling factor parameter space ranging $1.0005 \leq f \leq 1.1681$. Due to the observed trend of timescales, we conclude this set of simulations to span $\geq 99.997\%$ of the RLO phase. Within this parameter space, MT is distinctly subdivided into two regimes.

In the first regime, spanning the majority of the system evolution time, MT is distinctly stable. This spans from the wind RLO of model A into subsequent full RLO in low overfilling factors. Due to our methodology of piecewise evolving by a step of 1% of the Roche timescale, this regime is only spanned by two of our models. As such, we do not attempt to characterize this regime with a fit equation. However, Roche timescales remained of the order 10 Myrs regardless of L1 tidal stream formation, which is of the same order as the main sequence evolution of the donor star. This regime spans $\sim$ 540 kyrs of piecewise evolution.

Taken collectively, this first regime is comparable to the nuclear timescale of the donor star. This suggests the donor star may generate a radial response to MT during the stable regime that could stabilize it indefinitely or accelerate the shift into the second regime. Future research may benefit from incorporating a stellar evolution code such as MESA into our piecewise evolution methodology to examine the impact of the stellar response during this regime (Paxton et al. 2010).

The second regime is characterized by the onset of unstable MT. The first three models of this regime (C-E) are arguably stable, spanning $\sim$ 70 kyrs with Roche timescales within an order of the evolutionary timescale. Unlike the first regime, however, this somewhat stable MT process occurs on a steep gradient of dynamic evolution shown in Figure 5.2. Beyond these first few, the remaining models (F-T) exhibit distinctly unstable MT across a span of only $\sim$ 2 kyrs. This feedback loop propels the system into ever-increasing overfilling factors and L1 mass fluxes that exceed $10^{-2} M_\odot$/yr or $10^{24}$g/s within our simulation set.

This unstable MT regime occurs more rapidly than the thermal timescale of the donor star. Existing research suggests this brevity and extreme rate of L1 mass flux may preclude radial response from the donor star until the end of main sequence evolution (Passy et al. 2012; Quast et al. 2019; Klencki et al. 2022). Therefore, should the system enter this regime, it is very likely to follow the progression we model here without needing to take into account additional considerations of stellar response.

In our analysis of MT efficiency in Section 4.2.3, we subdivided this second regime into the second, third, and fourth regimes of Figure 4.5. The dataset as represented by Figure 5.2 in this section exhibits a substantially smoother transition between these regimes, such that their exact distinctions are not relevant to this discussion. However, the threshold at which the first regime of stable MT transitions into the onset of unstable MT in the second regime is distinct and quantifiable in the map of L1 flux and Roche timescale across overfilling factor parameter space.

The separation between the first and second regimes is delineated by an approximate threshold of $f \sim 1.012$. We represent this threshold with a cyan vertical line on Figures 4.5 and 5.2. To obtain this threshold, we first fit the curvature of the unstable MT regime for both L1 mass flux ($\dot{M}_{L1}$) and Roche timescale ($\tau_{Roche}$). With $R^2 > 0.99$ for both datasets, we derived the fit

$$\dot{M}_{L1} \quad \propto \quad \tau_{Roche}^{-1} \quad \propto \quad \log_{10}^{D_2}(f) \quad , \tag{5.1}$$

where $D_2 = 4.42$ in the test system, though that value may not be general. We match this trend to our dataset with constants of proportionality $1.9 \times 10^{29}$ and 0.002, respectively without requiring an additional variable for intercept offset. This trendline would exceed the Roche timescale of models A and B at $f \sim 1.012$, which we therefore take as a lower threshold of the second regime. This threshold, like the fit exponent $D_2$, may vary by system geometry.

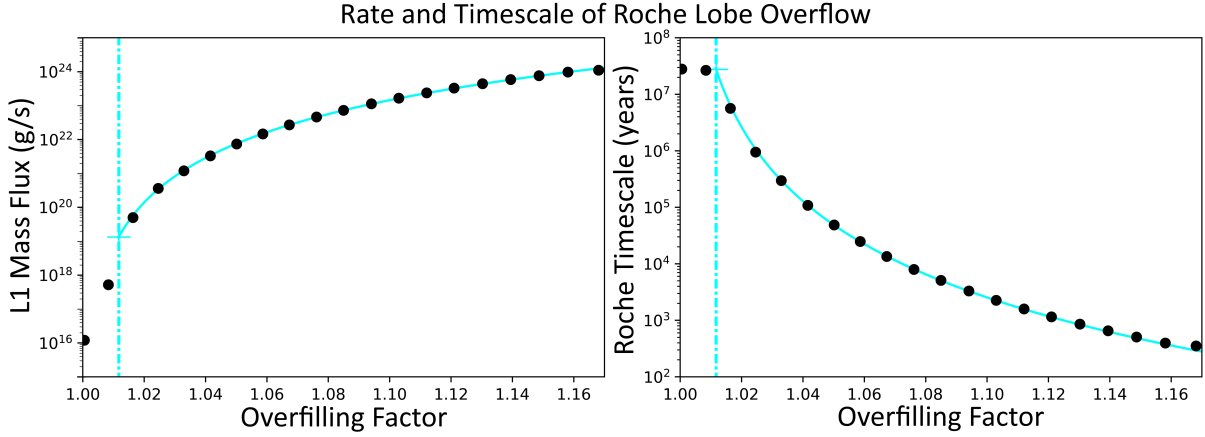

**Figure 5.2** L1 Mass Loss Rate and Roche Timescale vs. Overfilling Factor. The cyan trendline corresponds to our modified logarithmic solution for RLO. The solution begins from the inferred unstable RLO threshold, marked by the vertical line around $f \sim 1.012$.

This is in keeping with the results of Savonije (1978) for low mass binaries of comparable mass ratios and periods. When their set of simulations was allowed to develop with a simple prescription for RLO and a calculated angular momentum response to re-synchronize donor stars to the rotation of the binary, they found similar curvature in the evolution of the MT rate with respect to time. While their model was not equipped to track overfilling factor, the substantial similarity of their results suggests the proportionality expressed in Equation 5.1 may be applicable to low mass binaries.

This threshold corresponds to an L1 mass flux of $\dot{M}_{L1} \sim 10^{-6} M_\odot/\text{yr}$. This is comparable to the maximum mass flux of wind RLO given by Mohamed and Podsiadlowski (2007) in a general case as well as the total mass flux of the donor wind in the case of M33 X-7. Either or both of these similarities in magnitude may be coincidental, and future research across a range of systems is required to provide detailed constraints on the mass flux and overfilling factor thresholds of unstable MT onset.

The latter models in our set extend past two critical points discussed previously. Model S passes the eclipse inflection point, meaning that for models S and T, the L1 tidal stream completely eclipses the donor star from the sightline of the accretor. Beyond the eclipse inflection, the eclipse radius of the donor is directly determined by the L1 region. This potentially compounds sensitivity to truncation method and tidal stream opacity assumptions. These conditions are only met in the final regime of fast unstable MT, however. As such, our model's choice of initialization from thermal equilibrium may significantly differ from the reality of the system, limiting the extent to which we can speak on the impact of these critical points.

Models S and T only vary from the trend of previous models very slightly, showing a slightly increased rate of change in the binary separation $d$. This result may not be statistically significant, given the limited number of models involved, but may be the result of the reduced rate of change of effective potential embodied by each corresponding step in overfilling factor. This shift is shown in Figure 1.2.

The donor surface of model T also surpassed the equipotential of the L2 point on the far side of the accretor; due to the change of mass ratio and orbital separation, this occurred at a higher overflow than the eclipse inflection. This theoretically enables gas with initially low velocities on the stellar surface to escape the accretor lobe at much higher rates. As model T only marginally exceeded the L2 potential for its geometry, the full effects of this excess potential may not be encapsulated by our system geometry. However, model T shows no significant variation from the trends of the previous models, including in efficiency. This may suggest MT invariance with respect to the relative potential of the Lagrange point on the far side of the accretor.

# CHAPTER 6

# CONCLUSIONS

This chapter integrates and adapts some work previously published in Dickson (2024) in Section 6.3.

In order to model the formation of the merging binary black holes now being explored by burgeoning gravitational wave observation techniques, we perform a detailed examination of Roche lobe overflow as a progenitor process. Not only is this process central to the most likely progenitor pathway of gravitational wave sources, but it is also critical to the progenation of common envelopes and the evolution of massive stars and stellar binaries as a whole. However, characterizing the mass and angular momentum loss of this highly dynamical and three-dimensional phase of binary evolution has proven computationally prohibitive. This limitation has led to a number of key assumptions and prescriptions that have yet to be thoroughly tested in contact binary simulation prior to this work. We present a methodology to resolve contact binaries in 3D without the use of prescriptive MT models.

While HMXBs and other contact binaries have been the subject of extensive theoretical and computational work, we here present the highest resolution model to date of both RLO and the associated tidally-fed disk dynamics. Our simulation holistically models the donor envelope, photosphere, wind, and L1 tidal stream, as well as the accretion disk surrounding the BH accretor, on a non-uniform grid.

We apply this method to the eclipsing HMXB M33 X-7 in two sets of simulations. We utilize a parameterized set to examine the L1 dynamics of the observed system parameters at two overfilling factors, $f = 1.01$ and $f = 1.1$. We also implement a sequential series of twenty simulations to examine system geometry, fluxes, timescales, and efficiencies across a piecewise evolution of non-uniform time increments in overfilling factor parameter space $1.0005 \leq f \leq 1.1681$. This was achieved by a piecewise evolution of timesteps equal to $0.01 \times \tau_{\text{Roche}}$, where $\tau_{\text{Roche}}$ is the Roche timescale characteristic to a given system state given by Equation 1.10.

## 6.1 Summary of Results

The astrophysical simulation tools we here present provide a computationally accessible method of obtaining 3D hydrodynamic solutions of binary systems such as HMXBs. This enables future research to more readily simulate binary dynamics corresponding to observations or theoretical systems to demystify these intricate astrophysical objects. We also obtain constraints on $\alpha_{\text{MT}}$ that enable future theoreticians to generate more physical and realistic simulations in their own research on binary stars, black holes, galactic enrichment, or accretion disks.

With this method, we provide greater constraints on the applicability of the computational approaches of Lubow and Shu (1975) and Marchant et al. (2021) to differing extremes of Roche lobe overflow. These semi-analytical models, and those closely related to them, are widely used in a range of relevant fields of research without sufficient validation from 3D hydrodynamic models allowed to freely vary.

We additionally test previous observations of the sample system M33 X-7, primarily those made most recently by Ramachandran et al. (2022). By a thorough analysis of the physical consequences of their observed overflow and mass accretion rate, we provide constraints on the plausible range of the observable HMXB population. We describe this observational window primarily through the differing durations of RLO stability regimes.

For the first time, we here quantify the efficiency, rate, and timescale of mass transfer mechanisms in simulated HMXBs. These quantities, as well as the associated efficiencies, rates, and timescales of angular momentum transport, are mapped across variations in overfilling factor. We vary overfilling factor across nearly the entirety of the MT phase from wind RLO prior to full RLO onset to within 15 years of phase transition into a common envelope, merger, or core collapse.

## 6.2 Primary Conclusions

1. We provide the first model of the efficiency of MT and the associated angular momentum as they vary with overfilling factor. We find unstable MT to be fully conservative in both mass and angular momentum transport onto the accretion disk and stable MT to be efficient (> 50%) at tidal stream formation and become fully conservative as it approaches $\dot{M}_{L1} \sim \dot{M}_{wind}$. We observe stable MT to be initially less efficient in transporting angular momentum, as shown in Figure 4.5. This calculation excludes the efficiency of transport after deposition onto the disk.

2. Stable MT can form at a mass ratio $q = 3.3 \gg 0.83$, well above the conventional stability limit. We observe both stable and unstable MT occurring at identical mass ratios, suggesting its invalidity as a predictor of MT stability. We find MT to transition between stability regimes across the evolution of a single system, occurring on timescales $1-2$ orders of magnitude shorter than $\tau_{Roche}$ as defined by Equation 1.10.

3. We provide the first quantitative model of the timescale of RLO, through which we identify wind, stable, and unstable RLO phases to be successive rather than alternative MT modes. Both wind RLO and stable full RLO occurred on roughly nuclear timescales of $\sim 300$ kyrs. For the system parameters of M33 X-7, the transition from wind RLO to stable full RLO occurred at an overflow of $1.0005 < f \leq 1.0083$ when the overflow became optically thick at L1. RLO remained stable until $f \sim 1.012$, and subsequently became increasingly unstable. Unstable RLO began on thermal timescales and became exponentially faster with higher overflow, for a total duration of $\sim 70$ kyrs.

4. At a threshold of approximately $f \sim 1.012$ and $\dot{M} \sim 10^{-6} M_\odot/\text{yr}$ in our system, the onset of unstable MT begins. This threshold corresponds to the point at which $\dot{M}_{donor}$ predominates the Roche timescale, as given by Equation 1.10. At $f \geq 1.012$, increasingly unstable MT generates increasingly deep tidal-feeding outflows in the donor envelope, differing from the assumption of e.g. Marchant et al. (2021) that the tidal stream is primarily fed by surface flows. These deep envelope outflows induce eddy currents that exhibit complete turnover by $f = 1.1$.

5. We provide a novel relation for orbital geometry during RLO. We find the relation $q \propto d \propto f^{D_1}$ to hold across the entire series of piecewise evolution of MT in M33 X-7 with $D_1 = -0.815$, where $q$ and $d$ respectively denote the mass ratio and binary separation, and $f$ denotes the overfilling factor defined by Equation 1.7.

6. We characterize the relationship between overfilling factor and both timescale and mass flux of RLO-induced MT for the first time. In the $f \geq 1.012$ regime, we find $\dot{M}_{\text{L1}} \propto \tau_{\text{Roche}}^{-1} \propto \log_{10}^{D_2}(f)$, where $D_2 = 4.42$ in our test system. This relation is invariant across the eclipse inflection and potential in excess of the accretor-side L2.

7. We find the $f \geq 1.1$ regime to be exceptionally brief and distinctly unstable, such that RLO phase could not persist for timescales greater than 100 years after reaching $f = 1.1$. As such, M33 X-7 is unlikely to be as substantially overflowing as predicted by Ramachandran et al. (2022). The formation of L2/L3 secondary tidal streams predicted by Marchant et al. (2021) is similarly unlikely outside of temporary transient states in physical systems comparable to our simulations.

8. We find the semi-analytical model of Lubow and Shu (1975) to be valid to within an order of magnitude in the low overflow ($f \lesssim 1.01$) regime. At high overflow, two of their widely-used prescriptions appear to remain valid; mass crossing the L1 point does so at Mach 1 and carrying angular momentum roughly equivalent to that of a circular orbit of the system center of mass. Their model geometry otherwise breaks down in the high overflow regime as the tidal stream forms a conical funnel.

9. The updated model of Marchant et al. (2021) remains valid to within an order of magnitude in both low and high overflow regimes we modeled except as noted in points 4 and 7. This includes their prediction that high overflow strips the donor of significant angular momentum and therefore induces orbital hardening without requiring a CE phase.

10. We find the accretion disk to spontaneously clear a lower-density disk gap around $R_{\text{gap}} \sim R_\odot$, forming a banded structure at both high and low overflow. Mass flux into the inner band of the disk is still predominantly fed by the outer band of the disk.

11. Wind RLO can form and dominate binary interaction at low overfilling factors $f > 1$. We observe wind RLO at $f = 1.0005$ launch with a system MT efficiency of $\leq 3\%$. While this wind RLO exhibits stable MT, it directly induced the transition into full RLO and unstable MT onset on the order of the nuclear timescale.

12. M33 X-7 is likely to undergo stable MT on nuclear timescales followed by a transition into a subsequent unstable MT phase on thermal timescales, unless interrupted by donor evolution. Regardless of present stability, the system appears to be near its Eddington luminosity, and potentially in the super Eddington range.

## 6.3 Future Work

Future work may improve upon our method by applying additional physics to the dynamics and simulating a wider range of RLO systems. Our radiative envelope solution does not account for stellar evolution due to the short timescale of the simulation; future research could incorporate MESA or a similar stellar evolution code into system initialization to model varied stellar evolutionary stages (Paxton et al. 2010). At the transition from the isothermal photosphere to the linearly-increasing temperatures of the envelope, we see winds launch naturally from the star. These winds follow the Sobolev approximation for linearly-driven stellar winds, but we acknowledge the asymmetry of our donor star limits the applicability of this approximation. Further research could expand upon this by more physically constraining the wind dynamics. The greatest limitation of our work is its lack of accounting for X-ray feedback and relativity in the near-BH region. The applicability of this model could be expanded by incorporating a more complete radiative transport implementation. Particularly to near-Eddington and super-Eddington systems, the role of relativistic jets, X-ray heating, and radiation pressure could significantly impact the accretion dynamics. At the short timescale of the 1.1-case Roche evolution, the donor may not maintain corotation with the binary system. This could impact the effective potentials and therefore the MT rates, efficiencies, and geometry. Further research may improve constraints upon the high overflow regime by examining the system without the assumption of a corotating donor.

# APPENDICES

# APPENDIX

# A

# VARIABLES

**Table A.1** List of Variables Used in this Work. We list those variables we explicitly defined with the equation number of their defining equations. We universally apply dot notation to represent time derivatives. Arrows indicate vectors, while hats indicate unit vectors.

| Symbol | Variable | Equation |
|:---:|:---:|:---:|
| $a_{\text{eff}}$ | Acceleration due to effective gravitational potential | 3.4 |
| $A_{\text{fringe}}$ | Total surface area of the fringe zones | |
| $A_{\text{non-fringe}}$ | Total surface area of the non-fringe zones | |
| $c$ | Speed of light | |
| $C_{\text{fringe}}$ | Fringe weighting coefficient | 3.6 |
| $C_{\text{schw}}$ | Schwarzschild variable scaling coefficient | 3.12 |
| $c_s$ | Sound speed | |
| $d$ | Binary separation | 1.3 |
| $D_1$ | Exponent relating separation and mass ratio to $f$ | 4.1 |
| $D_2$ | Exponent relating mass loss and timescale to $f$ | 5.1 |
| $E$ | Total energy | 3.1 |

**Table A.1** (continued)

| | | |
|---|---|---|
| $e_i$ | Internal potential energy per unit mass | |
| $f$ | (Over)filling factor | 1.5 & 1.7 |
| $G$ | Gravitational constant | |
| $g$ | Surface gravity | |
| $H_p$ | Local pressure scale height | |
| $J_{tot}$ | Angular momentum of the total system | 1.9 |
| $L$ | Luminosity | |
| $L_{acc}$ | Angular momentum of the accretor | |
| $L_{donor}$ | Angular momentum of the donor | |
| $\dot{L}_{L1}$ | Angular momentum flux of the L1 tidal stream | |
| $\dot{L}_{stream,out}$ | Angular momentum flux of the accretor outflow stream | |
| $m$ | Mass enclosed | 3.9 |
| $M_{acc}$ | Mass of the accretor | |
| $M_{donor}$ | Mass of the donor | |
| $\dot{M}_{L1}$ | Mass flux of the L1 tidal stream | |
| $\dot{M}_{stream,out}$ | Mass flux of the accretor outflow stream | |
| $M_{tot}$ | Total mass of binary system | |
| $M_\odot$ | Mass of the sun | |
| $P$ | Pressure | |
| $q$ | Mass ratio | 1.4 |
| $r$ | Radius enclosed | 3.9 |
| $R^2$ | Coefficient of determination | |
| $R_1$ | Distance from a point to the donor | |
| $R_2$ | Distance from a point to the accretor | |
| $R_3$ | Distance from a point to the center of mass | |
| $R_a$ | Characteristic radius of accretion | 1.6 |
| $R_b$ | Inner radius of convective zone | |
| $R_{donor}$ | Radius of the donor | |
| $R_{ecl}$ | Eclipse radius | |
| $R_{norm}$ | Radius normalized to the binary separation | |
| $R_{RL}$ | Radius of the (donor) Roche lobe | 1.2 |
| $R_{vol}$ | Volume-equivalent radius | |
| $R_\odot$ | Radius of the sun | |

**Table A.1** (continued)

| Symbol | Description | Ref |
|---|---|---|
| $\mathfrak{R}$ | Gas constant | |
| $\mathfrak{R}_o$ | Rossby number | 2.3 |
| $SH$ | Photospheric scale height | 3.7 |
| $t$ | Time | |
| $T$ | Temperature | |
| $T_{\text{orbit}}$ | Orbital period | |
| $T_{\text{rot}}$ | Period of rotation | |
| $\Delta t_{\text{settling}}$ | Hydrodynamic grid settling time | |
| $v$ | Velocity | |
| $v_{\text{conv}}$ | Velocity at half-height of the convective zone | |
| $v_{\infty}$ | Terminal velocity of donor wind | |
| $X$ | Mass fraction | |
| $\alpha_{\text{AM(L1)}}$ | Efficiency of L1 tidal stream angular momentum transfer | 1.12 |
| $\alpha_{\text{AM(tot)}}$ | Efficiency of total system angular momentum transfer | 1.11 |
| $\alpha_{\text{MLT}}$ | Mixing length parameter | |
| $\alpha_{\text{MT(L1)}}$ | Efficiency of L1 tidal stream mass transfer | 1.12 |
| $\alpha_{\text{MT(tot)}}$ | Efficiency of total system mass transfer | 1.11 |
| $\gamma$ | Adiabatic Index | |
| $\Theta_{\text{crit}}$ | Smallest angular resolution represented on-grid | |
| $\kappa$ | Opacity | 3.8 |
| $\mu$ | Mean molecular weight | |
| $\rho$ | Density | |
| $\tau$ | Optical Depth | |
| $\tau_{\text{conv}}$ | Convective turnover time | 2.2 |
| $\tau_{\text{dyn}}$ | Timescale of adiabatic dynamical flow | 2.1 |
| $\tau_{\dot{L}}$ | Timescale of system angular momentum loss | |
| $\tau_{\text{KH}}$ | Kelvin-Helmholtz timescale | 1.13 |
| $\tau_{\text{rad}}$ | Timescale of radiative diffusion | 2.1 |
| $\tau_{\text{Roche}}$ | Timescale of change in donor Roche lobe radius | 1.10 |
| $\phi_{\text{eff}}$ | Effective gravitational potential, offset by $\phi_{\text{L1}}$ | 1.1 |
| $\phi_{\text{L1}}$ | Effective gravitational potential at L1 | |
| $\Omega$ | Orbital angular velocity | |

# APPENDIX B

# ADDITIONAL FIGURES

We here present all twenty sequential models as fit by the same isodensity surface ($10^{-11}$g/cm$^3$) also employed in Figures 4.2 and 4.6. Unlike model A, model B is optically thick at the L1 point; however, its stream does not remain optically thick throughout its approach to the BH, necessitating additional visualization in the ecliptic plane given in Figure B.1.

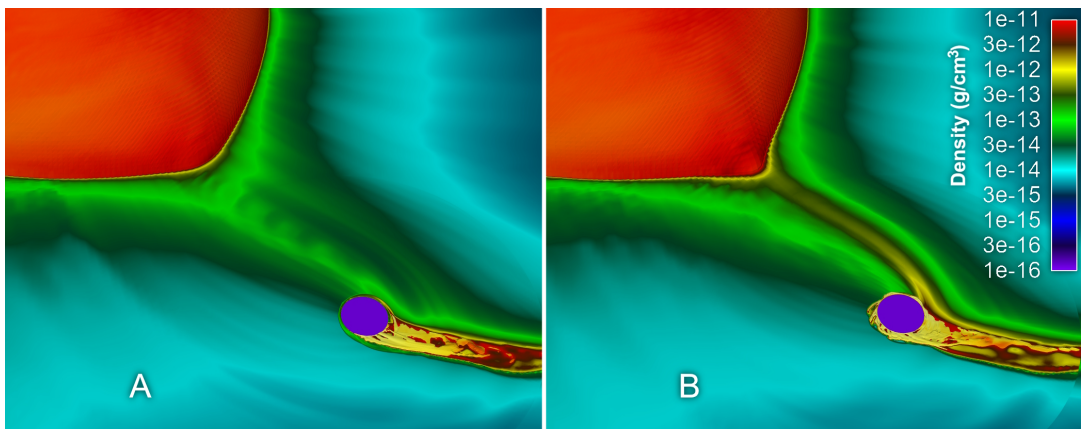

**Figure B.1** Binary System Visualized for Models A-B.

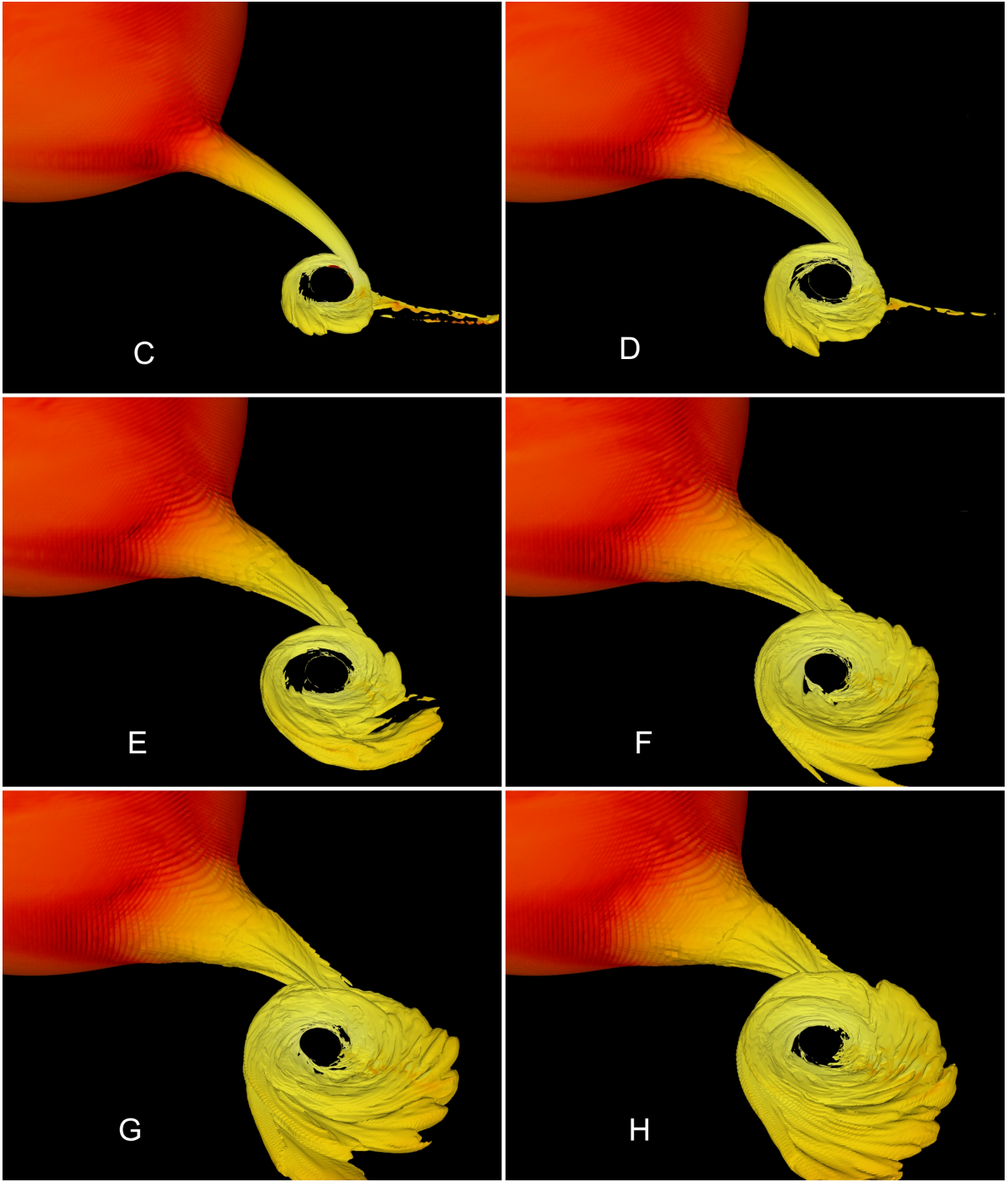

**Figure B.2** Binary System Visualized for Models C-H.

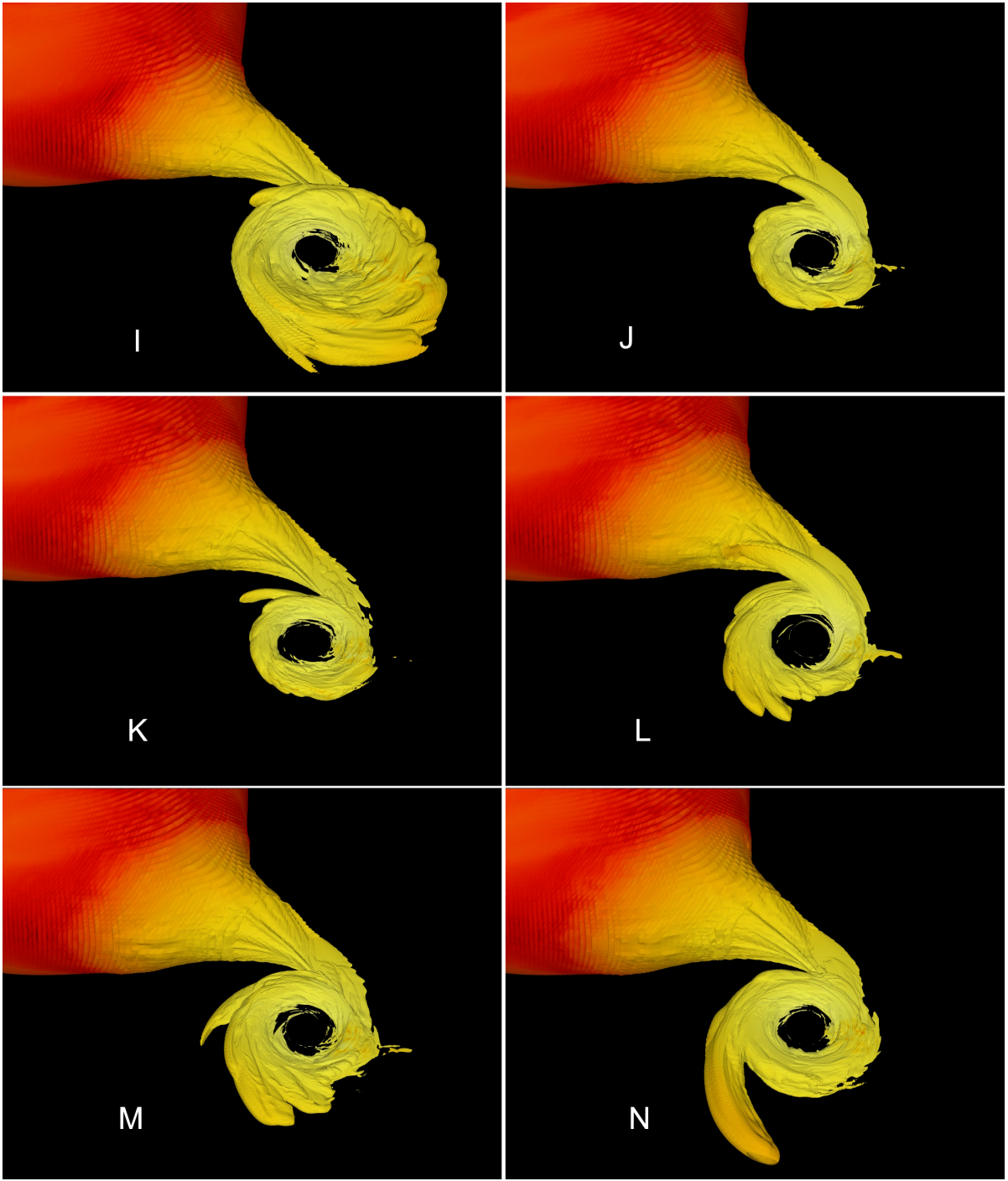

**Figure B.3** Binary System Visualized for Models I-N.

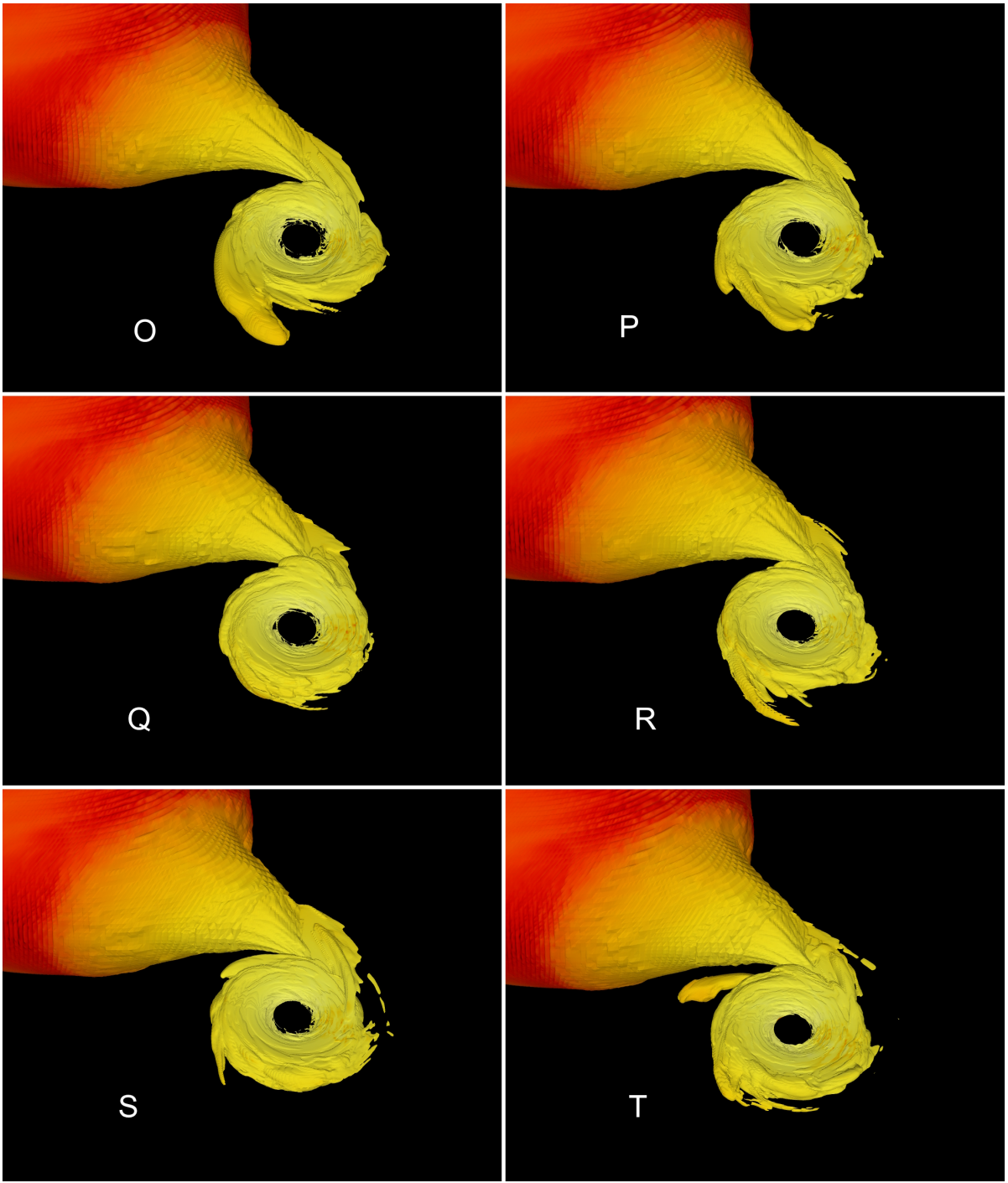

**Figure B.4** Binary System Visualized for Models O-T.

# APPENDIX

# C

# ACRONYMS

Table C.1  List of Acronyms Used in this Work.

| Abbreviation | Acronym |
|:---:|:---:|
| AM | Angular Momentum |
| BBH | Binary Black Hole |
| BH | Black Hole |
| CE | Common Envelope |
| CEE | Common Envelope Evolution |
| He-WD | Helium-rich White Dwarf Stars |
| HMXB | High-Mass X-ray Binary |
| MESA | Modules for Experiments in Stellar Astrophysics |
| MPI | Message Pathing Interface |
| MT | Mass Transfer |
| RLO | Roche Lobe Overflow |
| SPH | Smoothed Particle Hydrodynamics |
| VH-1 | Virginia Hydrodynamics One |